\documentclass[usenatbib,a4paper]{mn2e}
\pdfoutput=1
\usepackage[letterpaper,total={17.8cm,24.0cm},centering]{geometry}
\usepackage{graphicx}
\usepackage{amsmath}
\usepackage{url}

\newcommand{\hi}{\mbox{H\,{\sc i}}}
\newcommand{\msun}{$M_{\odot}$}

\newcommand{\kms}{~km\,s$^{-1}$}
\newcommand{\apj}{ApJ}
\newcommand{\apjl}{ApJL}
\newcommand{\apjs}{ApJS}
\newcommand{\aj}{AJ}
\newcommand{\aap}{A\&A}
\newcommand{\aaps}{A\&AS}
\newcommand{\nat}{Nature}
\newcommand{\mnras}{MNRAS}
\newcommand{\pasj}{PASJ}

\begin{document}

% For mn2e:
\title[KAT-7: Cold Gas, Star Formation, Substructure in Antlia]%
    {KAT-7 Science Verification: Cold Gas, Star Formation, and Substructure in the Nearby Antlia Cluster}
\author[K. M. Hess et al.]%
    {Kelley M. Hess,$^{1,2,3}$\thanks{hess@astro.rug.nl}
    T. H. Jarrett,$^1$ Claude Carignan,$^{1,4}$ Sean S.~Passmoor,$^5$ 
    \newauthor Sharmila Goedhart$^{5,6}$ \\
    $^1$Astrophysics, Cosmology, and Gravity Centre (ACGC), Department of Astronomy, University of Cape Town, Private Bag X3, \\ Rondebosch 7701, South Africa \\
    $^2$Kapteyn Astronomical Institute, University of Groningen, PO Box 800, 9700 AV Groningen, The Netherlands \\
    $^3$ASTRON, the Netherlands Institute for Radio Astronomy, PO Box 2, 7990 AA Dwingeloo, The Netherlands \\
    $^4$Laboratoire de Physique et Chimie de l'Environnement (LPCE), Observatoire d'Astrophysique de l'Universit{\'e} de Ouagadougou \\ (ODAUO), Ouagadougou, Burkina Faso \\
    $^5$Square Kilometre Array South Africa, The Park, Park Road, Pinelands 7405, South Africa \\
    $^6$School of Physics, North-West University, Potchefstroom campus, Private Bag X6001, Potchefstroom, 2520, South Africa
    }

\maketitle

\begin{abstract}

The Antlia Cluster is a nearby, dynamically young structure, and its proximity provides a valuable opportunity for detailed study of galaxy and group accretion onto clusters.  We present a deep \hi\ mosaic completed as part of spectral line commissioning of the Karoo Array Telescope (KAT-7), and identify infrared counterparts from the \textit{WISE} extended source catalog to study neutral atomic gas content and star formation within the cluster.  We detect 37 cluster members out to a radius of $\sim$0.9 Mpc with $M_{HI}>5\times10^7$\msun. Of these, 35 are new \hi\ detections, 27 do not have previous spectroscopic redshift measurements, and one is the Compton thick Seyfert II, NGC~3281, which we detect in \hi\ absorption.  The \hi\ galaxies lie beyond the X-ray emitting region 200 kpc from the cluster center and have experienced ram pressure stripping out to at least 600 kpc. At larger radii, they are distributed asymmetrically suggesting accretion from surrounding filaments.  Combining \hi\ with optical redshifts, we perform a detailed dynamical analysis of the internal substructure, identify large infalling groups, and present the first compilation of the large scale distribution of \hi\, and star forming galaxies within the cluster.  We find that elliptical galaxy NGC~3268 is at the center of the oldest substructure and argue that NGC~3258 and its companion population are more recent arrivals.  Through the presence of \hi\ and on-going star formation, we rank substructures with respect to their relative time since accretion onto Antlia.
\end{abstract}


\section{Introduction}

Galaxy clusters are the product of collapse, assembly, and merging of dark matter haloes over cosmic time in a $\Lambda$ cold dark matter universe \citep{Springel05}, and we know that hierarchical structure formation is at work in early times as galaxy clusters are well established by at least $z=1$ (e.g.~\citealt{Fassbender11,Gettings12} and references therein).  Galaxies may be accreted on to clusters individually or as members of galaxy groups.  On the outskirts of clusters, galaxy transformation occurs on relatively rapid time-scales, while within the cluster, dynamical friction smoothes out galaxy distributions, so the evidence of cluster assembly decays with time.  None the less, nearly all clusters display an element of youth through the presence of substructure: a signature of their merger history.  Understanding the assembly history of clusters through cosmic time is important for both understanding the evolution of large scale structure, and the properties of the galaxy population within it.  

Theoretically, 30\% of all systems should contain substructure, although observations report a range of 27--73\% for massive groups (more than 20 members) and clusters \citep{Hou12,Bird94,Dressler88,Ramella07}.  Substructure tends to be anti-biased with respect to the dark matter distribution, preferentially surviving on the outskirts of systems \citep{Oguri04}, but it is unclear how long in-fallen substructure may survive as both intermediate and low redshift clusters can host density enhancements and multiple massive galaxies (e.g.~\citealt{Beers83}).
Even in the interior of clusters, cD galaxies typically reside at the center of galaxy density enhancements rather than at the center of the clusters \citep{Beers91,Bird94,Zabludoff90}, suggesting themselves originated from the merger of groups \citep{Zabludoff98}.  

In general, substructure is identified through line-of-sight velocities, and the spatial distribution of small collections of galaxies relative to the overall cluster population \citep{Bird94,Dressler88,Serna96,Jaffe12,Pranger13}.  
Despite a bevy of statistical tools at our disposal, we are still striving to understand to what degree clusters are dynamically relaxed or are continuing to accrete from their environment, even at low redshift \citep{Hou14}.

A spectroscopic census of the stellar content of clusters illuminates these substructures, however the presence of atomic gas, measured directly by \hi\ 21 cm observations, not only provides a complimentary evidence of substructure \citep{BravoAlfaro00,Jaffe12}, but also indicates which galaxies have most recently accreted from the surrounding environment. 
Late-type galaxies in clusters are \hi\ deficient with respect to their counterparts in the field (e.g.~\citealt{Haynes84,Solanes01}).  They have shrunken \hi\ discs where X-ray emitting hot intracluster gas fills the cluster potential well \citep{BravoAlfaro00,Chung09} and high resolution \hi\ images of Virgo cluster members show that the hot intracluster medium is responsible in many cases for removing the gas through ram pressure stripping \citep{Kenney04,Crowl05,Chung09,Jaffe15}.  In most cases, \hi\ objects avoid the centers of clusters all together.

The removal of the atomic gas, in the long run, leads to the shut down of star formation and a color transformation \citep{Schawinski14}, and environment is responsible for quenching star formation in galaxies across a range of parent dark matter halo masses (e.g.~\citealt{Woo13,Wetzel12}).
However, the impact of the cluster environment on star formation within an individual galaxy is not straight forward: the \textit{fraction} of star forming galaxies in clusters increases with radius until reaching the field value at 2--3 virial radii, but the mean star formation \textit{rate} shows no radial dependence \citep{Rines05}.  Similarly, despite cluster galaxies being \hi\ deficient, CO, which traces the cold H$_2$ out of which stars form, shows no environmental dependence \citep{Kenney89,Casoli91}.  Thus, quenching may be delayed if galaxies have a sizeable reservoir of molecular gas.  These complications persist despite the fact that the distance at which a galaxy resides from the cluster center is correlated with the time since infall \citep{deLucia12}.  Combining \hi\ and star formation proxies, a cluster member which has star formation, but is not detected in \hi\ may indicate an intermediate age cluster member, stripped of its \hi\, but converting molecular gas to stars.

The infall region of clusters is an interesting volume in which to study galaxy evolution because the rapid change in galaxy density and properties of the intergalactic medium between the field and cluster environment have a strong impact on the stars and interstellar medium.  However, a number of studies show that local density is a more important factor determining the present day star formation and gas content of galaxies than large-scale environment suggesting that ``pre-processing'' is important in intermediate density environments before galaxies arrive in clusters (e.g.~\citealt{Rines05,Lane07,Blanton07,Hess11}). Disentangling the relative importance of the ``pre-processing'' that happens to galaxies in groups before they are accreted (e.g.~\citealt{Zabludoff98,Hess13,Berrier09}), versus the impact of the cluster environment is an ongoing challenge.

\citet{Hou14} tackle this by combining optical group catalogs with the Dressler-Shectman statistical test \citep{Dressler88} to identify sub-haloes which they classify as infall, backsplash, or virialized.  They find evidence for ``enhanced quenching'' of the star formation in infalling subhaloes, which they attribute to ``pre-processing'' in group haloes, but that it is only important for the most massive accreting systems, $>10^{14.5}$\msun.   
This is unlikely to be the complete picture, but has provided insight into the most massive systems, and the mergers of haloes with the highest mass ratios. 
Complementarily, studies of the \hi\ content of galaxies and semi-analytical models show that environmental effects become important in groups with $>10^{13}$\msun\ of dark matter \citep{Jaffe12,Hess13,McGee09} and \citet{Jaffe15} show gas-poor galaxies in the infall region of Abell 963 were likely pre-processed in their previous environment.

Estimates for the contribution of cluster stellar mass that has been accreted through galaxy groups (and the redshift at which this is important) range from 12\% to 50\% \citep{Berrier09,McGee09,deLucia12}.  
When galaxies fall into clusters as groups, pre-processing which has already occurred in the group environment likely influences the rate at which quenching or morphological transformation occurs.  However, if galaxies fall in as individuals from the field, the interaction with the intracluster medium may induce quenching on more rapid time scales.  Understanding these different evolutionary paths may help us further understand the resulting cluster populations and the morphological transformation towards early-type and S0 galaxies \citep{Schawinski14}.

The number of nearby galaxy clusters in which we can perform a detailed, resolved study of both the stellar and gaseous component of galaxies is limited, but include systems with a range of apparent dynamical ages from quite old such as Coma, through Virgo and Fornax, to quite young such as Abell 1367.
At $\sim$40 Mpc, the Antlia Cluster (Abell S0636) is the third most nearby cluster after Virgo and Fornax, yet it has been poorly studied due to its relatively low Galactic latitude ($b=19^{\circ}$) and southern declination 
%($10^h30^m03.6^s -35^{\circ}19^{\prime}24^{\prime\prime}$).  
($\delta=$ -35.3$^{\circ}$).  It was not covered by the \hi\ Parkes Zone of Avoidance Survey \citep{KraanKorteweg02}, and only recently has there been a significant survey of the stellar content of the cluster \citep{SmithCastelli08,SmithCastelli12}.  \citet{Vaduvescu14} provide an excellent review of the optical photometric, spectroscopic, and X-ray campaigns of the Antlia Cluster, which are very incomplete.
  
From what we do know, Antlia is a compelling target for further study.  \citet{Ferguson90} found that the galaxy density is 1.7 times higher than Virgo and 1.4 times higher than Fornax. \citet{SmithCastelli08} note that S0s outnumber elliptical galaxies by a factor of three, contrary to Virgo and Fornax where ellipticals dominate.  Together, this suggests that Antlia is the youngest of the three systems.  Further, if ellipticals and S0s have different formation mechanisms, it may indicate that the clusters formed differently, if not only on different time-scales.

Overall, the cluster contains an estimated 375 galaxies \citep{Ferguson90}, including at least two dozen confirmed early-type dwarfs \citep{SmithCastelli12}.
At the center of Antlia are two concentrations of galaxies focused around the massive ellipticals, NGC~3268 and NGC~3258.  They have the same systemic velocity, but these ellipticals each host an extended globular cluster population \citep{Dirsch03}, and hot X-ray halo \citep{Nakazawa00,Pedersen97}, suggesting that the galaxies are the product of past mergers, and they sit at the center of their own subhalo.  Antlia itself lacks a central X-ray excess that is typical of poor clusters, but the X-ray emission and globular cluster systems around NGC~3268 and NGC~3258 are extended along the line connecting the two galaxies, suggesting that now the two systems reside within the same dark matter halo.

Previous studies of Antlia detail the properties of the oldest merged components within the cluster: early-type galaxies at the center embedded in hot X-ray haloes, and their early-type dwarf companions.
In this work we combine \hi\ spectral line commissioning observations from the Karoo Array Telescope (KAT-7) with infrared proxies of the stellar mass and star formation rate of galaxies from the Widefield Infrared Survey Explorer (\textit{WISE}), and optical spectroscopic redshifts to disentangle the history of mass assembly.  The \hi\ indicates recent infall to the cluster, and star formation persists even in galaxies where the gas content has been exhausted or dropped below current detection limits.  A deeper understanding of Antlia, and its evolutionary state provides an important addition to the detailed knowledge of other nearby clusters.

Throughout this paper we adopt for the cluster a heliocentric velocity of 2797\kms\ ($z=0.00933$; \citealt{SmithCastelli08}). We assume a concordance cosmology with $H_0 = 73$ \kms\ Mpc$^{-1}$, $\Omega_M = 0.27$, $\Omega_{\Lambda}=0.73$, putting Antlia at a distance of $38.1\pm2.7$ Mpc, corrected for Virgo, the Great Attractor, and the Shapley supercluster.

\section{KAT-7 Observations \& Commissioning}

The seven dish Karoo Array Telescope (\citealt{Carignan13}, Foley et al., submitted) is a test bed instrument for the South African precursor to the Square Kilometre Array (SKA), MeerKAT.  We observed the Antlia Cluster with KAT-7 using a seven pointing mosaic centered on NGC~3268 ($\rmn{RA}(2000)=10^{\rmn{h}}30^{\rmn{m}}03.6^{\rmn{s}}$, $\rmn{Dec.}(2000)=-35\degr 19\arcmin 23\farcs 88$).  The pointings were arranged in a discrete hexagonal pattern in the style of the NRAO VLA Sky Survey (NVSS; \citealt{Condon98}) with pointing centers separated by $1/\sqrt{2}$ times the half power beam width\footnote{\url{https://science.nrao.edu/facilities/vla/docs/manuals/obsguide/modes/mosaicking}}.  The total area of the mosaic covers approximately 4.4 deg$^{2}$.  

These observations were done as part of commissioning the ``c16n25M4k'' \hi\ spectral line correlator mode of KAT-7 which features 4096 channels covering 25 MHz, with 6.104 kHz spectral resolution ($v=1.29$\kms\ at $z=0$).  We centered the band at 1406.7 MHz (2922\kms) and cycled through the full mosaic with four minutes per pointing between three minute observations of the complex gain calibrator, 1018-317.  The bandpass calibrators, 3C138 and PKS 1934-638, were observed once every two hours. The data were taken between March 2013 and July 2013 over 19 observing sessions. Sessions typically spanned 9.5--12.75 hours for a total of 192 hours, including 147 hours on the mosaic with 21 hours per pointing.  Table \ref{obs} provides a summary of the observing and imaging parameters.  

\begin{table}
\caption{KAT-7 Observations \& Imaging}
\begin{tabular}{ll}
\hline
Observing Parameter & Value \\
\hline
Date of Observations & March--July 2013 \\
Total integration time ($7\times21$ h) & 147 h \\
Total bandwidth & 25 MHz \\
Central frequency & 1406.7 MHz \\
Spectral resolution & 6.104 kHz \\
Flux/Bandpass calibrator & 3C138, PKS 1934-638 \\
Phase/Gain calibrator & 1018-317 \\
\hline
Image Property & \\
\hline
Mosaic area & $\sim$4.4 deg$^2$ \\
Velocity coverage & 1159--4628\kms\\
Velocity resolution & 15.5\kms \\
Synthesized beam & $3.7^{\prime}\times3.0^{\prime}$ \\
Final RMS & 0.97 mJy beam$^{-1}$ \\
\hline
Velocity resolution & 31.0\kms \\
Synthesized beam & $3.3^{\prime}\times3.1^{\prime}$ \\
Final RMS & 1.2 mJy beam$^{-1}$ \\
\hline
\end{tabular}
\label{obs}
\end{table}

\subsection{Data Calibration}
\label{datacalib}

Data calibration was done in CASA 4.1.0 \citep{McMullin07} using standard methods.  Each observing session was flagged and calibrated individually at full spectral resolution.  We split off the central 2700 channels, averaged every three channels, and flagged again where necessary.  KAT-7 does not doppler track, so we applied a velocity correction to the barycentric reference frame as described in \citet{Carignan13}.  We combined the 19 data sets and averaged a further 4 channels, such that the spectral resolution of the final data is 15.5\kms.

Over the course of the five months in which we observed, Antlia went from night time observations, to day time observations.  In this period we noticed an increase in the amount of flagging required, particularly along short baselines, due to the impact of solar interference.  However, the fraction of the data flagged was low and although the solar interference was broadband, it primarily affected data on the lower frequency edge of the band, which was mostly discarded when we split the data.

We performed continuum subtraction with a first order polynomial to the visibilities using the {\sc uvcontsub} task in CASA.
We imaged the cluster using the ``mosaic'' imager mode in the {\sc clean} task with robust parameter of 0, and produced two data cubes of 15\kms and 31\kms\ spectral resolution.  The final mosaics cover a roughly circular region 155 pixels in diameter, corresponding to $\sim1.7$ Mpc at the distance of Antlia, and spanning 228 and 114 channels, respectively.  The final restoring beams are $3.7 \times 3.0$ and $3.3 \times 3.1$ arcminutes corresponding to a spatial resolution of $41 \times 33$ kpc and $36 \times 34$ kpc, respectively.

\subsection{Additional Processing: $u=0$}

Despite careful inspection and flagging of the visibilities, after significant averaging, low level artefacts remained in the image cubes.  Spatially, we saw horizontal stripes in every channel whose amplitude was comparable to the noise.  The dominant mode had a characteristic size of order 1440 arcseconds peak-to-peak, corresponding to $\sim$30 m projected baseline.  In the spectral dimension they appeared as a sinusoidal variation whose phase varied as a function of position in the mosaic.  The period of the ripples was of order 4 MHz.  We believe the artefacts are two manifestations of the same issue, known as the ``$u=0$ problem''.

Interferometer baselines sample varying regions of the UV-plane throughout an observation.  When baselines cross the $u=0$ axis, the fringe rate is equal to zero so incoming signal from the surrounding environment, which is usually added incoherently because of the geometric delay between antennas, is correlated.  Anything that is bright enough to be seen in the visibility data is flagged during calibration, so what is left in the image cube is very low level radio frequency interference (RFI) that adds up to be visible after a significant amount of integration time or averaging.  This problem affects short baselines more than long baselines, and can be seen in KAT-7 observations after as little as about 8 hours.  The source of RFI can be internal or external to the telescope itself, and it is now under investigation how to best mitigate it in MeerKAT while maintaining both interferometric and beam forming mode capabilities.

The solution to remove these artefacts was to Fourier Transform (FT) each of the image cubes, mask the $\pm2$ pixels on either side of the u=0 axis in the UV plane, and perform the reverse Fourier Transform back to the image plane.  The first FT produces a real and an imaginary UV cube which corresponds to the real and imaginary part of the gridded visibilities.  The reverse FT, after masking, produces an amplitude and a phase image cube, but the phases are expected to be zero in the planar approximation because emission from the sky is real.  In the resulting image cubes, the horizontal stripes were successfully removed, and the sinusoidal variation of the bandpass was significantly reduced.  The rms noise improved by 26\%.

\subsection{Bandpass Fitting, Cleaning, \& Imaging}

Due to the detection nature of our observations, we did not know {\it a priori} which channels were line-free.  Therefore, we performed a second fit to the bandpass in the image plane.  The fit was performed by masking pixels in the cubes whose absolute values were greater than $2\sigma$, thereby removing emission and absorption features from influencing the fit. The pixel columns were then fit with a spline function which was subtracted from the image, removing residual structure in the bandpass.  

Finally, bright sources were cleaned in the standard way to remove sidelobes.  Overall, the rms noise improved by 29\%.  In the 15\kms cube, we achieved a final value of 0.97 mJy beam$^{-1}$ channel$^{-1}$, corresponding to a 3$\sigma$ \hi\ mass sensitivity of $4.49\times10^7$\msun\ over 45\kms.  In the 31\kms cube, the rms noise improved by 32\% and we achieved a value of 1.2 mJy beam$^{-1}$ channel$^{-1}$, corresponding to $7.4\times10^7$\msun\ over 60\kms.

\section{Ancillary Data}

To compliment the radio observations we compiled spectroscopic redshifts from the literature and online for all objects in the field, and used the NASA Extragalactic Database (NED)\footnote{\url{http://nedwww.ipac.caltech.edu/}} to crossmatch \hi\ detections with optical, infrared, and ultraviolet sources.  In addition, we have extracted fluxes from the \textit{WISE} survey \citep{Wright10} for extended sources in the KAT-7 mosaic.

\subsection{Optical Velocity Measurements}
\label{vel}

The velocities for the brightest optical cluster members come from \citet{Ferguson90} who compiled the work of \citet{Lauberts82}, \citet{Hopp85}, and \citet{Sandage87}.  Velocity information on an additional 25 early-type galaxies comes from \citet{SmithCastelli08} who added velocities from 6dF Galaxy Survey (6dFGS; \citealt{Jones04}) and their own spectroscopic measurements.  \citet{SmithCastelli12} observed an additional 22 dwarf elliptical and dwarf spheroidal galaxies from the central region of Antlia.  Finally, we have included 28 more unique redshifts from the final 6dFGS catalog\footnote{\url{http://www-wfau.roe.ac.uk/6dFGS/}} \citep{Jones09} within a roughly 1.4 degree radius from NGC~3268 in the redshift range of the cluster, 1200--4200\kms\ \citep{SmithCastelli08}.

In total, we find 97 cluster members with optical redshift measurements in the literature and online.  For sources with multiple redshifts measurements, we use those from \citet{SmithCastelli08} or 6dFGS since they provide the largest uniform sample.  This results in a slightly narrower velocity dispersion derived for the cluster in Section \ref{veldisp} than if we preferentially use \citet{Ferguson90} redshifts. The estimated uncertainty for individual 6dF redshifts is better than 55\kms\ \citep{Jones09}.

\subsection{\textit{WISE} Data}
\label{wise}

\textit{WISE} is a confusion limited survey which observed the entire sky in four mid-infrared bands: 3.4 $\mu$m, 4.6 $\mu$m, 12 $\mu$m, and 22 $\mu$m, conventionally known as \textit{W1}, \textit{W2}, \textit{W3}, and \textit{W4} \citep{Wright10}.  The \textit{W4} band performance has recently been updated \citep{Brown14} and we apply the new bandwidth performance in which the central wavelength is closer to 23 $\mu$m.  In the local (rest) universe the \textit{WISE} bands are sensitive to: evolved stars, silicate absorption, PAH molecular bands associated with star formation, warm ISM dust from reprocessed star formation and AGN disc accretion, respectively (\citealt{Cluver14}, and references therein). The 3.4-4.6 $\mu$m color has proven to be a powerful tool to measure stellar mass in the nearby Universe \citep{Jarrett13,Cluver14}, and in particular the 12 $\mu$m flux is a relatively useful estimate of star formation rate in star forming galaxies (see also \citealt{Donoso12}).  It is most reliable at high rates of dust-obscured star formation, but shows correlation to low star formation rates (relative to H$\alpha$), albeit with increasing scatter.

Large image mosaics ($\sim$3 sq degrees) of the KAT-7 field were constructed from \textit{WISE} individual image frames and using the ``drizzle'' method described in \citet{Jarrett12}.  The resulting images are more sensitive to resolved galaxy detection and characterization compared to the resolution degraded public-release \textit{WISE} mosaics.  Given the close proximity of the Antlia Galaxy Cluster and the $\sim$6 arcsecond angular resolution of the \textit{W1} mosaics, we expect most cluster members to be resolved by \textit{WISE}. 

The \textit{WISE} detection sensitivity depends on the number of orbit crossings. In the case of the Antlia field, the \textit{W1} and \textit{W2} bands have a coverage of $\sim$27 orbits, and the longer bands have about half of that amount.  Accordingly, in the 3 sq.~degree area \textit{WISE} detects approximately 480 extended sources in the most sensitive \textit{W1} band using the pipeline developed for \textit{WISE} galaxy characterization \citep{Jarrett13}.   In the area covered by the KAT-7 mosaic, \textit{WISE} detects approximately 240 resolved sources.  These include both Antlia and background galaxies.  The \textit{W1} detections are limited to $S/N = 10$, reaching a limiting Vega magnitude of 16.05 (117 $\mu$Jy).  The other bands, carried by the \textit{W1} detection, have extractions that are limited to S/N = 3, and the corresponding limiting magnitudes are respectively (\textit{W2}, \textit{W3}, \textit{W4}):  15.8 (82 $\mu$Jy), 12.9 (200 $\mu$Jy) and 8.9 (2.3 mJy).  At the bright end, the largest and brightest galaxy detected is the early-type NGC~3268, with a \textit{W1} 1-sigma isophotal diameter of 6.83 arcmin and \textit{W1} integrated flux of 7.76 mag (241 mJy).  We use this resolved galaxy catalogue to position crossmatch with the KAT-7 \hi\ detections, and the All\textit{WISE} point source catalog for faint Antlia galaxies not resolved by \textit{WISE}.

\section{Results}

\begin{figure*}
\includegraphics[scale=0.82]{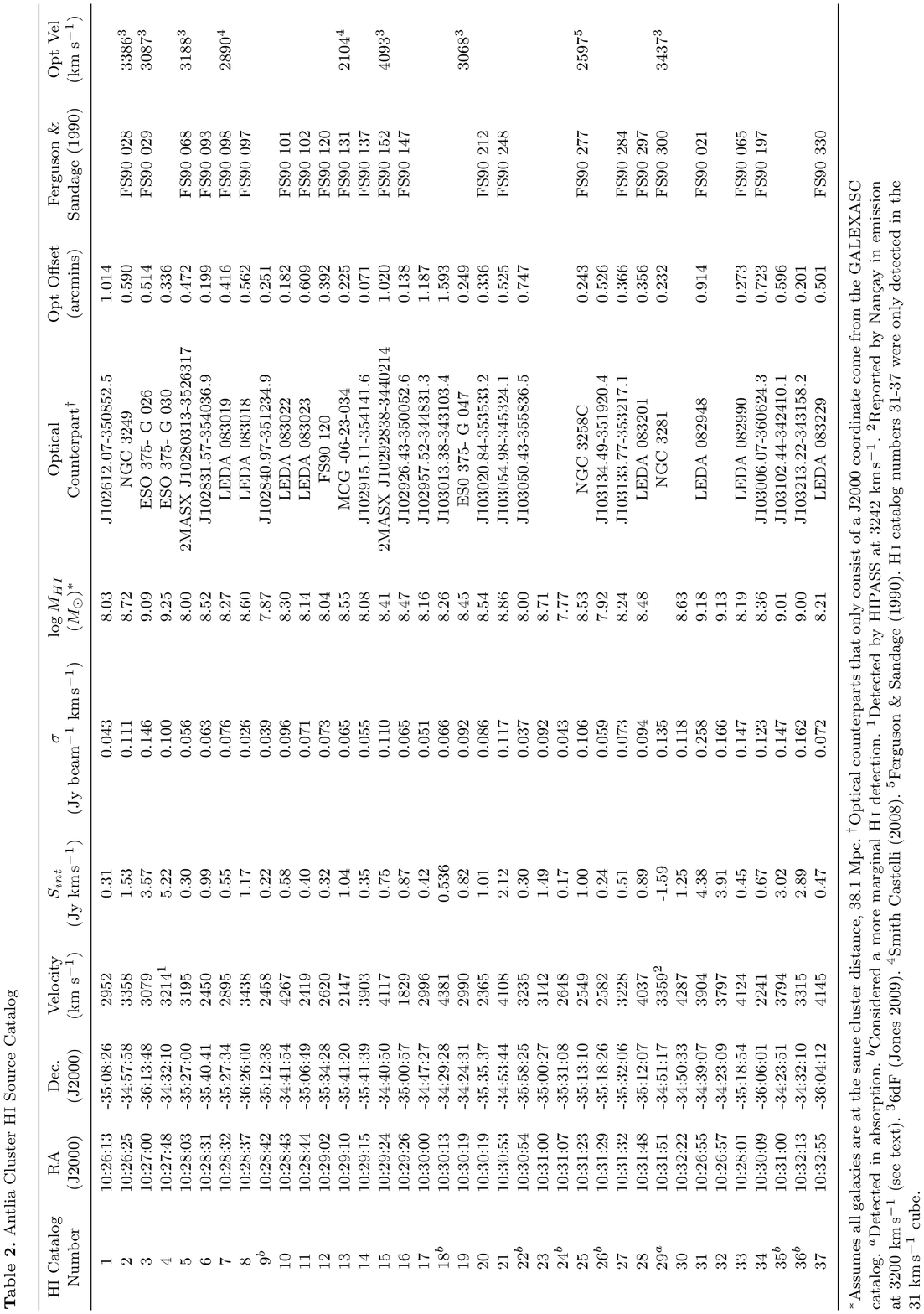}
\end{figure*}

\begin{figure*}
\includegraphics[scale=1.0,clip,trim=50 70 0 35]{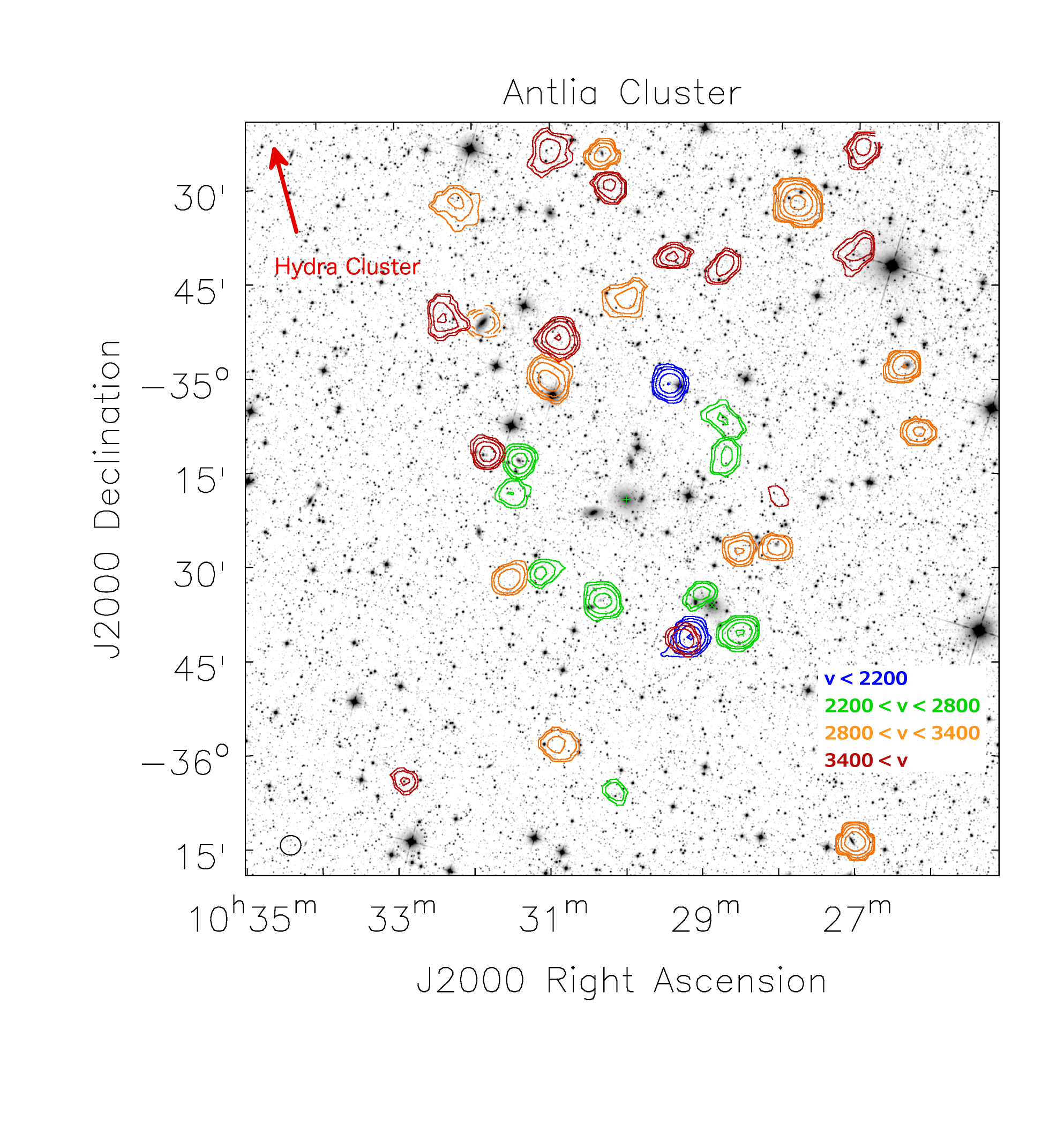}
\caption{\hi\ contours on a \textit{WISE} 3.4 $\mu$m image.  The contours correspond to column densities of 2, 4, 8, 16, 32, $64\times\sigma$ listed in Column 6 of Table 2 and are color coded by the systemic velocity of the galaxy.  The center of the image coincides with the center of the \hi\ mosaic and the center of the cluster, NGC~3268.  Dashed contours represent the negative \hi\ absorption around NGC~3281, northeast of the cluster center.}
\label{HImosaic}
\end{figure*}

\begin{figure*}
\includegraphics[scale=0.95,clip,trim=50 140 0 135]{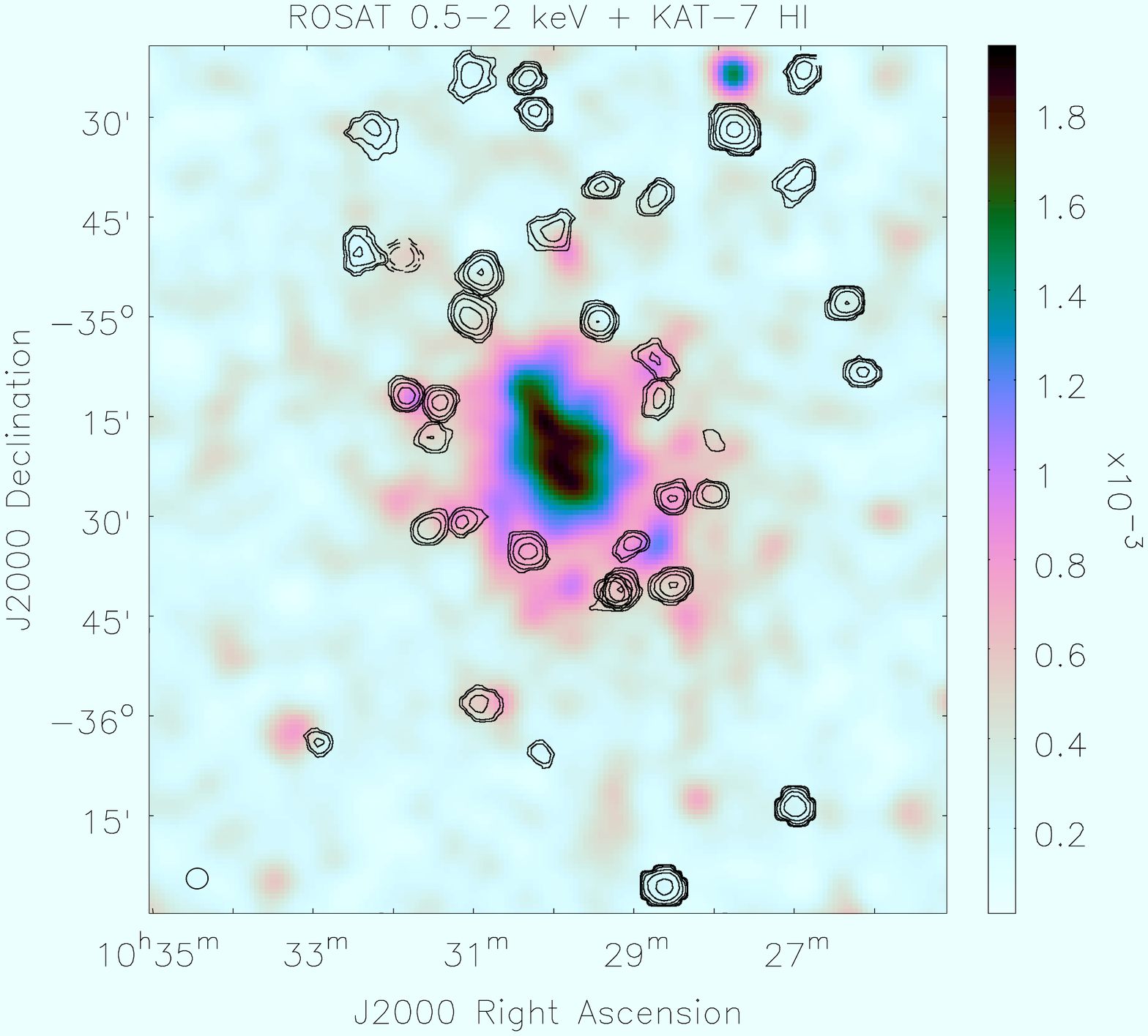}
\caption{ROSAT 0.5--2 keV hard X-ray intensity image from \textit{SkyView}\protect\footnotemark \citep{McGlynn96}, convolved with a $5\times5$ arcminute Gaussian, and overlaid with \hi\ contours. The X-ray image has had no point source removal performed.}
\label{rosat}
\end{figure*}

We report 37 detections in \hi\ with KAT-7 in the Antlia Cluster.  Of these 35 are new \hi\ detections, and 27 are new redshift detections at any wavelength.  Among the \hi\ sources, 24 are strongly detected in \hi\ emission in both the 15\kms and 31\kms cubes, one is strongly detected in absorption, five are considered more marginal detections because they are relatively low signal-to-noise and appear in only 3--4 channels in the 15\kms\ cube, and seven were only detected in the 31\kms\ cube.  The \hi\ detections are completely or mostly unresolved.  From comparison with DSS optical images, most of our sources do not appear to suffer from blending or confusion.  Only one detection (KAT7HI J103009-360601) has two good optical galaxies within the KAT-7 beam, neither of which have an alternate redshift measurement, but both of which were identified as potential cluster members by \citet{Ferguson90}.  We list the brighter of the two optical galaxies as the most likely optical counterpart.

Candidate detections were identified by examining cubes channel-by-channel, and then confirming their profile in position-velocity slices.  Detections were required to be at least three sigma and spatially contiguous at that level in at least three adjacent velocity channels in the 15\kms\ cube and two adjacent velocity channels in the 31\kms\ cube.  Additionally, we searched optical (DSS), ultraviolet (\textit{GALEX}), and infrared (\textit{WISE}) images for each candidate \hi\ detection to determine the most likely stellar counterparts.  We rejected marginal candidates if they both lacked a stellar counterpart, and had low \hi\ signal in noisy channels.  Two objects (KAT7HI J103100-350027, KAT7HI J102657-342309) lack a stellar counterpart in NED or \textit{WISE}, but the first has a strong, well-defined \hi\ profile in clean channels (see Appendix), and the second is spatially convincing albeit with a weaker profile.  KAT7HI J103222-345033 lacks a counterpart entry in NED or \textit{WISE}, but has a faint apparently star-forming optical counterpart in DSS images.  Unfortunately it is confused by a bright star diffraction spike in \textit{WISE} \textit{W1} images.

Table 2 describes the \hi\ properties of the detections and their best known stellar counterpart.
Column (1) is our catalog number, and superscripts indicate a note on the \hi\ detection. 
Columns (2) and (3) are the \hi\ centroid in J2000 coordinates.  The centroid was determined by the peak of a two dimensional Gaussian fit to the total intensity map of the source.
Column (4) is the intensity weighted velocity of the galaxy at the \hi\ centroid.  The uncertainty on these values is roughly 22\kms.
Column (5) is the total integrated flux.
Column (6) is the rms in the \hi\ total intensity (moment 0) map.
Column (7) is the \hi\ mass, assuming all galaxies lie at the distance of the cluster, $D=38.1$ Mpc.
Column (8) is the best optical counterpart, determined by manual inspection of DSS and \textit{GALEX} images.
Column (9) is the optical offset between the \hi\ centroid and the optical counterpart in arcmins.
Column (10) is the candidate cluster member from the \citet{Ferguson90} Antlia catalog.
Column (11) is the velocity measured from optical spectroscopy.

%\begin{table*}{{\em n\/}}
%  \vbox to 646pt{\vfil \caption{{\bf Table {\em 2\/}.}
%    Landscape table to go here.}\vfil}
%  \label{HIsources}
%\end{table*}

\subsection{Spatial Distribution of \hi\ Detected Objects}

The overall \hi\ distribution of galaxies reflects the effect of the cluster environment \citep{BravoAlfaro00}, and reveals the most recent accretion of galaxies and groups by the cluster.
Figure \ref{HImosaic} shows contours of the 37 \hi\ detections from the KAT-7 mosaic overlaid on a \textit{WISE} \textit{W1} image centered on NGC~3268.  To give an impression of the three dimensional structure the contours are color coded by the systemic velocity of the galaxy.

Several points are immediately evident from Figure \ref{HImosaic}.  First, \hi\ detections avoid the immediate volume around NGC~3268.  The closest \hi\ detections appear in a ring surrounding the elliptical galaxy at projected distances greater than 200 kpc.  This is in contrast to the elliptical galaxy NGC~3258 to the southeast which has four \hi\ detected galaxies projected within 80 kpc.  Second, we observe that galaxies which are blue-shifted with respect to the cluster velocity preferentially lie in the ring, whereas redshifted galaxies are distributed asymmetrically and preferentially to the north.  In the context of the large scale structure, this is in the same direction as the Hydra Cluster which, at a distance of 55 Mpc ($v=3777$\kms) and a projected $\sim$8 degrees from Antlia, lies $\sim$18 Mpc away.  In the following sections, we argue that the distribution of \hi\ is due to asymmetry in the accretion from the surrounding environment.  It may be suggestive of accretion along a filament connecting the two clusters.
%(17^2+6^2)^0.5 = 18.03

Figure \ref{rosat} shows that the ring of \hi\ detections is coincident with the extent of the X-ray halo around NGC~3268.  The X-ray intensity shows complex structure.  On the large scale, it is elongated in the direction between NGC~3268 and NGC~3258, which has its own extended Xray emission.  The \hi\ detections avoid the brightest X-ray regions.  We conclude that NGC~3268 and its large X-ray halo are the dominant and central structure of the Antlia Cluster, while NGC~3258 and its globular cluster and dwarf galaxy population are likely to be more recently arrived.  Consistent with observations of other nearby clusters (Virgo, \citealt{Chung09}; Coma, \citealt{BravoAlfaro00}), the center of Antlia is very \hi\ deficient, and suggests the extended hot X-ray halo is, in part, responsible for removing \hi\ from galaxies.

\subsection{Velocity Distribution and Infall}
\label{veldisp}

\footnotetext{\url{http://skyview.gsfc.nasa.gov/current/cgi/titlepage.pl}}

\begin{figure}
\includegraphics[scale=0.62,clip,trim=30 0 0 0]{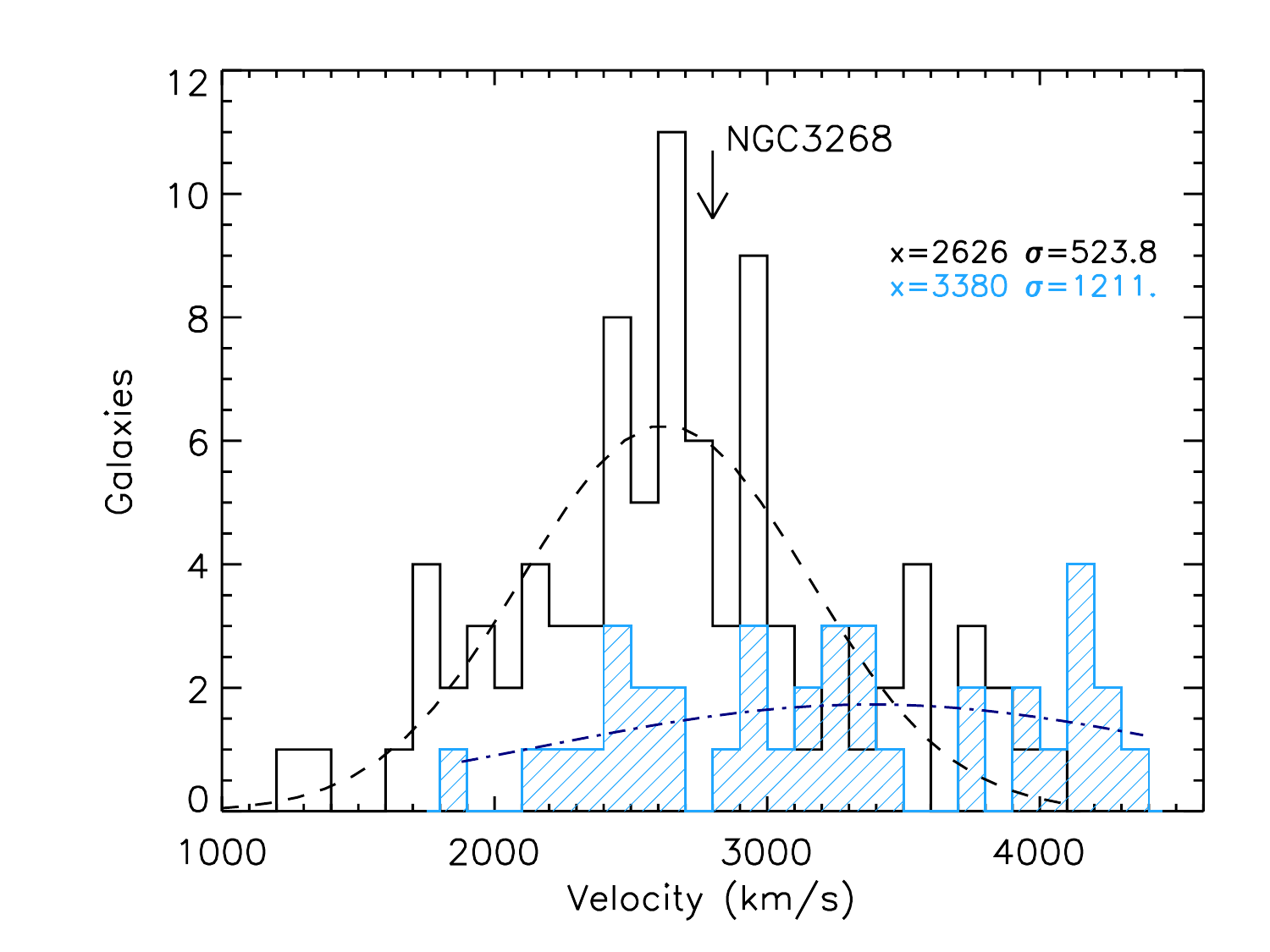}
\caption{Velocity histograms of 97 galaxies with only optical spectroscopic redshifts (black) and 37 \hi\ spectroscopic redshifts (blue).  The optical sample is dominated by early-type galaxies, while the \hi\ sample is primarily late-type galaxies. The legend gives the center and dispersion of each velocity distribution.  The velocities of late-type galaxies are systematically offset and more broadly distributed than early-type galaxies.}
\label{hist}
\end{figure}

Figure \ref{hist} shows a velocity histogram of cluster members with only optical redshifts (predominantly early-type galaxies), versus those with \hi\ redshifts (predominantly late-type galaxies).  We find that the \hi\ detected galaxies are by no means well fit by a Gaussian and may be better described as multi-modal \citep{Scodeggio95}.  None the less, as a population they have a broader velocity distribution than the optical sample, by a factor of 2.31, and are centered at a systematically higher redshift by 700\kms.
%1211/523.8  = 2.31

Assuming isotropic orbits, the ratio between the velocity dispersion of an infalling galaxy population and a virialized population is predicted to be $\sigma_{infall} / \sigma_{vir}=\sqrt{2} = 1.4$ \citep{Conselice01}.  For example, the ratio of velocity dispersion between Coma late-type and early-type galaxies is 1.4, while the ratio between Virgo late- and early-type galaxies is 1.64 (Table \ref{veltab}).  Virgo also shows greater asymmetry in the overall distribution of galaxies and in its X-ray contours than Coma, suggesting it is further from being virialized.  Similar ratios have been seen in larger samples of galaxy clusters when characterizing galaxies by color or morphology (e.g.~\citealt{Adami98} and references therein).  The value of Antlia is much larger than Virgo or Coma and is consistent with the \hi\ galaxies being a younger cluster population with regards to their relative age since accretion than those galaxies which are not detected in \hi.

The offset in the \hi\ velocity distribution from the optical population is likely due to a real asymmetry in the accretion of galaxies from the surrounding environment.
As an example, \citet{Gavazzi99} used the Tully-Fisher relation to show that redshifted galaxies in Virgo are closer than blue-shifted galaxies: galaxies on the far side of the cluster are falling inward--towards us--and thus are blue-shifted with respect to the cluster center.  If the change in observed mass sensitivity between the front and back of the Antlia cluster were \textit{significant}, we would expect a paucity of galaxies at blue-shifted velocities.  However, generously assuming Antlia is 2 Mpc in diameter translates to only a 10\% difference in \hi\ mass sensitivity.  Thus, we argue that the accretion onto Antlia is intrinsically asymmetric, rather than observational bias.  Further, given the larger scale structure around Antlia, we suggest that the asymmetry is due to a higher rate of accretion of galaxies from behind the cluster, notably towards Hydra, than from in front of it.

\begin{table}
\caption{Summary of velocities for Antlia and other clusters}
\begin{tabular}{lcc}
\hline
Description & $v$ (\kms) & $\sigma$ (\kms) \\
\hline
NGC~3268 & 2800 & \\
NGC~3258 & 2792 & \\
Antlia opt galaxies & 2626 & 524 \\
Antlia \hi\ galaxies & 3380 & 1211 \\
\textbf{All Antlia galaxies} & \textbf{2747} & \textbf{656} \\
Virgo early-type galaxies & 1134 & 573 \\
Virgo late-type galaxies & 1062 & 888 \\
\textbf{All Virgo galaxies} & \textbf{1112} & \textbf{757} \\
Coma early-type galaxies & 6807 & 1017 \\
Come late-type galaxies & 7375 & 1408 \\
\textbf{All Coma galaxies} & \textbf{6917} & \textbf{1038} \\
\hline
\end{tabular}\\
Virgo \citep{Binggeli87}; Coma \citep{Colless96}.
\label{veltab}
\end{table}

\subsection {Infrared Counterparts \& Properties}
\label{wisedescrip}

\begin{figure*}
\includegraphics[scale=1.0]{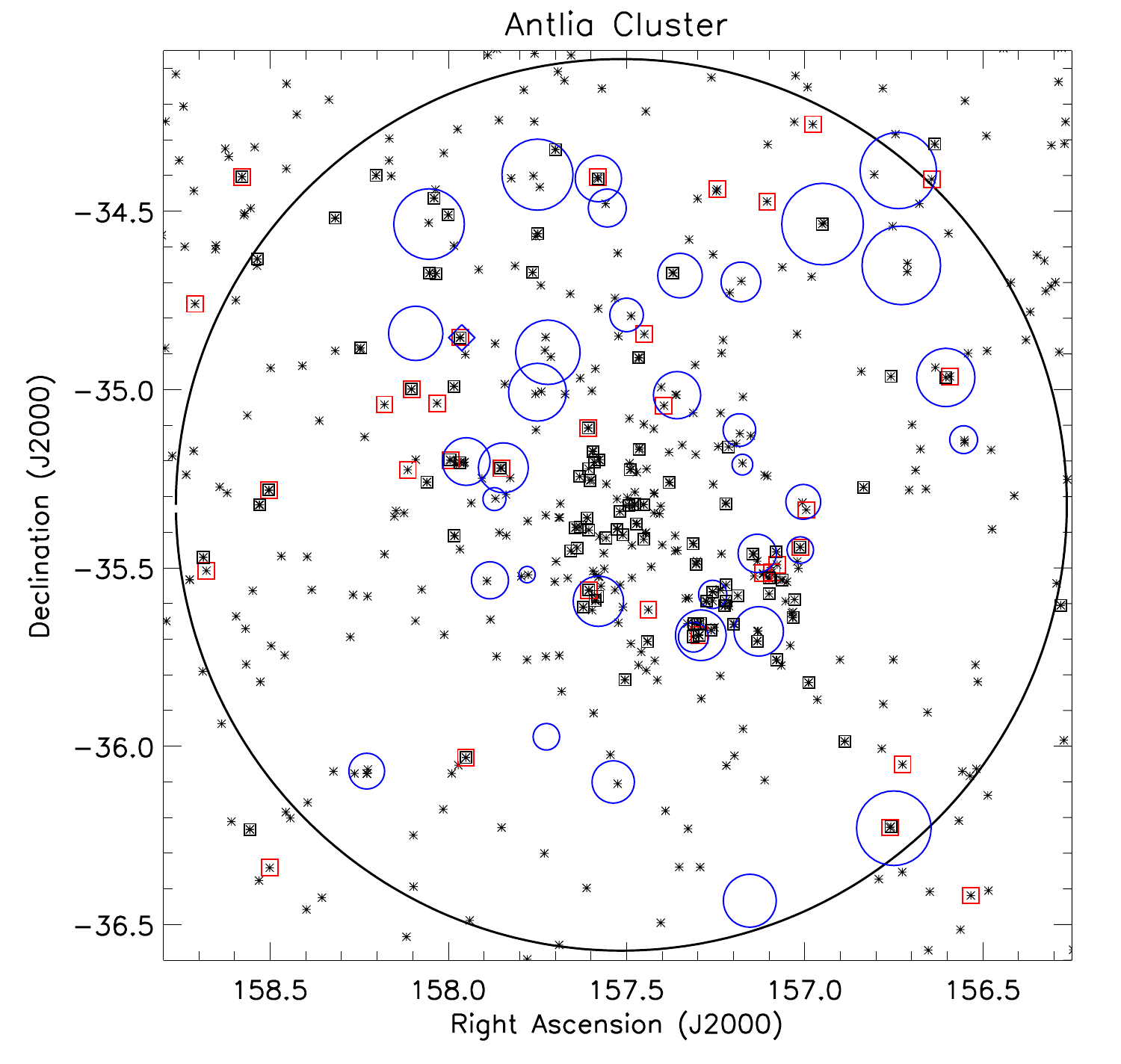}
\caption{The spatial distribution of \textit{WISE} extended sources in and around the Antlia Cluster.  Red symbols highlight ``star-forming'' over passive galaxies: those with $W2-W3>1.5$ and $\text{SFR}>0.05$\msun\ yr$^{-1}$.  Black squares indicate known cluster members with optical redshifts.  Blue symbols indicate \hi\ cluster members where the radius is scaled by $\log(M_{HI})$.  The blue diamond is NGC~3281, detected in \hi\ absorption.  The blue symbols and black squares make up the subset of galaxies in Figure \ref{dstest}. The large circle corresponds to the extent of the KAT-7 \hi\ mosaic.}
\label{spatial}
\end{figure*}

Using a radius of $\sim$1 arcminute, positional cross-matching between the Antlia \hi\ detections and the \textit{WISE} resolved extractions gives rise to 23 confident matches.  Of these, eight are below the standard S/N = 10 limit for the \textit{W1} extended source catalog, but are measured \textit{a posteriori} based on the \hi\ detection.  Another four \hi\ detections were matched with unresolved \textit{WISE} sources from the All\textit{WISE} catalog \citep{Cutri13}, although the chance for a false match is much greater since these WISE sources may be foreground stars or background galaxies.  Hence, the resolved-galaxy match rate is 23/37 (62\%) and the total match rate is 27/37 (73\%).

We use \textit{WISE} extracted photometry to estimate the host stellar mass and the obscured star formation rate.  For the stellar mass, we use the ``resolved galaxy'' (i.e., nearby galaxies) $M/L$ relations of \citet{Cluver14}, which in combination with the \textit{W1} 3.4 $\mu$m ``in-band'' luminosities \citep{Jarrett13} and the $W1-W2$ color (which roughly accounts for metallicity and morphology differences), infers a stellar mass and formal uncertainty through error propagation.   Accordingly, for the Antlia matches the minimum stellar mass is found to be $\sim10^{7.8}$\msun\ (WXSC J10280155-351900.4; or considering point sources, $\sim10^{6.6}$\msun\ for WISEA J102956.70-344834.1) and the most massive source at $10^{11.1}$\msun\ is the barred lenticular galaxy, NGC~3271.

The star formation rate is estimated from both the 12 $\mu$m and 22 $\mu$m measurements using the \citet{Cluver14} relations.  The most sensitive mid-infrared band of \textit{WISE} is the \textit{W3} band, whose rest bandpass captures the 11.3 $\mu$m PAH band, a star-formation tracer, although exhibiting more scatter in comparison to other indicators \citep{Cluver14}.  Comparatively, \textit{WISE} 22 $\mu$m band is less sensitive, but is dominated by dust-emitting continuum in thermal dynamic equilibrium and thus a more reliable SF tracer (e.g.~\citealt{Jarrett13}).  Most of the Antlia sources have very low star formation activity--as traced by the obscured star formation--with most inferred SFR values $<0.01$ \msun yr$^{-1}$, consistent with galaxy cluster members in the local universe.  At the high end, a few sources do have rates greater than unity, notably NGC~3281 with an inferred value of $\sim$10 and 10.7 \msun yr$^{-1}$ in \textit{W3} and \textit{W4}, respectively.  It should be noted that since this host galaxy has an AGN (type II Seyfert) and thus mid-infrared emission arising from the accretion disc (see Figure \ref{ngc3281w} for a visual demonstration of powerful nuclear emission), the host galaxy star formation luminosity is over-estimated and should be treated with caution.  In total, 6/37 (16\%) of the \hi\ detections have star formation rates greater than 0.05 \msun yr$^{-1}$.

Table 4 lists the properties of \textit{WISE} counterparts to \hi\ detected sources:
Column (1) is the \hi\ catalog number from Table 2;
Column (2) is the \textit{WISE} detection;
Columns (3--6) are the \textit{W1}, \textit{W2}, \textit{W3}, and \textit{W4} Vega magnitudes;
Column (7) is the stellar mass;
Columns (8--9) are the star formation rate calculated from the 12 $\mu$m and 22 $\mu$m bands;
Column (10) is the \hi\ mass-to-stellar mass ratio.

%\begin{table*}{{\em n\/}}
%  \vbox to 646pt{\vfil \caption{{\bf Table {\em 4\/}.}
%    Landscape table to go here.}\vfil}
%\label{WISEcounter}
%\end{table*}

\begin{figure*}
\includegraphics[scale=0.82]{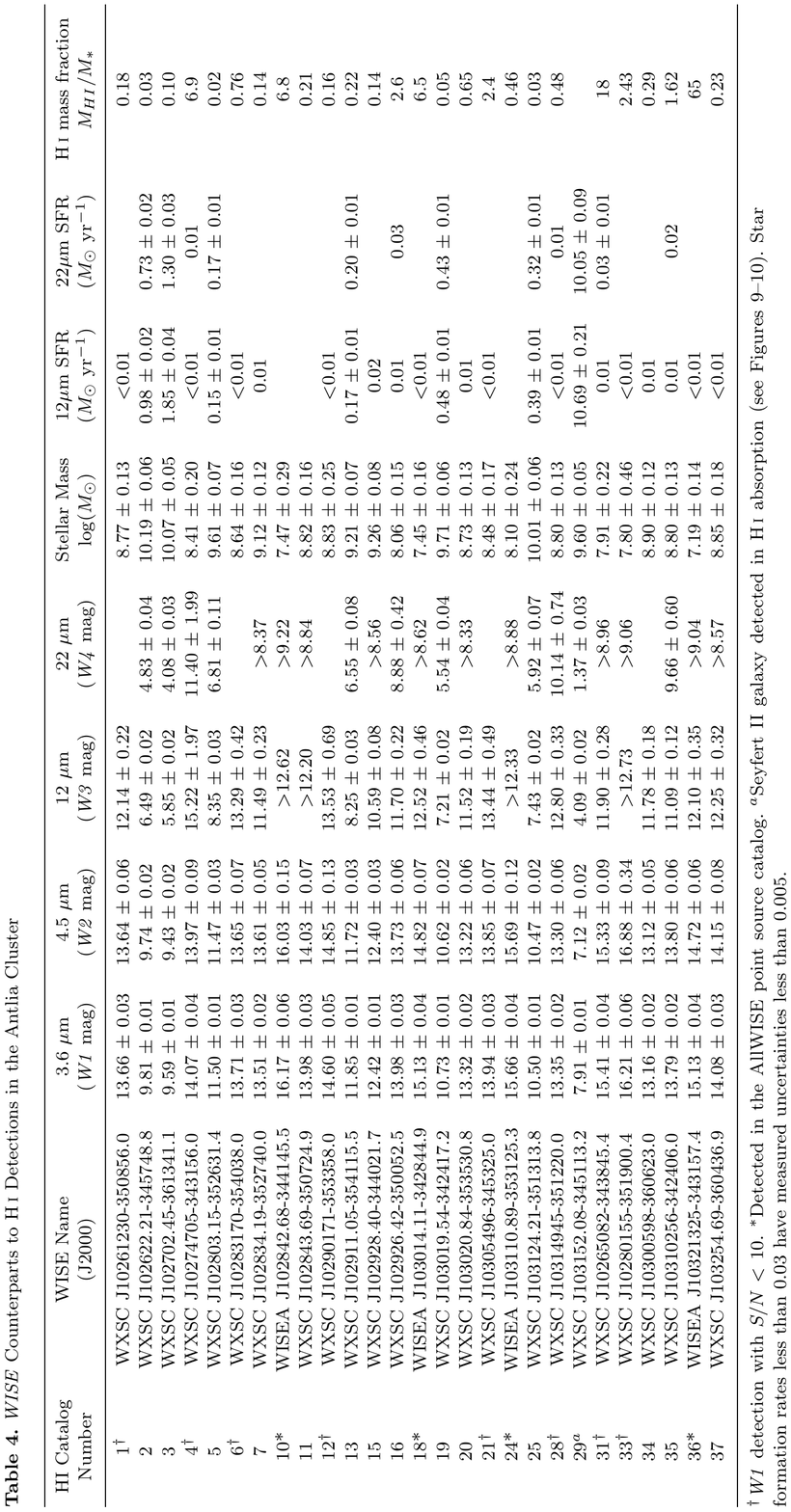}
\end{figure*}

\subsection{Stellar Masses, Star Formation Rates, and Gas Fractions}
\label{sf}

Combining quantitative measures of the gaseous and stellar components of cluster members provides the most complete context for studying the overall cluster dynamics and galaxy evolution with environment.  From cross-matching to infrared sources and Table 4, we find that the \hi\ population is dominated (28/37) by dwarf galaxies with $M_*<10^9$\msun, and star formation rates below 0.1\msun\ yr$^{-1}$.  Further, most objects have $M_{HI}/M_{*}<1$ and the lowest stellar mass objects have the highest gas fractions, consistent with what is seen on average across all environments \citep{Huang12,Maddox15}.  

Figure \ref{spatial} shows the spatial distribution of all \textit{WISE} extended sources in and around the Antlia Cluster, including background galaxies. (There are no known foreground galaxies below $v=1200$\kms).  
The figure conveys a sense of the over-density of galaxies within Antlia in relation to the \hi\ detections, and we highlight ``star-forming'' galaxies, relative to the majority of passive cluster members, if they have $W2-W3>1.5$ and $\text{SFR}_{12\mu m}>0.05$\msun\ yr$^{-1}$.

\begin{figure}
\includegraphics[scale=0.59,clip,trim=15 0 0 0]{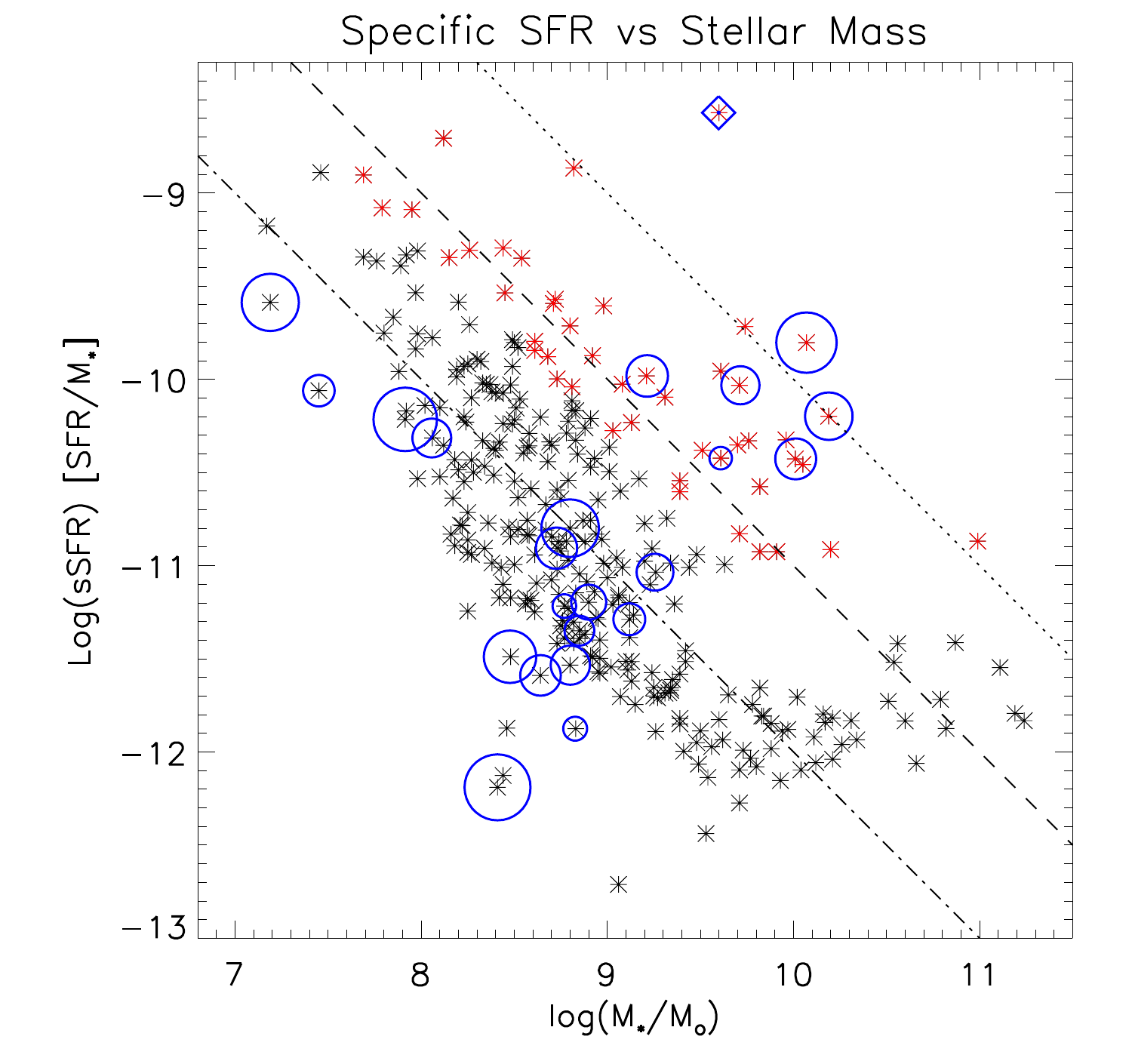}
\caption{The specific star formation versus stellar mass for all \textit{WISE} extended sources, including Antlia and background galaxies.  Star-forming galaxies are in red, and \hi\ detected objects are in blue as in the previous figure. Lines of constant star formation (0.01,0.1,1 \msun yr$^{-1}$) are shown by dot-dashed, dashed, and dotted lines, respectively.}
\label{ssfr}
\end{figure}

We find 13 of the 124 spectroscopically confirmed cluster members (10\%) and 6/37 \hi\ detections (16\%) have star formation rates greater than 0.05\msun\ yr$^{-1}$.  At least seven cluster members have appreciable star formation without detected \hi. Only one, WXSC J103025.79-350628.7 (MCG -06-23-043; v=1781\kms, 6dF), is projected within the 200 kpc \hi\ ring: forming stars at a rate of 0.18\msun\ yr$^{-1}$ in a $10^{9.8}$\msun galaxy. Figure \ref{ssfr} shows the specific star formation rate versus stellar mass for all \text{WISE} extended sources: most massive galaxies are passive, non-star-forming systems.  There are no obvious trends between \hi\ mass, stellar mass, or star formation rate.

\begin{figure}
\includegraphics[scale=0.59,clip,trim=15 0 0 0]{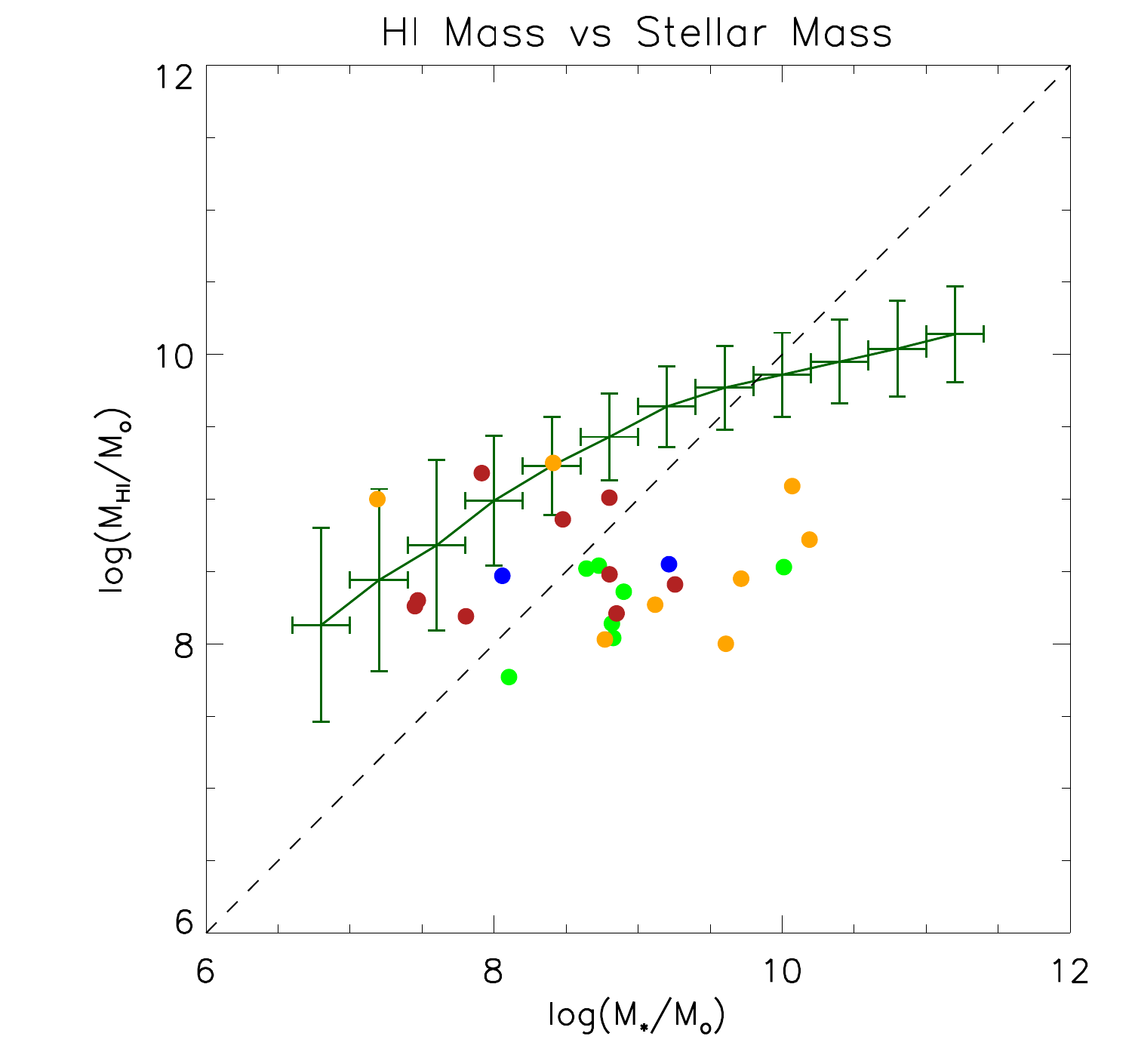}
\caption{The \hi\ mass versus stellar mass for all \hi\ Antlia galaxies with WISE counterparts.  The dashed black line is the one-to-one gas fraction relation.  The green line is the median values for the ALFALFA survey from \citet{Maddox15} where the horizontal error bars represent bin size, and the vertical errors bars represent the 1$\sigma$ distribution.  The filled circles are the Antlia \hi\ detections where the colors are the same as Figure \ref{HImosaic}.}
\label{bpe}
\end{figure}

Finally, Figure \ref{bpe} shows the \hi\ mass versus stellar mass of the Antlia galaxies compared to the mean values of the entire ALFALFA sample from \citet{Maddox15}.  The ALFALFA sample is dominated by field galaxies \citep{Hess13}, so this provides a baseline for the impact of the cluster environment on the gas content of individual galaxies compared to the field.  Compared to galaxies of the same stellar mass, all Antlia galaxies, except a handful of the dwarfs with $M_*<10^{8.5}$\msun, are \hi\ deficient and fall well below the 1$\sigma$ distribution.  The most massive galaxies by stellar mass are also the most gas deficient compared to the field, by 2-3 orders of magnitude in \hi\ mass.  Meanwhile, the most \hi\ rich, both by mass and gas fraction, are dwarf galaxies.  From Figure \ref{spatial}, we see that these dwarfs reside on the extreme outskirts of the cluster.  Their reservoir of \hi\ gas will not survive once they have fallen into the cluster environment.

We conclude that gas stripping is occurring in the cluster, and is happening at appreciable levels outside the X-ray emitting halo: out to at least $\sim$600 kpc in projected radius from the cluster center.  The \hi\ mass is also a measure of future star formation potential.  So in addition to removing \hi\ the cluster environment must impact future star formation, but it is unclear whether accretion on to the cluster enhances star formation before quenching it, and whether quenching occurs slowly \citep{Kauffmann04} or rapidly \citep{Rines05}.  
Galaxy transformation may begin at several virial radii from the cluster center, but star forming galaxies can still exist within the cluster X-ray halo \citep{Caldwell93}.  We show that the tracers of star formation are still visible even when the atomic gas has been exhausted or stripped below detectable limits, with star forming galaxies potentially drawing from a reserve of dense molecular gas \citep{Kenney89,Casoli91}.

\section{Discussion}
\label{discussion}

From the spatial distribution of \hi\ detections we have established that, despite the presence of two massive early type galaxies at its center, Antlia's oldest and most dominant structure is centered on NGC~3268.  The velocities of the NGC~3268 and NGC~3258 at $2800\pm21$\kms\ and $2792\pm28$\kms, respectively, give no hints as to how a merger between the two subclusters may have proceeded.  Further, we have shown that there is little active star formation going on in the center of the cluster, and that the \hi\ galaxies within at least 600 kpc of the center have undergone significant stripping even though they lie outside the X-ray halo.

Statistical dynamical tests in combination with the \hi\ and star formation content of galaxies will help us to further disentangle the assembly history of the cluster.  In the following subsections, we analyse the three dimensional structure of Antlia, investigate substructure in the context of their stellar and neutral gas content, calculate the total cluster mass, and put Antlia into an evolutionary context with respect to other nearby, well-studied clusters.  Finally, we present the serendipitous detection of \hi\ absorption in the Compton thick Seyfert II galaxy, NGC~3281.

\subsection{Substructure in the Antlia Cluster}
\label{substructure}

\begin{figure*}
\includegraphics[scale=0.63,clip,trim=0 0 -23 10]{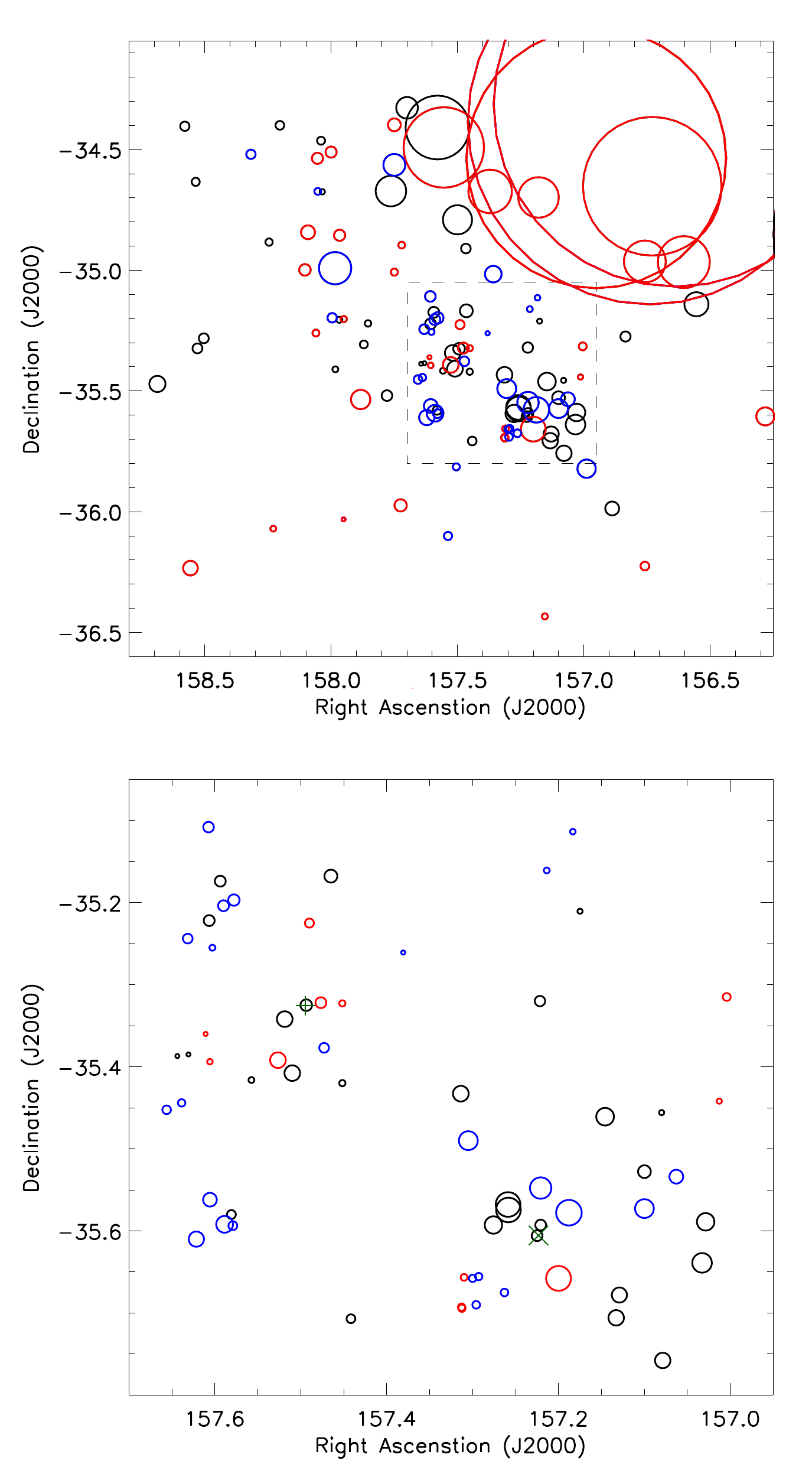}\hspace{-0.3in}
\includegraphics[scale=0.592,clip,trim=90 12 90 0]{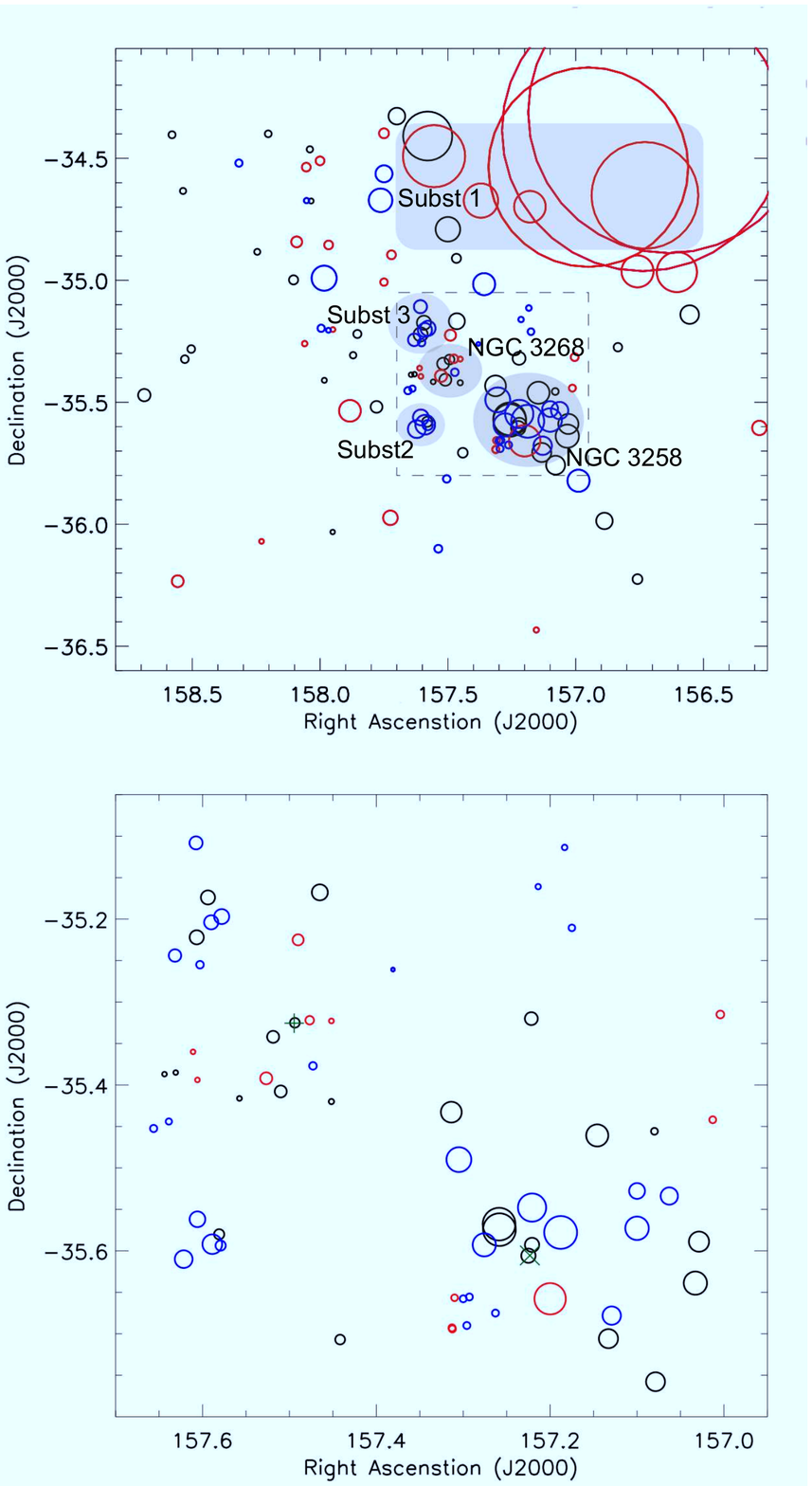}
\caption{\citet{Dressler88} ``bubble plot'' for 124 galaxies with redshifts, where the center of each circle corresponds to the position of a galaxy, and the size scales with exp($\delta_i$).  Left: the kinematic deviation has been calculated from the derived mean velocity of the cluster $\bar{v}=2747$\kms.  Right: the kinematic deviation is calculated assuming $\bar{v}=2800$\kms, the velocity of NGC~3268.  Black symbols are galaxies with velocities within $\pm0.5\sigma$ of $\bar{v}$; blue symbols are blue shifted to $v<-0.5\sigma$; red symbols are redshifted to $v>+0.5\sigma$.  Top plots are the entire cluster; bottom plots are zoomed in on the inset, and show how the choice in $\bar{v}$ brings out substructure within the cluster center.  
The ``$+$'' denotes the position of NGC~3268 and the nominal cluster center; the ``X'' denotes the position of NGC~3258. Substructure is labeled in gray in the upper right panel for clarity.}
\label{dstest}
\end{figure*}

The clumpy distribution of galaxies in space and velocity suggests the ongoing accretion and merger of groups with the cluster, in addition to the infall of individual galaxies.  The presence of \hi\ or star formation provides a rough timeline for the merger activity because these aspects of infalling galaxies are transformed by the cluster environment.  \citet{Hopp85} identified five nearby groups to Antlia, which lie at projected distances greater than 2.5 degrees and 1.75 Mpc.  We focus on the substructure within the KAT-7 \hi\ mosaic which covers the volume containing 96\% of the optical galaxies associated with the cluster (Figure \ref{spatial}) and extends to 4--5 times the X-ray extent of NGC~3268 subcluster.

Spatial separation is a intuitive quality of substructure, but the Dressler-Schectman (DS) test provides a measure of the relative kinematic deviation of substructure from the global kinematics \citep{Dressler88}.  The test and its application to both groups and clusters is well summarized in \citet{Hou12}.  In particular, this test has been used to understand whether infall enhances or quenches star formation and gas accretion on to systems (e.g.~\citealt{Pranger13}).  Recent results find that substructure in groups and clusters is correlated with an increase in active galaxies \citep{Hou12,Cohen14}.

For each galaxy, $i$, we calculate the mean velocity, $\bar{v}^i_{local}$, and velocity dispersion, $\sigma^i_{local}$, for it and its $N_{nn} = 10$ nearest neighbors compared to the mean velocity, $\bar{v}$, and velocity dispersion, $\sigma$, of the entire cluster:
\begin{equation}
\delta_i = \left(\frac{N_{nn}+1}{\sigma^2}\right)\left[(\bar{v}^i_{local}-\bar{v})^2+(\sigma^i_{local}-\sigma)^2\right]
\end{equation}
\begin{equation}
\Delta = \sum_i \delta_i /n_{members}
\end{equation}
where the threshold $\Delta=1$ indicates the distribution is close to Gaussian and the fluctuations are random.  
For Antlia, $\bar{v}=2747$\kms\ and $\sigma=656$\kms\ for 124 galaxies with \hi\ or optical redshifts, and we find $\Delta=1.4$, suggesting the presence of substructure.

Cohesive substructure is identified with the DS statistic by locating associations of galaxies which deviate from the overall cluster systematics in the same way, and are close to each other in spatial proximity and velocity.  In Figure \ref{dstest}, we plot the kinematic deviation, $\delta_i$, for each galaxy as a function of position, and substructures can be identified by collections of circles of the same size and color.  We can then compare these substructures with Figure \ref{spatial} to learn if they are gas rich and/or star forming.
In the following sections we discuss these plots in detail, considering first the inner virialized region, then the outskirts where infall is reflected in the kinematics.

\subsubsection{The Inner Cluster}
\label{inner}

The systemic velocities of NGC~3268 and NGC~3258 are 2800\kms\ and 2792\kms, respectively, and their similar velocities tell us nothing about the galaxies' relative roles within the cluster.  With the DS test we identify differences between their respective subclusters.  Figure \ref{dstest} shows DS test performed twice: once as prescribed above with $\bar{v}=2747$\kms\ (left panels), and once using the velocity of NGC~3268 as $\bar{v}=2800$\kms\ (right panels).  Quantitatively, $\Delta$ is the same for both choices of $\bar{v}$, but the change in $\delta_i$ for individual galaxies exposes local kinematic variation between the NGC~3268 and NGC~3258 subclusters.

In the left panels of Figure \ref{dstest}, we see that $\delta_i$ for galaxies around both NGC~3268 and NGC~3258 are similar in magnitude, but when we set $\bar{v}=2800$\kms, the galaxies centered around NGC~3268 have systematically lower $\delta_i$, while those around NGC~3258 are larger, even though the elliptical galaxies have effectively the same velocity.  The system of galaxies around NGC~3268 are better described by the velocity of the massive ellipticals, than the mean velocity of the entire cluster.   Meanwhile the system of galaxies around NGC~3258 are kinematically less consistent with the velocity of their central elliptical.  

As mentioned above, four galaxies within 80 kpc projection of NGC~3258 and within its extended X-ray halo \citep{Pedersen97} are detected in \hi.  Three of these are blue-shifted with respect to 2800\kms, consistent with the rest of the NGC~3258 subcluster.  We cannot rule out that the \hi\ population may be its own substructure seen in projection against NGC~3258, but we require redshift independent distances to disentangle the \hi\ and non-\hi\ detected galaxies.  None the less, the locality of \hi\ in projection to NGC~3258, in addition to its weaker X-ray halo, is further evidence that the NGC~3258 subcluster is a younger system that has merged with the NGC~3268 cluster.  

Lastly, while $\Delta \sim 1.4$ for the entire cluster for both choices of $\bar{v}$, if we consider only galaxies within the inset (bottom plots of Figure \ref{dstest}): for $\bar{v}=2747$\kms, $\Delta=0.89$; while for $\bar{v}=2800$\kms, $\Delta=0.73$.  Thus, it appears that galaxies on the outskirts of Antlia bias the calculation of the mean velocity of the cluster, and their $\delta_i$ values dominate the overall $\Delta_i$, the collective impact of which is to wash out evidence for substructure in the center.

\subsubsection{The Cluster Outskirts}
\label{outskirts}

In addition to the NGC~3268 and NGC~3258 subclusters, we consider three more galaxy associations in Antlia.  This is by no means a complete list, but we contemplate the most obvious collections of four or more galaxies which may have accreted on to the cluster together and not yet been disrupted by gravitational interactions.  Based on the presence of \hi\ and/or star formation, we present them in increasing order of time since accretion.  Alternatively, the ranking may be interpreted as one of increasing effect of the cluster environment on the evolution of the galaxies within the substructures:
\begin{itemize}
\item{Subst 1, ($\alpha$, $\delta$) = (157.3, -34.5):
%$10^h29^m -35^d34^m$ 
four galaxies, redshifted with respect to the cluster center velocity, have high $\delta_i$ values and are at similar systemic velocities.  All four of the galaxies are detected in \hi\ (Figure \ref{HImosaic}) and the northwestern most galaxy is one of the most \hi\ massive objects in the cluster (LEDA 082948).}  We postulate that collectively these make up the most recently arrived group to the cluster.
\item{Subst 2, ($\alpha$, $\delta$) = (157.6 -35.6):
%$10^h30.5^m -35^d36^m$ 
four galaxies which are closely associated in space, and vary little in $\delta_i$. One of them is detected in \hi\ and one is detected in star formation.  A fifth galaxy, seen in black, we consider unassociated.}
\item{Subst 3, ($\alpha$, $\delta$) = (157.6 -35.2):
%$10^h30.5^m -35^d12m$, but within the inset of Figure \ref{dstest} and in line with NGC~3258 and NGC~3268, 
seven galaxies in blue or black which behave similarly in $\delta_i$ when we change $\bar{v}$ from 2711 to 2800\kms.  None of them are detected in \hi\, but one is detected by \textit{WISE} to have star formation. These galaxies are quite close together, and stand out in $\delta_i$ from the nearby NGC~3268 subcluster.}
\end{itemize}

Coincidentally, this list also ranks the substructure in decreasing projected distance to the cluster center.  During infall and assembly, the mixing of galaxy populations is incomplete, thus our proposed timeline is also consistent with a correlation between accretion time and distance to the cluster center \citep{deLucia12}.

Subst 1 contains the most \hi\ bright galaxies and, based on their velocities, it could be either a foreground or background group falling into the cluster.  The Hydra Cluster lies beyond the bounds of the KAT-7 mosaic to the north and east at higher redshifts \citep{RadburnSmith06,Courtois13}, and thus the galaxies may be in the background having come from a filament that connects Antlia with Hydra.  In any case, we argue that the asymmetry in the velocity and spatial distribution of \hi\ detected galaxies (and substructure) is an intrinsic characteristic of accretion on to the Antlia Cluster at the present epoch.

Considering the overall spatial distribution of the Antlia population, we note that the line connecting NGC~3258 to NGC~3268, if extended to the northeast in Figure \ref{dstest}, follows an over-density of galaxies which have small $\delta_i$.  It is unclear whether this whole linear feature is related to the merger of NGC~3258 with NGC~3268, and/or connected with filamentary structures on larger scales.  The small variation in $\delta_i$ would imply that much of the action in Antlia is occurring tangentially in the plane of the sky.  

\subsection{Dynamical Mass Estimates}
\label{dynest}

We calculate the radial mass profile and estimate the total dynamical mass of the cluster based on all available redshift measurements which includes 124 galaxies with unique spectroscopic measurements. 
\citet{Nakazawa00} use X-ray observations to estimate a total gravitating mass of $1.9\times10^{13}$\msun\ internal to 250 kpc.  By combining \hi\ and, in particular, 6dF redshifts we are able to probe out to a radius of 1 Mpc.

Mass estimates are commonly computed using the virial theorem, but can be biased. \citet{Heisler85} describe three alternative mass estimators for self-gravitating systems which provide consistent results, and work well even in systems where the number of galaxy members or available spectroscopic redshifts may be small, and the galaxies themselves have a range of masses.  The four estimators are: \\
a) virial mass,
\begin{equation}
  M_{VT} = \frac{3\pi N}{2G} \frac{\sum_{i}V^2_{zi}}{\sum_{i}\sum_{i<j}1/R_{\perp ij}}
\end{equation}
b) projected mass,
\begin{equation}
  M_{PM} = \frac{f_{PM}}{G(N-1.5)} \sum_{i}V^2_{zi}R_{\perp i} \\
\end{equation}
\begin{equation*}
  f_{PM}=32/\pi
\end{equation*}
c) median mass,
\begin{equation}
  M_{Me}=\frac{f_{Me}}{G} \text{med}_{i,j}(V_{zi}-V_{zj})^2 R_{\perp ij} \\
\end{equation}
\begin{equation*}
  f_{Me}=6.5
\end{equation*}
d) average mass,
\begin{equation}
  M_{Av}=\frac{2f_{Av}}{GN(N-1)} \sum_{i}\sum_{i<j}(V_{zi}-V_{zj})^2 R_{\perp ij} \\
\end{equation}
\begin{equation*}
  f_{Av}=2.8
\end{equation*}

\begin{figure}
\includegraphics[scale=0.58,clip,trim=16 0 0 0]{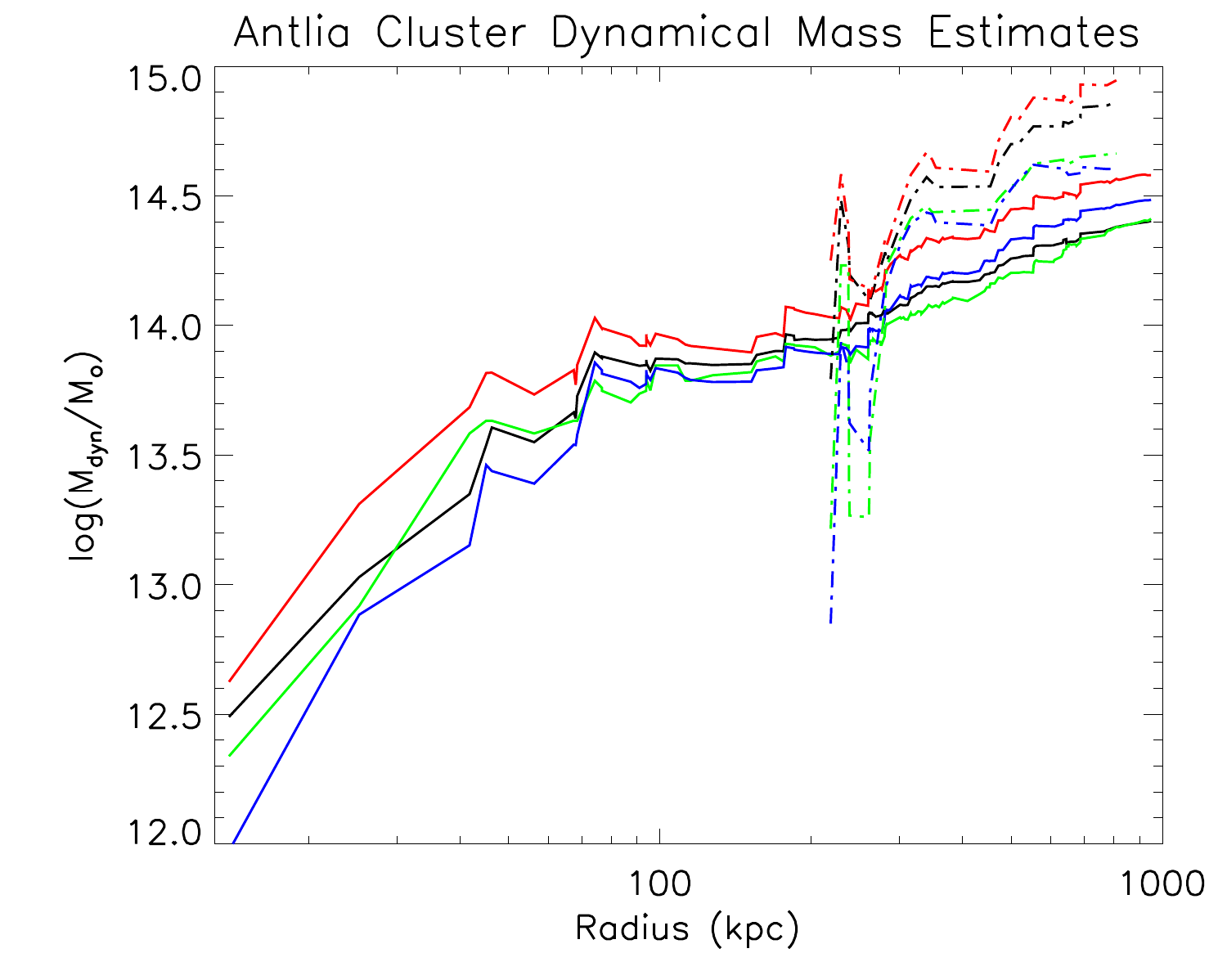}
\caption{Dynamical mass estimates as a function of projected radius for the virial mass (black), projected mass (red), median mass (green), average mass (blue).  The solid lines are calculated using the full sample of 117 galaxies for which redshifts are available.  Dot-dashed lines are calculated using only the 37 \hi\ detected galaxies.  These galaxies have a broader velocity dispersion and are not be virialized, and therefore over estimate the dynamical mass of the cluster.}
\label{dynmass}
\end{figure}

Figure \ref{dynmass} shows how the mass estimators vary as we include galaxies at increasing projected radius.  The cluster mass increases at a reasonable rate with projected radius, suggesting that the velocity selection criteria for cluster membership applied by previous authors reasonably defines the cluster, and that we have excluded outliers.

We use the average of these mass estimators to derive the cluster mass at various radii \citep{Puche88}.  
Within 250 kpc we find a cluster mass of $9.1\pm1.5\times10^{13}$\msun.  This is 4.8 times greater than that suggested by X-ray observations within the same radius.  Within 0.6--1 Mpc the total mass of the cluster grows to $2.6\pm0.6\times10^{14}$\msun. It may be interesting to note that virial mass and median mass track the closest with radius.  According to \citet{Heisler85}, the median mass estimate is the least sensitive to outliers.  

The mass estimates flatten at radii between 200--250 kpc.  Combining the mass estimate and the velocity dispersion of galaxies within 250 kpc with the virial theorem, we calculate a cluster virial radius of 0.9 Mpc.  Beyond 200--250 kpc, the mass estimators rise again out to 1 Mpc.  Assuming the cluster population extends to 1 Mpc, the mass and velocity dispersion suggest a much larger virial radius of 3.5 Mpc.  It is clear then that galaxies at greater distance from the cluster center are less likely to be virialized.  

Figure \ref{dynmass} shows that the \hi\ detected galaxies reside beyond at least 200 kpc from the cluster center, and systematically over-estimate the cluster mass for all estimators.  Mass estimates which rely on the \hi\ detected galaxies suggest a derived mass of $5.7\pm1.8\times10^{14}$\msun\ and virial radius of 4.6 Mpc.

This picture is consistent with what has been found in Virgo, Fornax, and Coma: late-type galaxies have a broader velocity dispersion than early-type galaxies because they are free falling into a virialized system, and \hi\ detected galaxies are on more radial orbits than those without \citep{Wojtak10}.  Beyond 250 kpc there is a sizeable population of older cluster members which share the same volume as where the infall of \hi\ objects is occurring from the surroundings.

\subsection{NGC~3281: AGN and Associated Absorber}

\begin{figure}
\includegraphics[scale=0.44]{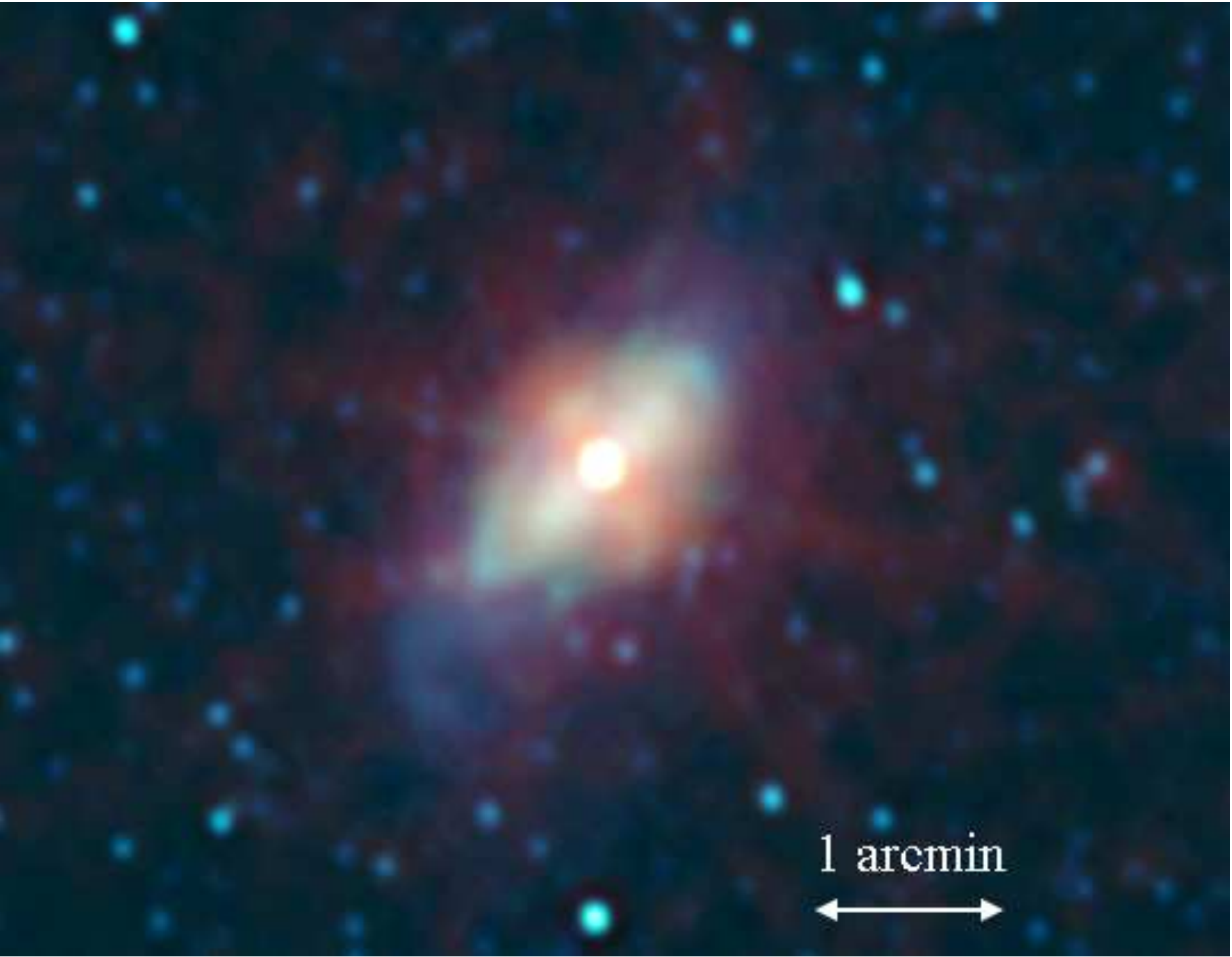}
\caption{WISE mid-infrared view of NGC 3281.   The image was constructed using the high-resolution sampling technique described in \citet{Jarrett12}, with color combination W1 (3.4 $\mu$m) is blue, W2 (4.6 $\mu$m) is green, W3 (12 $\mu$m) is orange and W4 (22 $\mu$m) is red.}
\label{ngc3281w}
\end{figure}

\begin{figure}
\includegraphics[scale=0.52]{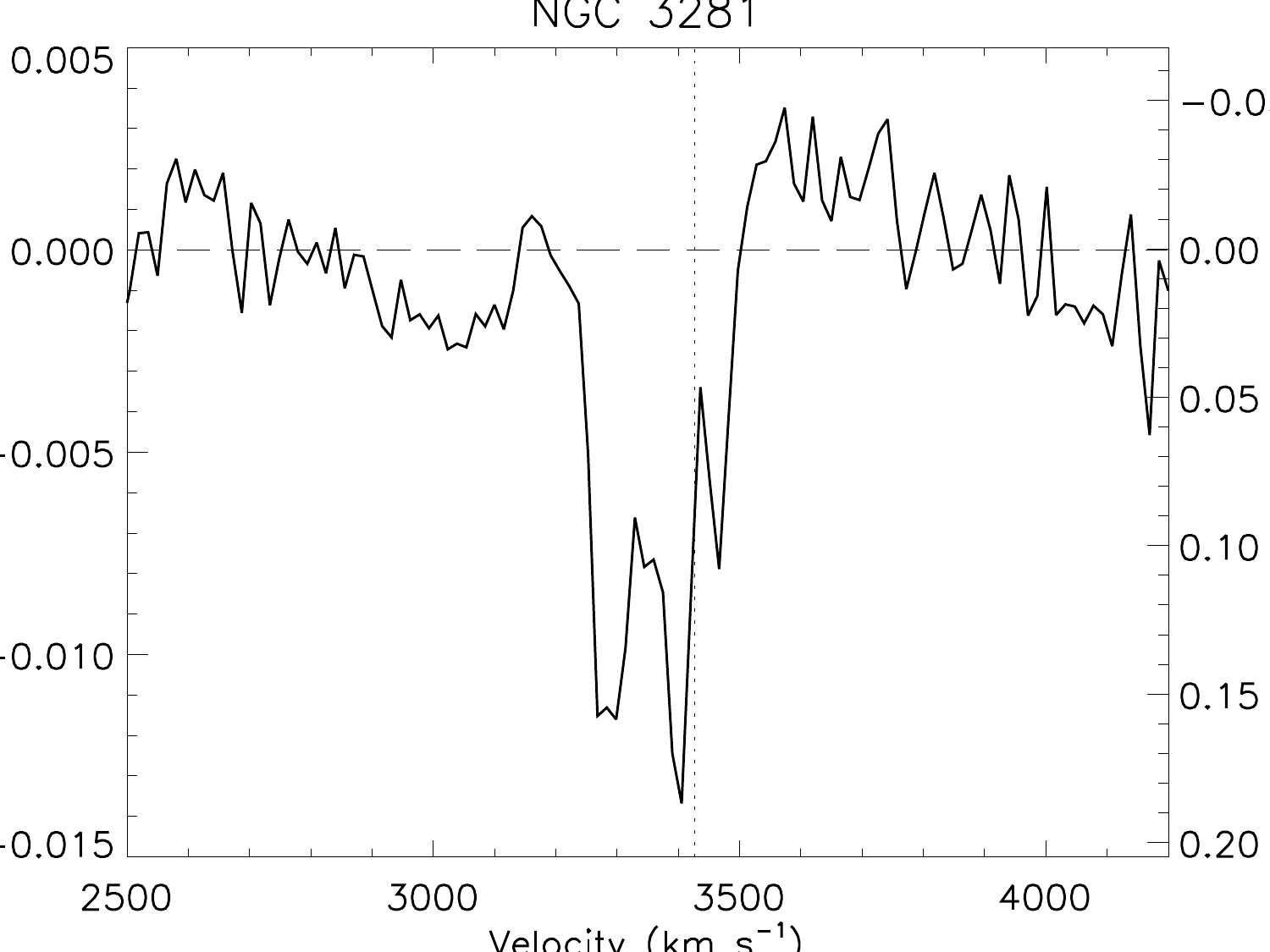}
\caption{Absorption line spectrum of NGC~3281.  The dotted line indicates the average redshift of results from the literature derived from optical emission lines.}
\label{ngc3281}
\end{figure}

NGC~3281 is the brightest object detected in \textit{WISE} and is projected at $\sim460$ kpc from NGC~3268 (Figure \ref{ngc3281w}).  The galaxy hosts one of the most nearby obscured, Compton thick active galactic nuclei (AGN; $N_H>10^{24}$ cm$^{-2}$), and has been a target of X-ray and infrared observations to investigate the nature of the obscuring torus \citep{StorchiBergmann92,Simpson98,Vignali02,Sales11}.  
The $3.4-4.6$ $\mu$m color is at the low end of the typical $W1-W2=0.8$ color cut indicative of AGN activity \citep{Jarrett11,Stern12}, but its 6dF emission line spectrum indicate it is a Seyfert II, and Compton thick AGN are known to span a range of \textit{WISE} colors (Figure 4; \citealt{Gandhi14}).  Redshift measurements in the literature from optical emission lines give an average systemic velocity for the galaxy of $3427\pm26$\kms \citep{Martin76,Sandage78,Rubin85,deVaucouleurs91,Jones09}. 

We find no \hi\ emission associated with the galaxy, but we detect it strongly in absorption (Figure \ref{ngc3281}).  The flux of the continuum source is 78.8 mJy/beam, and the peak of the absorption is -13.4 mJy/beam corresponding to a peak optical depth of 0.186.  By integrating the optical depth over the width of the line and using $N_{HI}=1.82\times10^{18}T_{spin}/f\int\tau(v)dv$ \citep{Wolfe75} we calculate an \hi\ column density of $5.05\times10^{19}\ T_{spin}/f$ cm$^{-2}$.  The absorption spans 230\kms\ and appears to be composed of three velocity components, two of which are blue-shifted with respect to the systemic velocity of the galaxy and the third of which is redshifted.

The canonical value for $T_{spin}/f$ is 100, but could be greater by a factor of 30 or more \citep{Holt06}.  Infrared observations give estimates for the dust temperature at the nucleus of NGC~3281 ranging from 300--1000 K or more \citep{Sales11,Winge00}.  The 300 K estimate is extracted from a 130 kpc region around the nucleus and NGC~3281 is unresolved in radio continuum down to 180 kpc \citep{Ulvestad89}.  Assuming the dust couples well to the neutral gas, this would imply \hi\ column densities of $N_{HI}=1.5-5\times10^{22}$ cm$^{-2}$.

Controversially, an observation from the meridian transit Nan\c{c}ay radiotelescope claims a redshift of 3200\kms\ for \hi\ in emission \citep{Theureau98}.  However, Nan\c{c}ay is not an imaging telescope and the beam (3.6 x 22 arcmin) is significantly elongated.  The determination of the instrumental baseline is likely the most uncertain aspect of accurately measuring the line strength and we speculate that the Nan\c{c}ay survey's polynomial baseline fitting may have fit the absorption, over-correcting, and mimicking emission in adjacent channels.  This is speculation as they do not report a conscientious search for absorption line objects in the survey, but the Nan\c{c}ay spectrum (NED) shows a narrow trough consistent with where we see the strongest absorption in the KAT-7 mosaic.

Redshifted \hi\ may be a source of fuel for the central AGN (e.g.~\citealt{vanGorkom89,Morganti09,Maccagni14}), while blueshifted gas may be the result of a jet driven outflow (e.g.~\citealt{Morganti05,Curran10,Mahony13}).  \citet{StorchiBergmann92} adopt a systemic velocity of 3396\kms\ and find evidence in NGC~3281 for an outflow in ionized gas from [\mbox{O\,{\sc iii}}] and H$\beta$ emission lines. They argue that from the combination of $-43$\kms\ blueshifted and $+125$\kms\ redshifted emission, the outflow is moving in a cone at 150\kms\ with an inclination of $64^{\circ}$.  Assuming a systemic velocity of 3427(3396)\kms, we find the blue-, redshifted absorption span $-174$($-143$)\kms, and $+55$($+86$)\kms, respectively. A detailed geometric analysis of NGC~3281 is outside the scope of this paper, but the optical emission lines are likely probing different size scales and different regions around the AGN than the \hi\ absorption lines. 

Alternatively, we cannot rule out that such an absorption line profile may be the result of \hi\ distributed in a flattened rotating disc \citep{Curran10,Gereb14}.  A jet in the plane of the sky would produce both red and blueshifted absorption on either side of the core \citep{Peck99}.

\section{Summary and Conclusions}

The Antlia Cluster is dynamically one of the youngest clusters in the local Universe, consisting of two substructures in the process of merging as revealed by the galaxy populations, diffuse X-ray emission, and globular cluster populations surrounding the two central dominant elliptical galaxies, NGC~3268 and NGC~3258. 

By combining KAT-7 \hi\ spectral line commissioning observations with \textit{WISE} broadband infrared and existing optical spectroscopic redshifts, we present a significant advance in our understanding of the dynamical state of the cluster.  From our \hi\ mosaic, X-ray image, and kinematic DS test, we conclude that NGC~3268 is at the center of an older, more dynamically mature structure, and nominally the core of the Antlia Cluster.
Surprisingly, we find four \hi\ detected galaxies in close projection to NGC~3258.  However, even if they are a chance alignment, the kinematics suggest the system is secondary in the substructure hierarchy.

We identify the infall region beyond 200 kpc based on the spatial distribution and dynamical mass estimates from the \hi\ detected galaxies compared to the predominantly early-type optical galaxy population.  Between 200 and 600 kpc radius, \hi\ detected galaxies show signs that they have undergone significant stripping of their atomic gas.  The largest cluster galaxies by stellar mass show the greatest \hi\ deficiency, down by 2-3 orders of magnitude compared to their counterparts of the same stellar mass in the field.  Beyond 600 kpc Antlia is still actively accreting galaxies from the surrounding environment, notably from the direction toward the Hydra Cluster.

Using the most complete sample of new and archival velocity redshifts, we estimate the cluster mass within a radius of 600 kpc to 1 Mpc to be $2.6\pm0.6\times10^{14}$\msun, and show that X-ray observations have underestimated the mass within the 200--250 kpc X-ray halo by nearly a factor of five.
We find that the velocity distribution of \hi\ selected galaxies compared to the predominantly early-type sample are consistent with what has been modeled in other galaxy clusters: that late-type galaxies as traced by the \hi\ content, are on primarily radial orbits, falling into a virialized system.  Further, this accretion on to the Antlia Cluster is strongly asymmetric.  

We use the Dressler-Schectman test to identify infalling substructure within Antlia and on the outskirts by deviations from the global cluster kinematics.  By examining galaxies within these substructures for their \hi\ content and star formation rate, we propose a relative sequence for their accretion on to the Antlia Cluster which is also consistent with their distance from the cluster center.  
We speculate that both the infall of individual galaxies from the field and the accretion of galaxy groups are important to the ongoing mass assembly of Antlia.  Additionally, there is evidence that kinematics and asymmetry of the cluster \hi\ distribution reflect the larger scale filaments and velocity flows of the super-cluster environment towards the Hydra Cluster, although we are unable to say anything conclusive.

Finally, we report on the \hi\ absorption associated with the Compton thick Seyfert II galaxy, NGC~3281.  The object belongs to a group of galaxies we identify from the DS test, but it is of further interest for studies of both AGN outflows and star formation driven winds.

Despite the story of cluster assembly we have presented, we cannot rule out that some of the substructure we identify may be foreground or background objects seen in projection.  Accurate distances require redshift independent measurements such as the Tully-Fisher relation.  Finally, to gain a greater understanding of how the infalling population is impacted by the cluster environment requires high resolution \hi\ observations.  These have the potential to reveal direct morphological evidence of interaction between the hot intracluster medium and cold interstellar medium of the galaxies through ram pressure stripping.

Antlia is an important addition to the collection of clusters in the local universe for which we are able to pursue detailed, high-resolution, multi-wavelength synthesis of the stellar, gas, and dark matter content, which in turn can shed light on mass assembly, cluster formation, and galaxy transformation.  MeerKAT will easily reach a factor of 10 deeper in \hi\ mass and resolving both the gas morphology in \hi\ and star formation in radio continuum.

Antlia's apparent youth provides a local snapshot of what more mature clusters may have looked like in the past, and in greater detail than what we can directly observe in intermediate redshift clusters.  Antlia in combination with other massive systems have the potential to show whether the way in which galaxies are accreted on to clusters, in groups or individually, impacts the resulting galaxy cluster population.  Although not the focus of this paper, the high fraction of S0 to elliptical galaxies in Antlia suggests that it may have a different formation history in contrast to other well-studied clusters.  Whether Antlia contains clues as to the different formation mechanisms of Es versus S0s, remains to be seen.

\section*{Acknowledgements}
We thank the anonymous referee for his/her helpful comments that improved the quality of this paper.  We thank Tom Oosterloo for his insightful discussions and sharing with us his continuum subtraction code, and Thijs van der Hulst for his useful comments.  The research of KH has been supported by a South African Research Chairs Initiative (SARChI), Square Kilometer Array South Africa (SKA SA) fellowship, and support from the European Research Council under the European Union's Seventh Framework Programme (FP/2007-2013) / ERC Grant Agreement nr. 291531.  The work of TJ and CC is based upon research supported by the SARChI program of the Department of Science and Technology (DST), the SKA SA, and the National Research Foundation (NRF). 

This work is based on observations obtained at the KAT-7 array which is operated by the SKA SA on behalf of the National Research Foundation of South Africa.

This publication makes use of data products from the Wide-field Infrared Survey Explorer, which is a joint project of the University of California, Los Angeles, and the Jet Propulsion Laboratory/California Institute of Technology, funded by the National Aeronautics and Space Administration (NASA).

This research has made use of the NASA/IPAC Extragalactic Database (NED) which is operated by the Jet Propulsion Laboratory, California Institute of Technology, under contract with the National Aeronautics and Space Administration.

\appendix
\section{Supplementary Figures}

Here we provide the \hi\ spectral line profiles for all the KAT-7 detections.  All profiles have the same velocity scale, but the intensity is adjusted to the individual detections (Figure \ref{hispecAppen}-\ref{hispecAppen3}).

\begin{figure*}
\begin{center}$
\begin{array}{ccc}
\includegraphics[width=2.3in,trim=20 0 0 0]{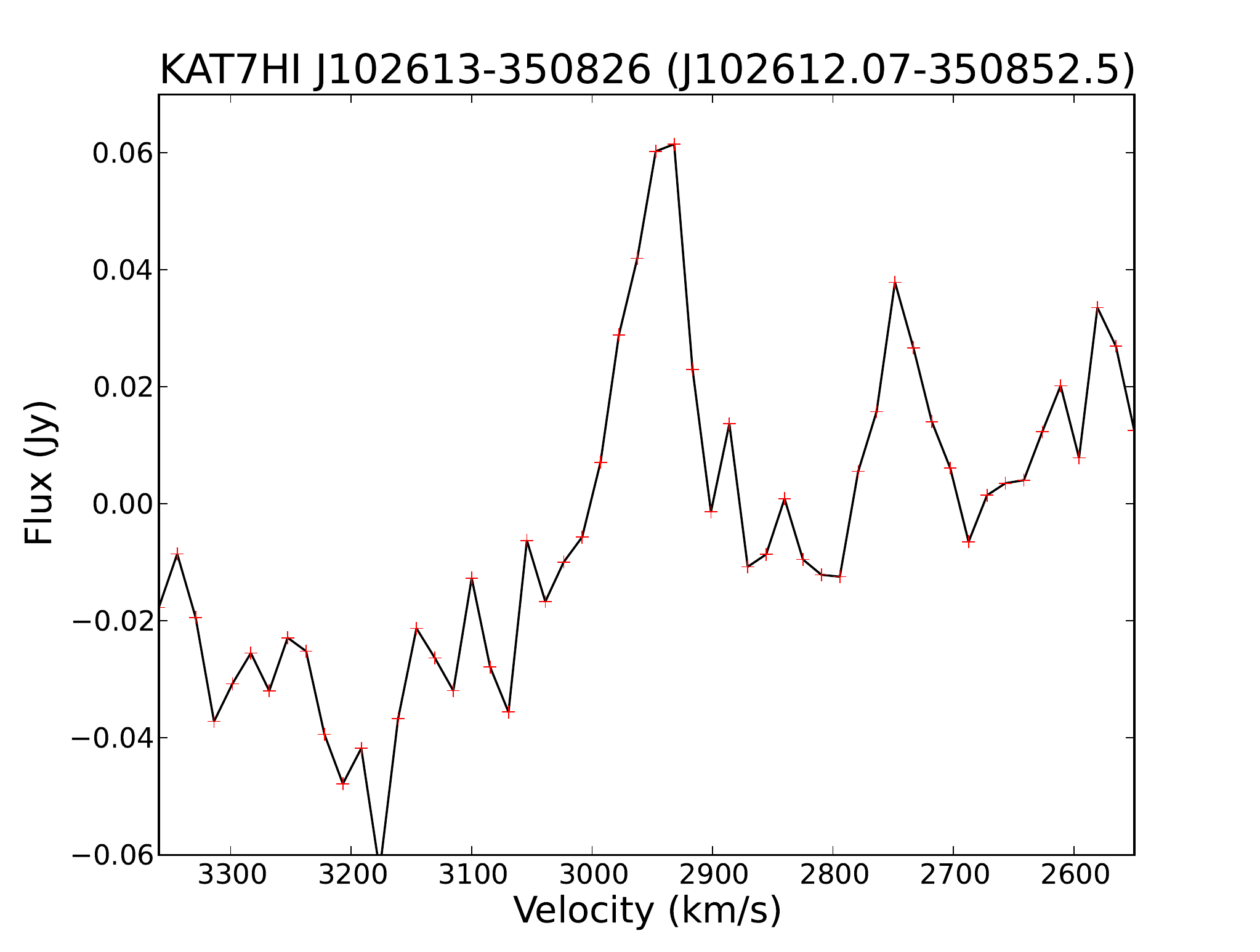} & \includegraphics[width=2.3in,trim=20 0 0 0]{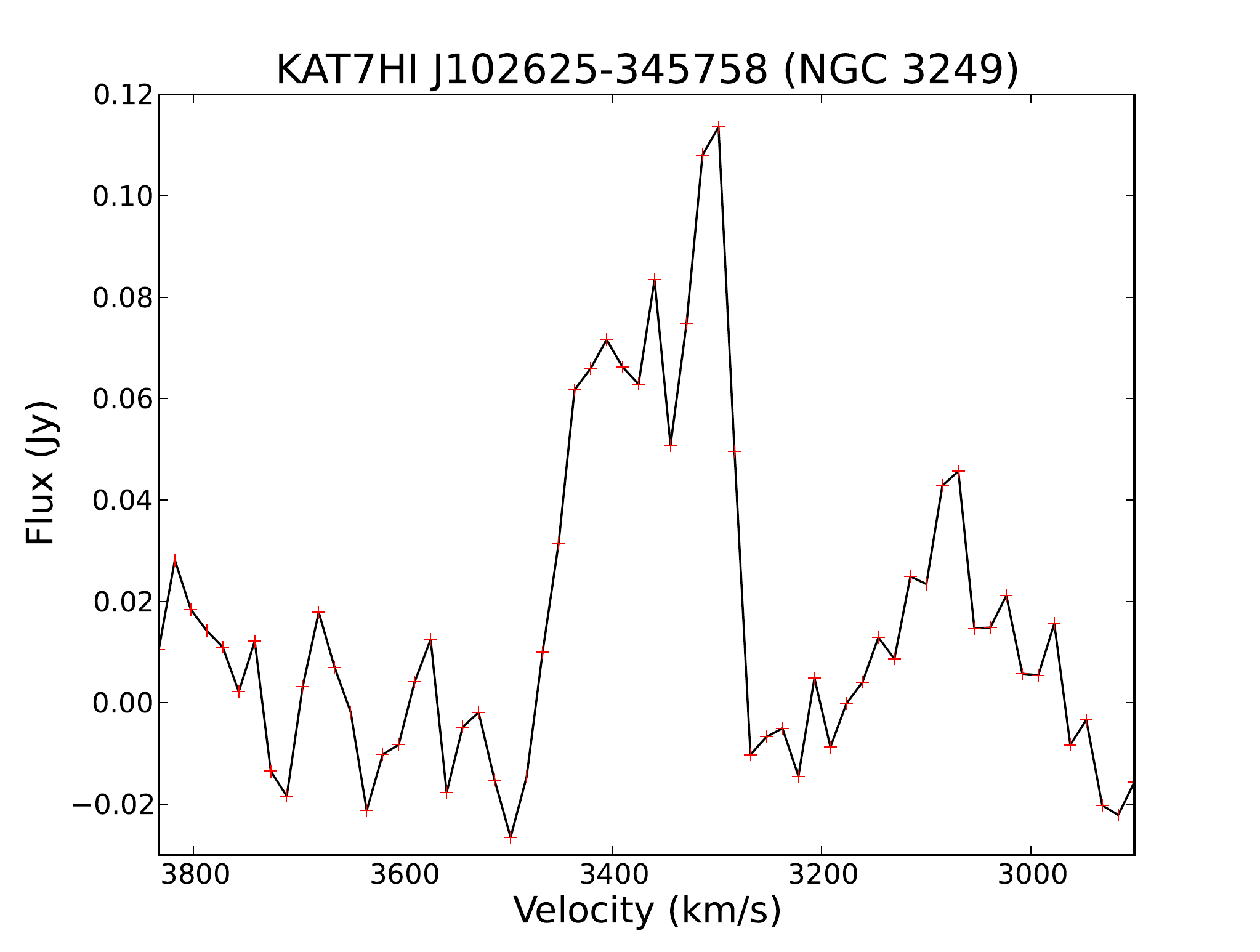} & \includegraphics[width=2.3in,trim=20 0 0 0]{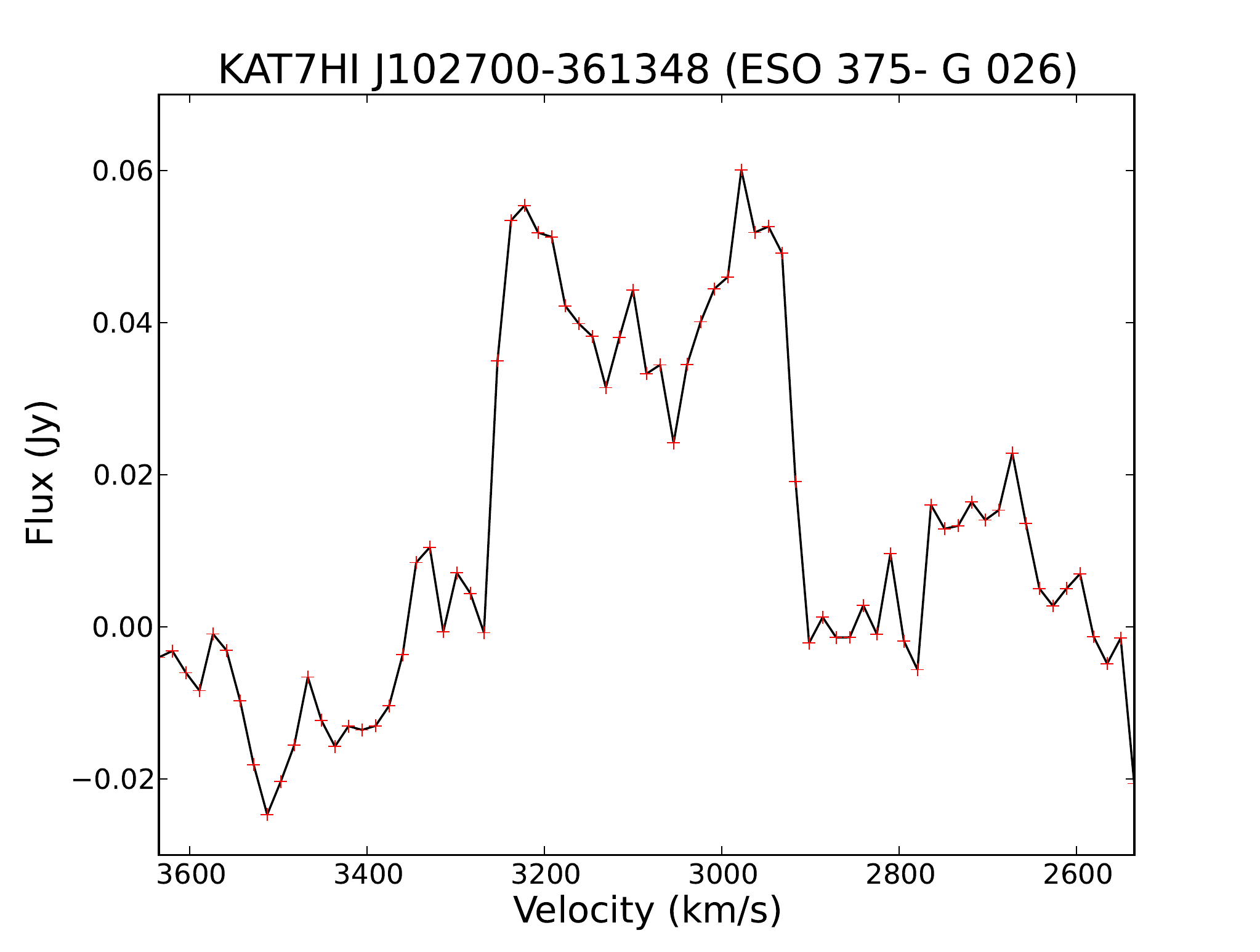} \\
\includegraphics[width=2.3in,trim=20 0 0 0]{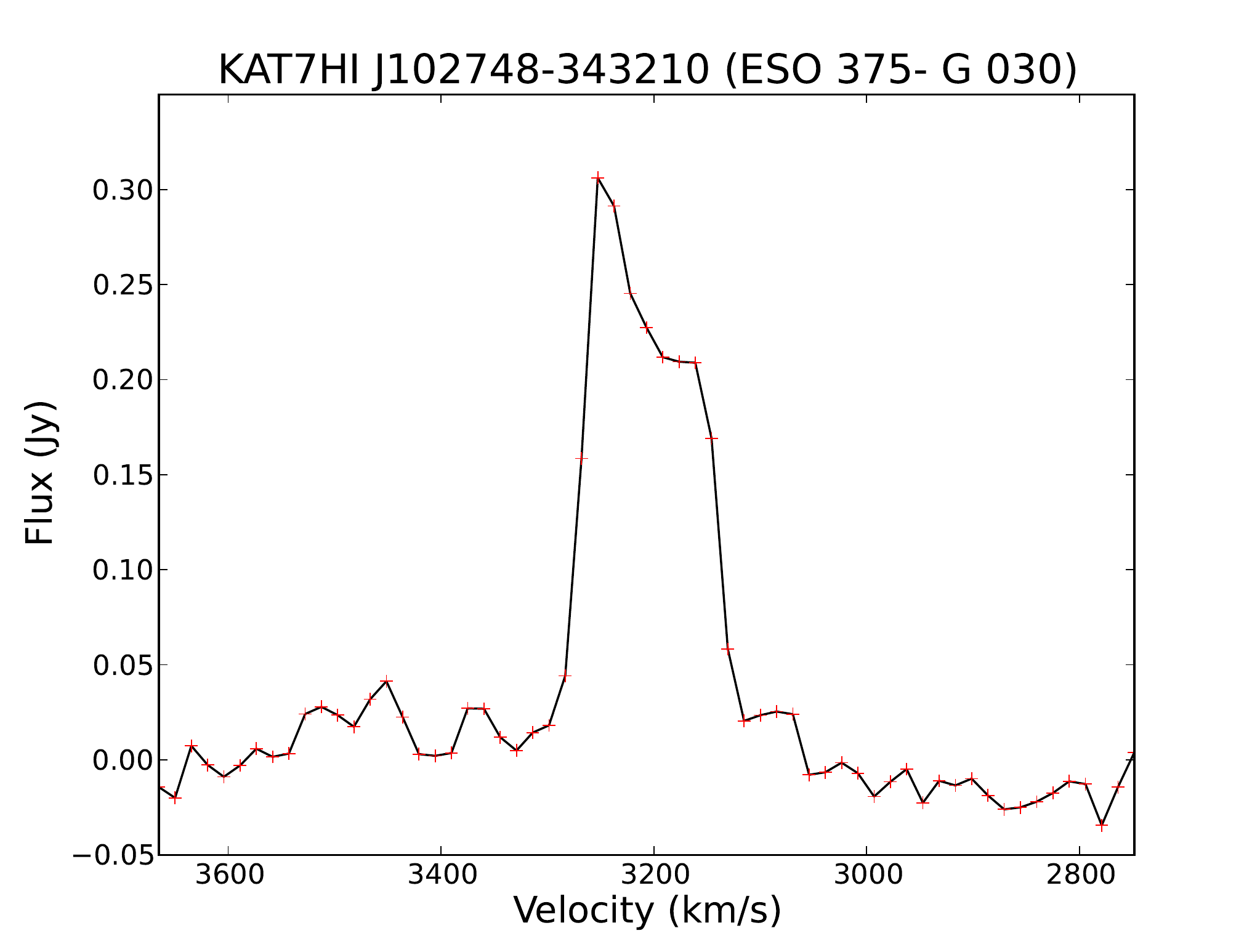} & \includegraphics[width=2.3in,trim=20 0 0 0]{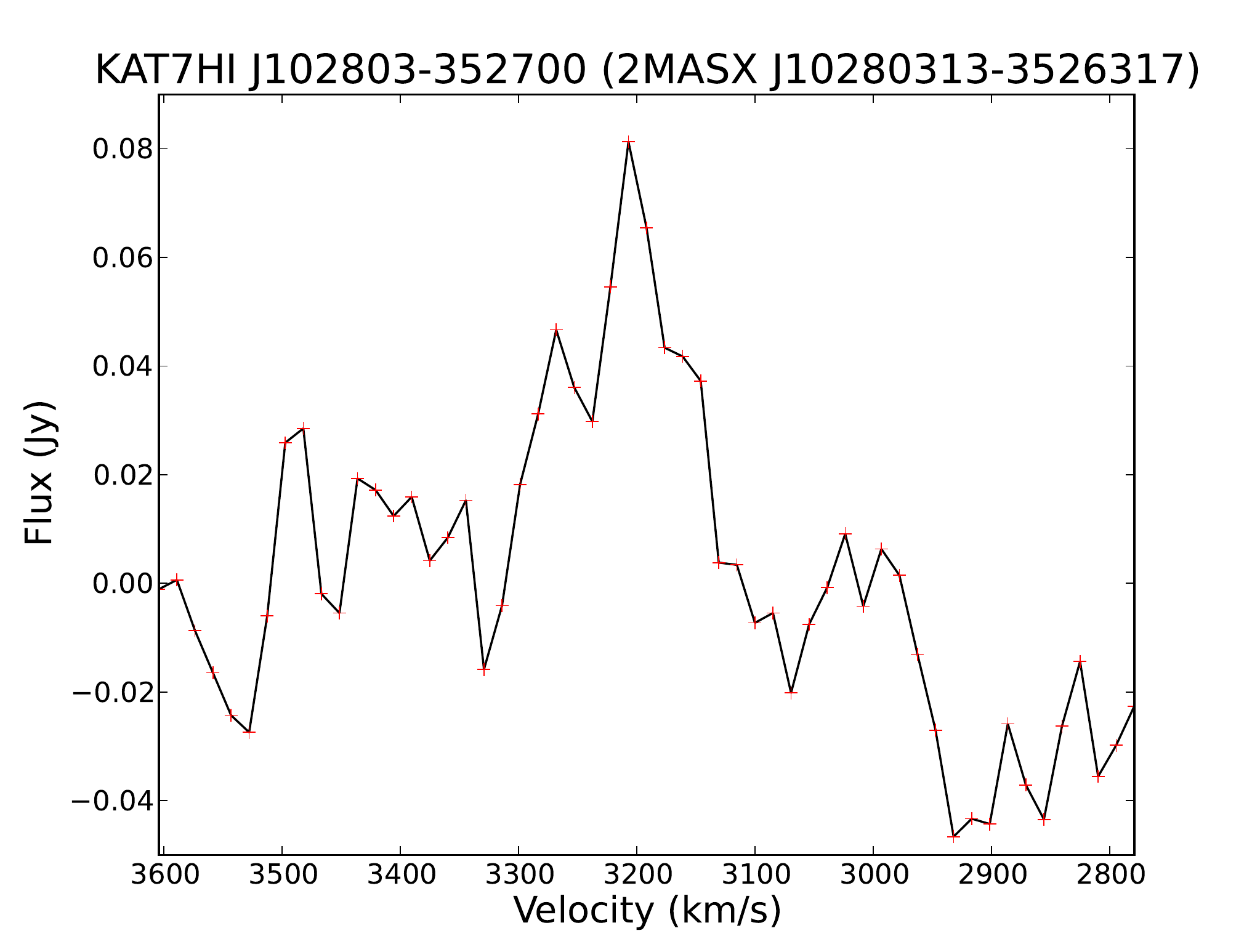} & \includegraphics[width=2.3in,trim=20 0 0 0]{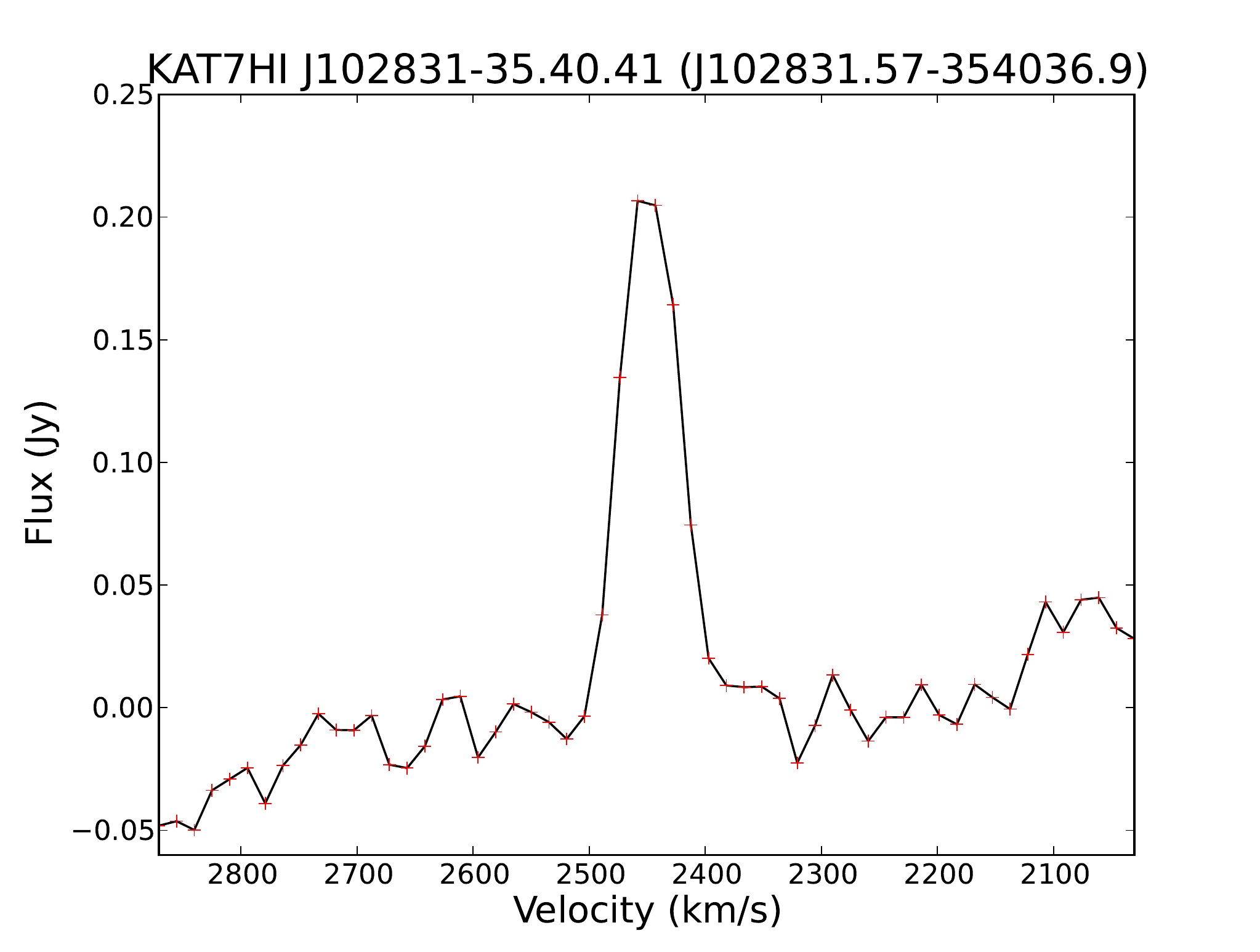} \\
\includegraphics[width=2.3in,trim=20 0 0 0]{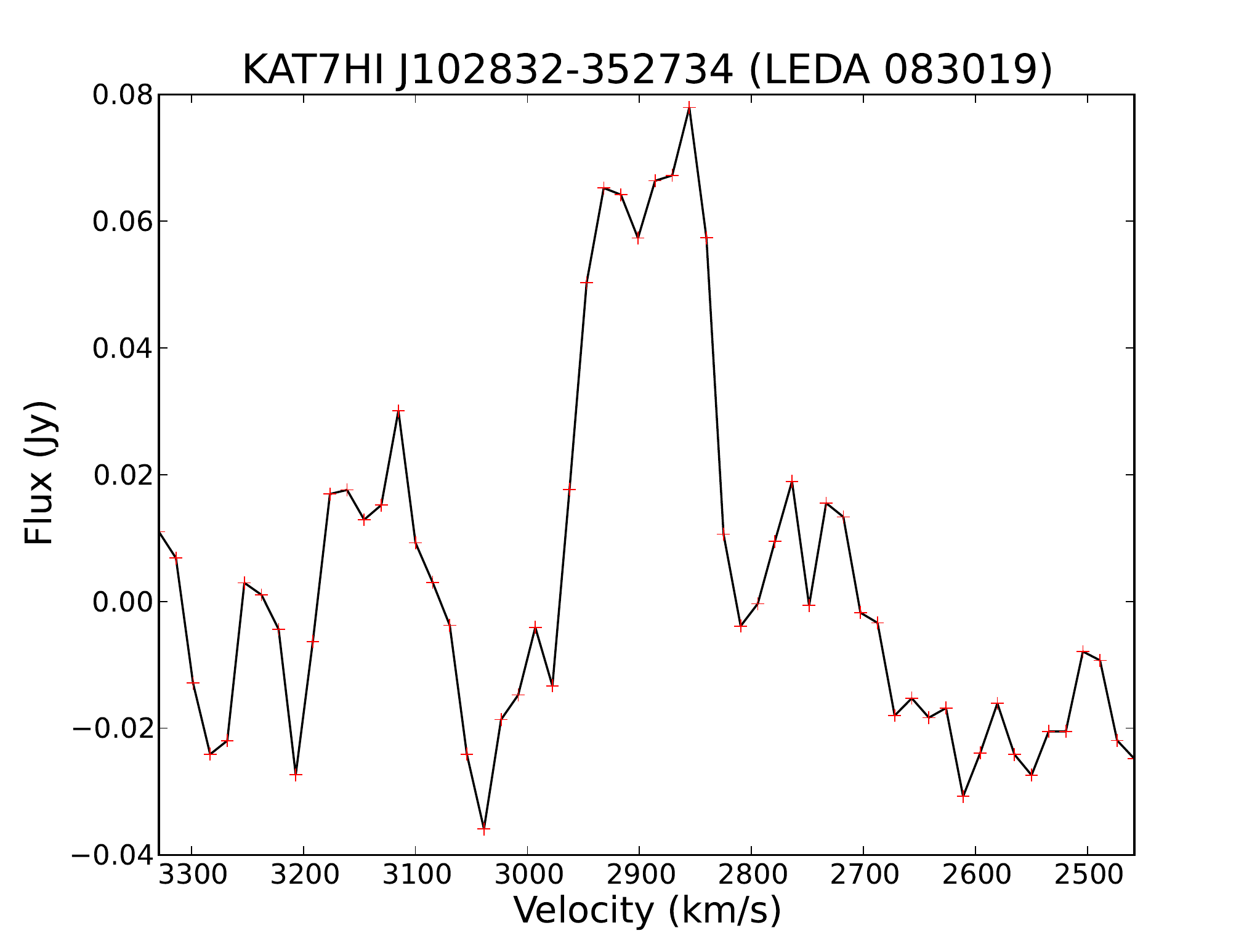} & \includegraphics[width=2.3in,trim=20 0 0 0]{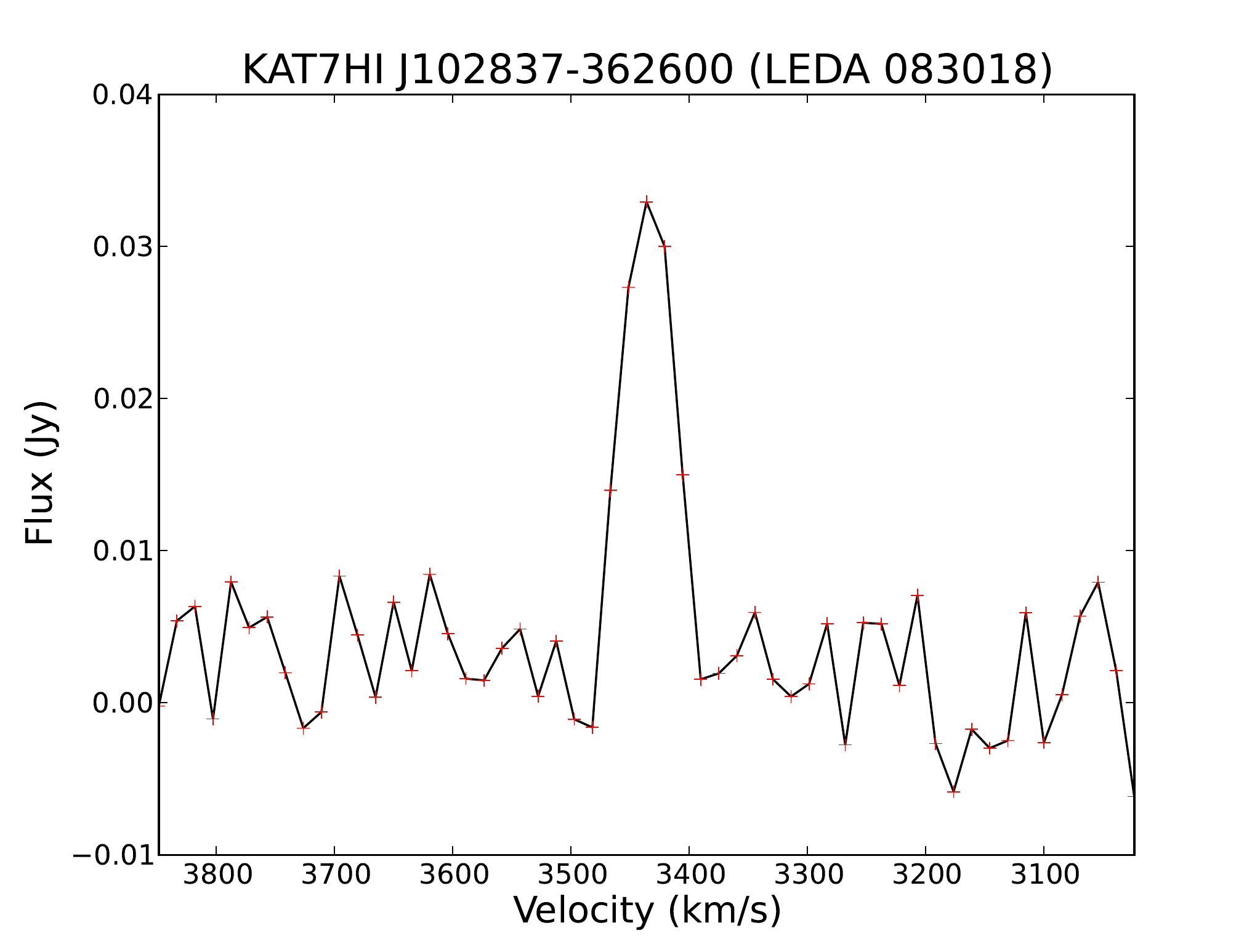} & \includegraphics[width=2.3in,trim=20 0 0 0]{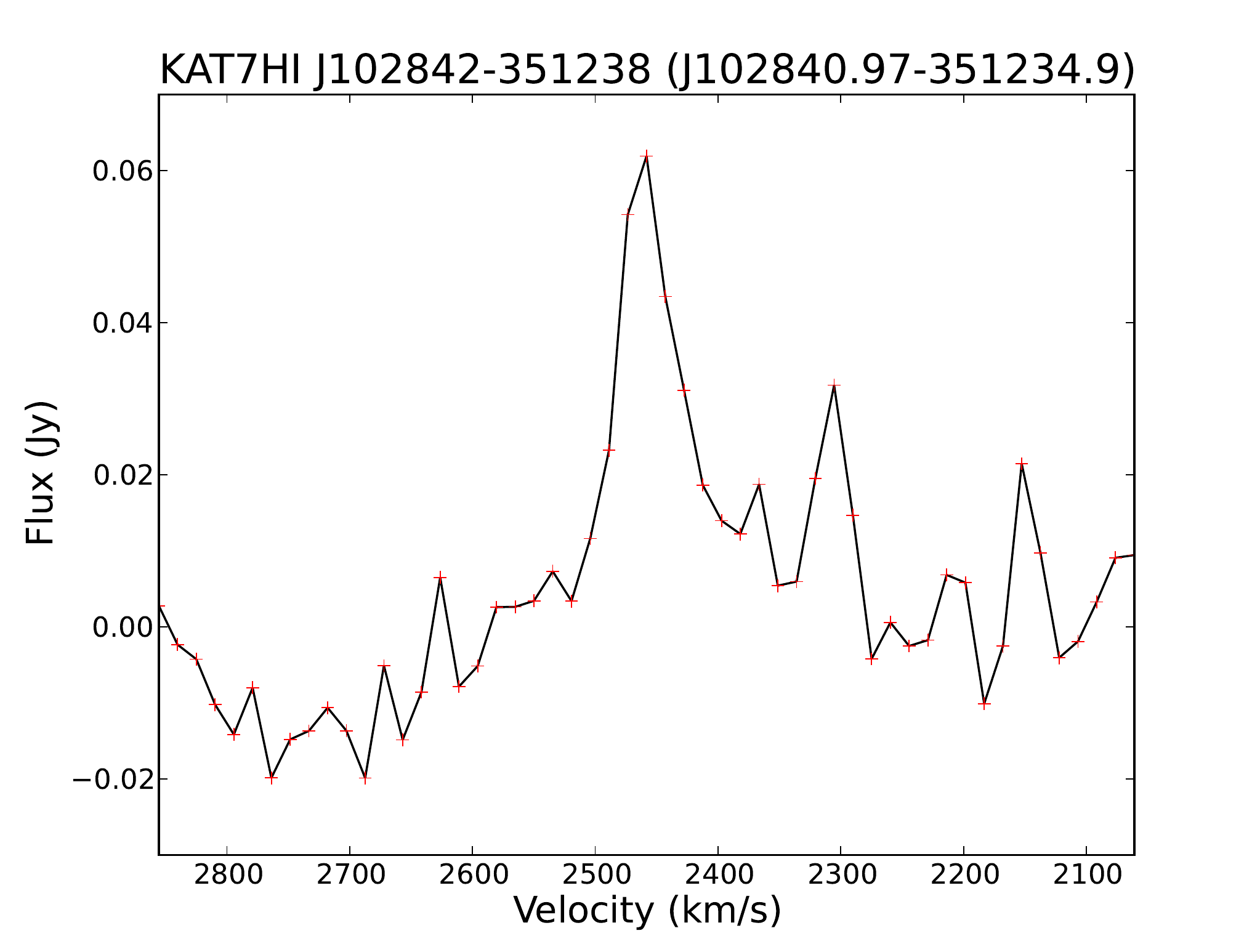} \\
\includegraphics[width=2.3in,trim=20 0 0 0]{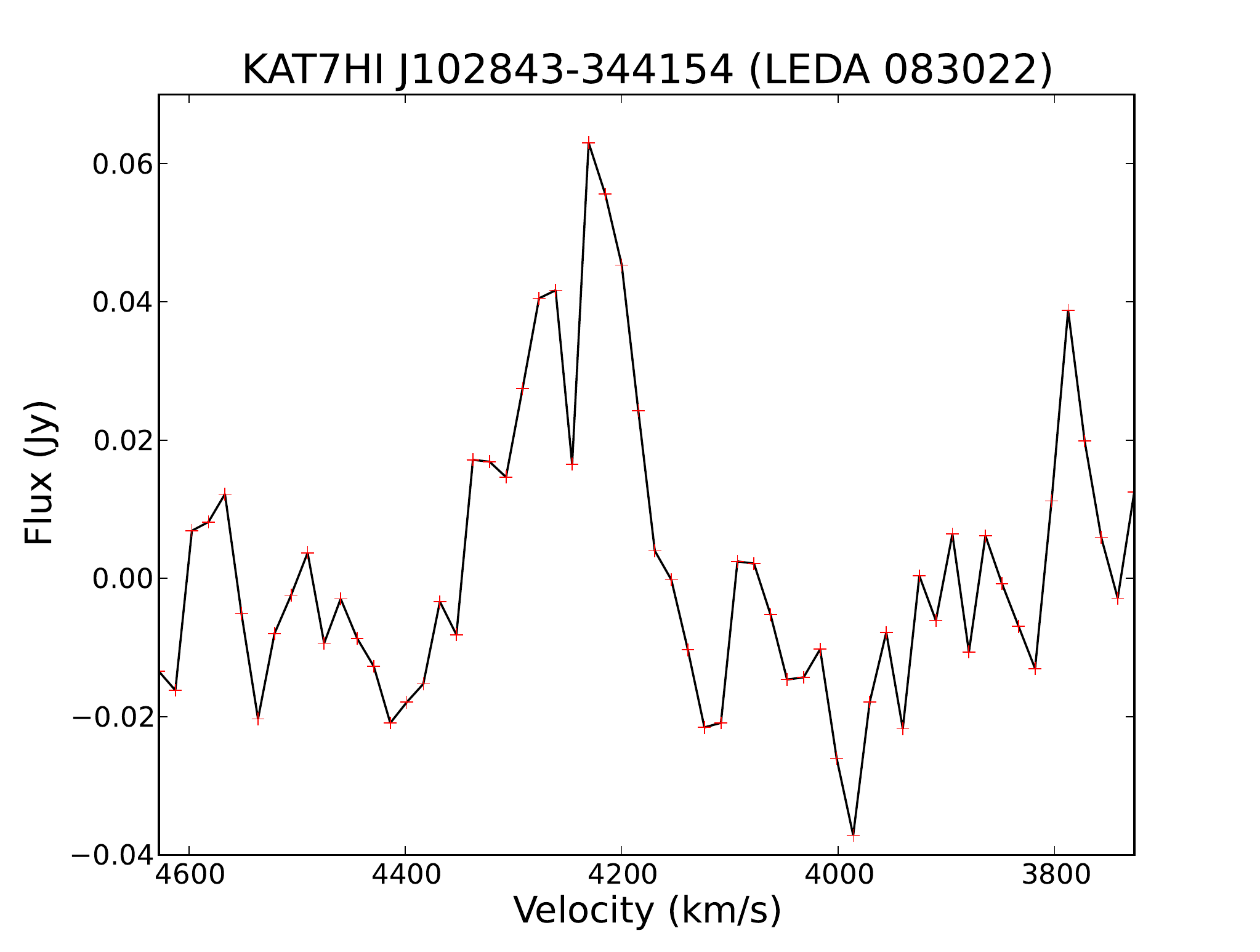} & \includegraphics[width=2.3in,trim=20 0 0 0]{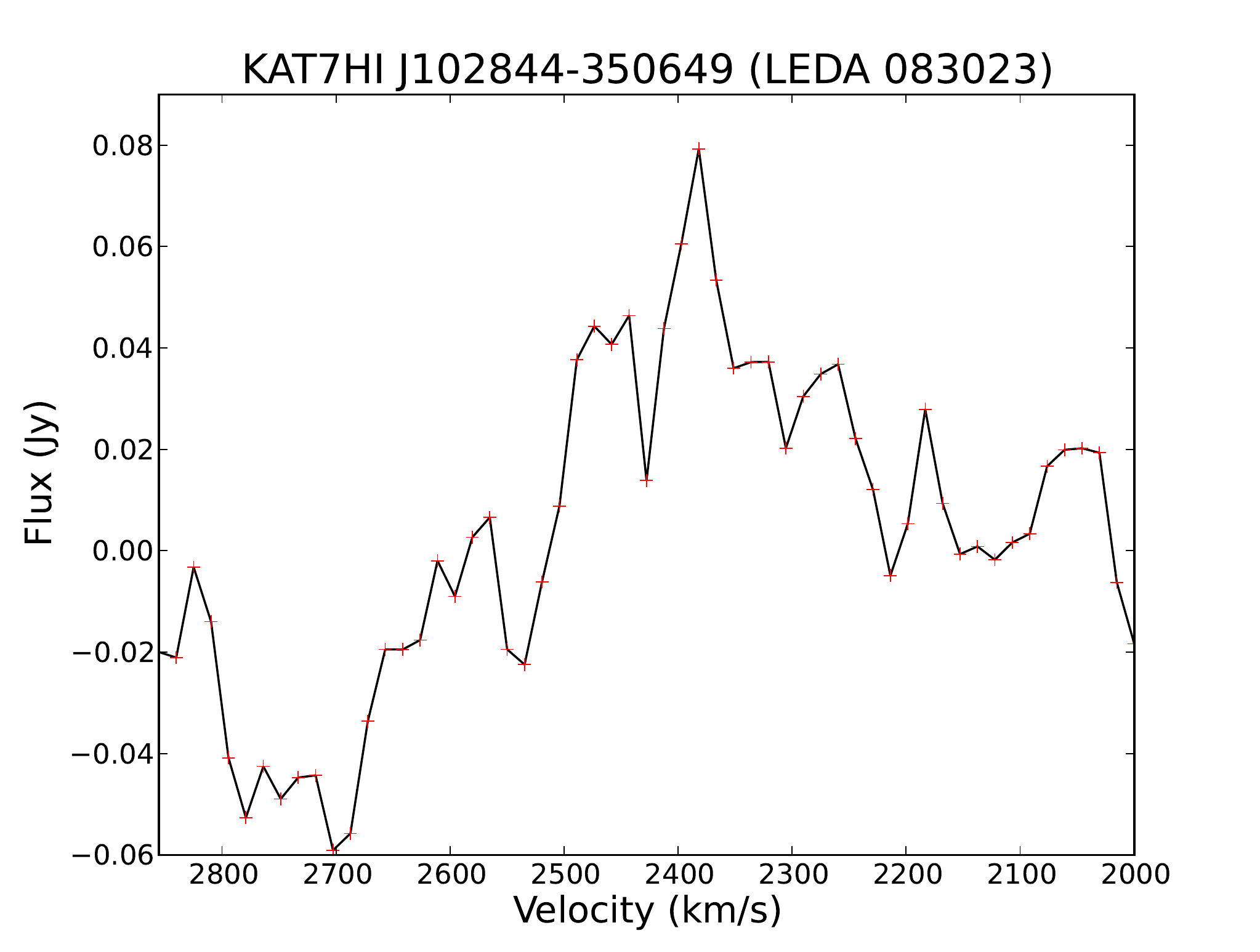} & \includegraphics[width=2.3in,trim=20 0 0 0]{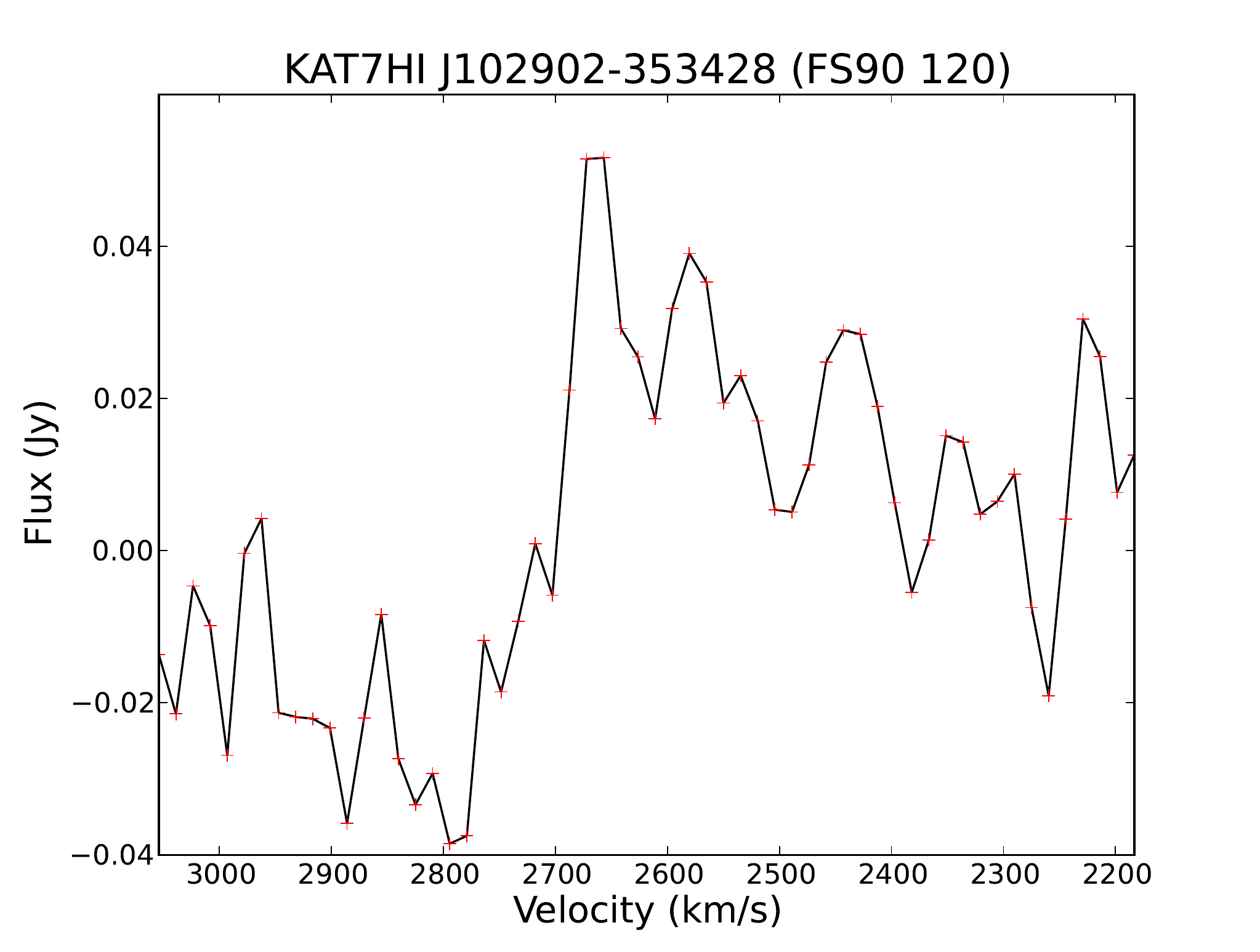} \\
\includegraphics[width=2.3in,trim=20 0 0 0]{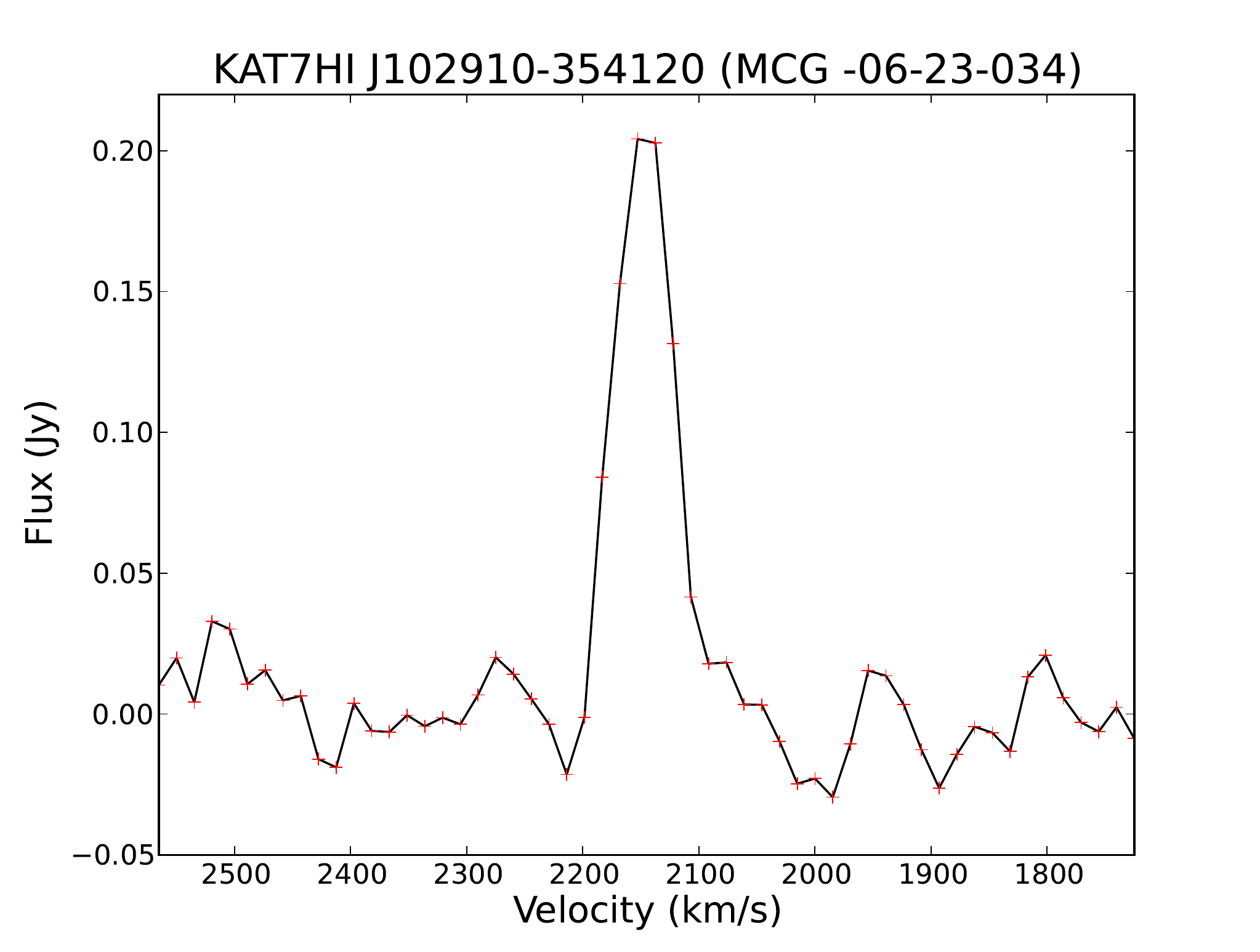} & \includegraphics[width=2.3in,trim=20 0 0 0]{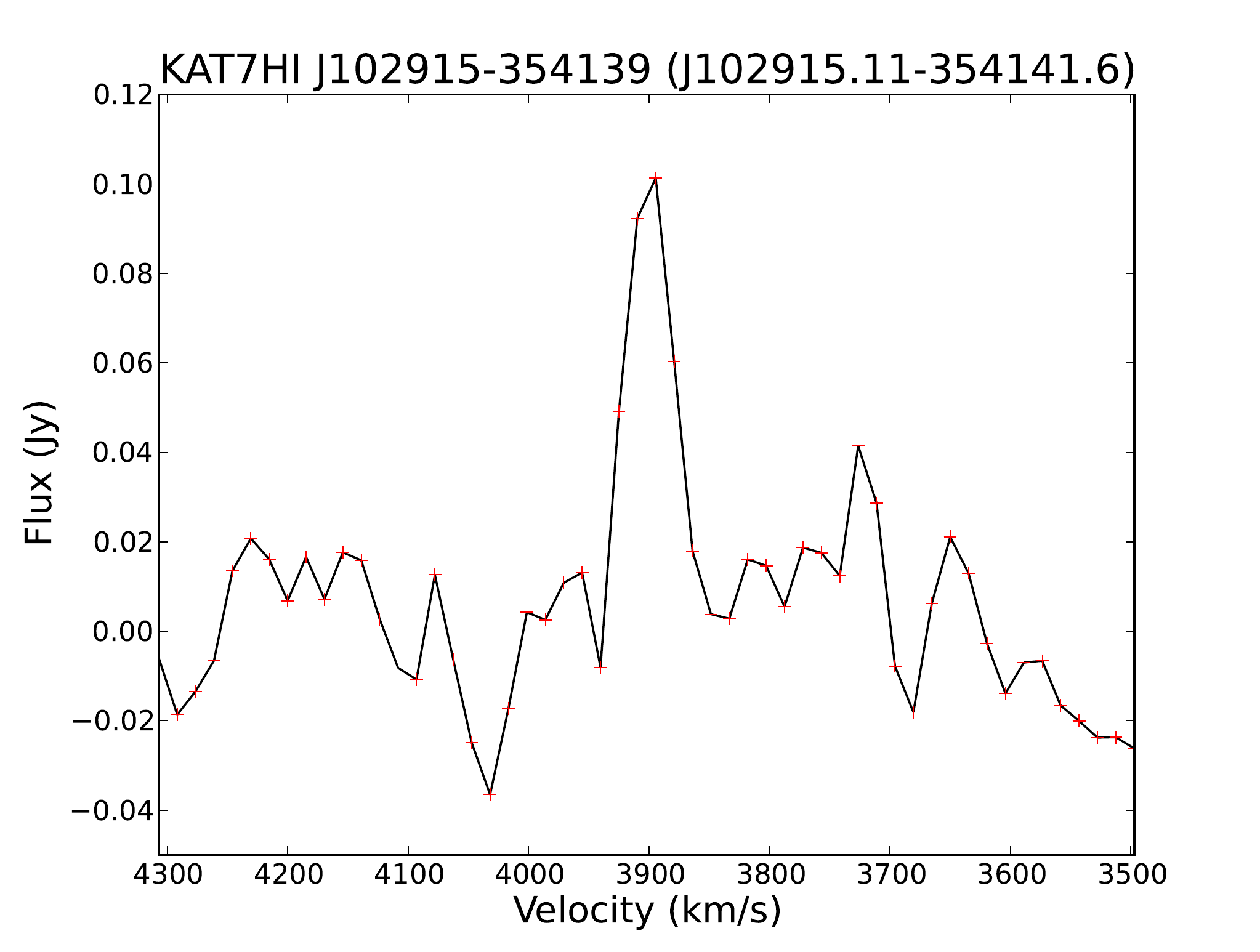} & \includegraphics[width=2.3in,trim=20 0 0 0]{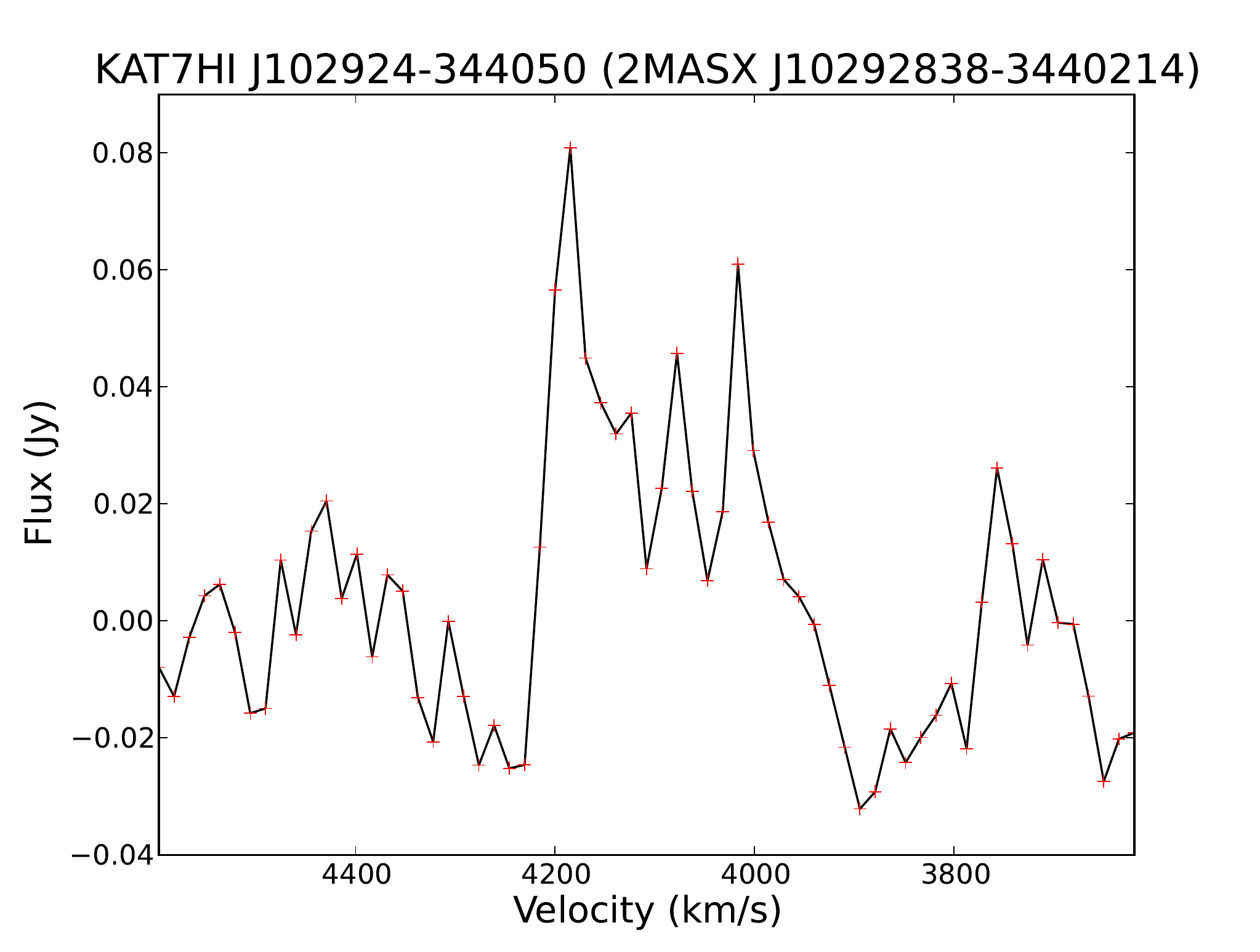} 
\end{array}$
\end{center}
\caption{\hi\ profiles of the KAT-7 detections detected in both the 15\kms\ and 31\kms\ cubes.  Objects 1-30 appear in the same order as Table 2.}
\label{hispecAppen}
\end{figure*}

\begin{figure*}\contcaption{}
\begin{center}$
\begin{array}{ccc}
\includegraphics[width=2.3in,trim=20 0 0 0]{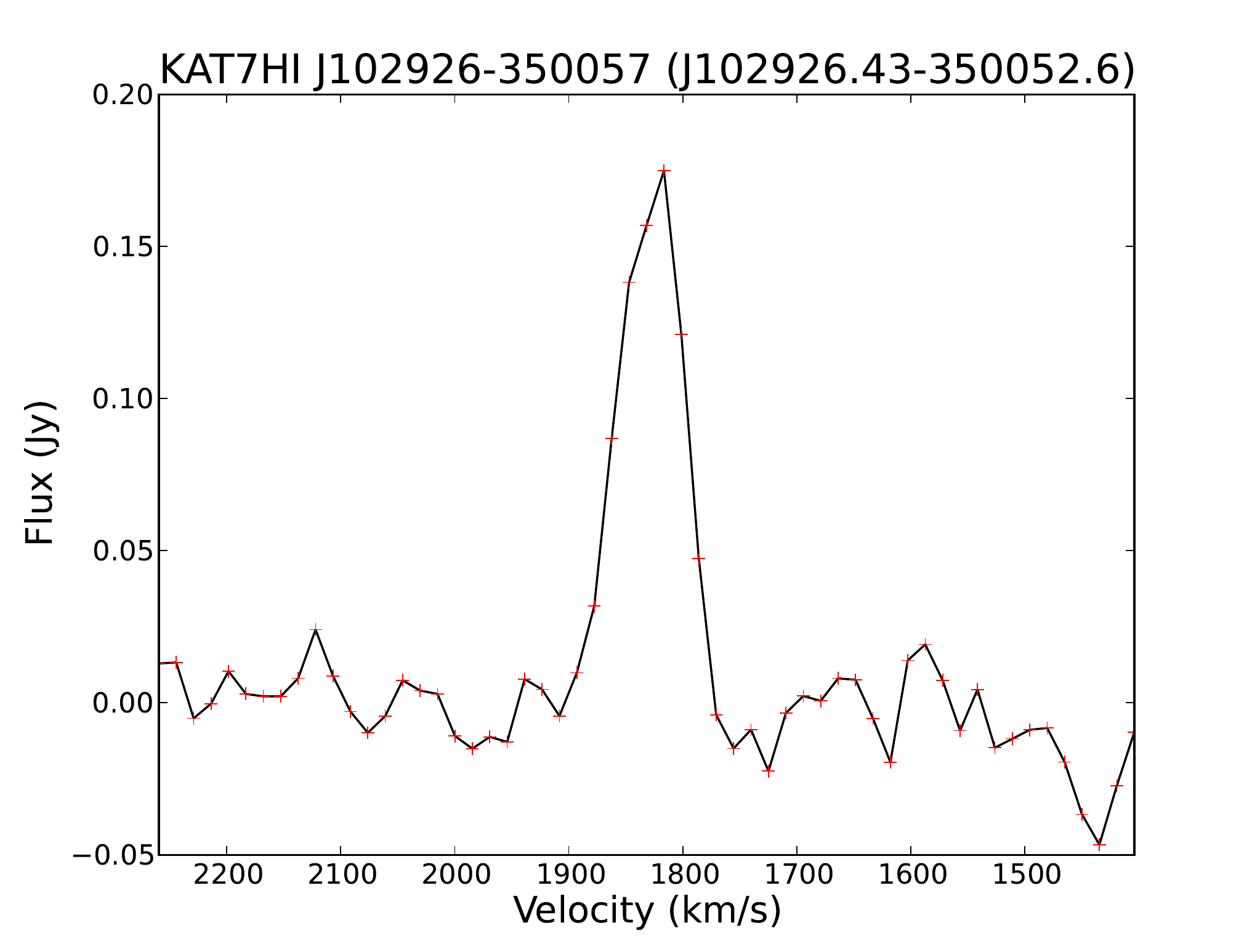} & \includegraphics[width=2.3in,trim=20 0 0 0]{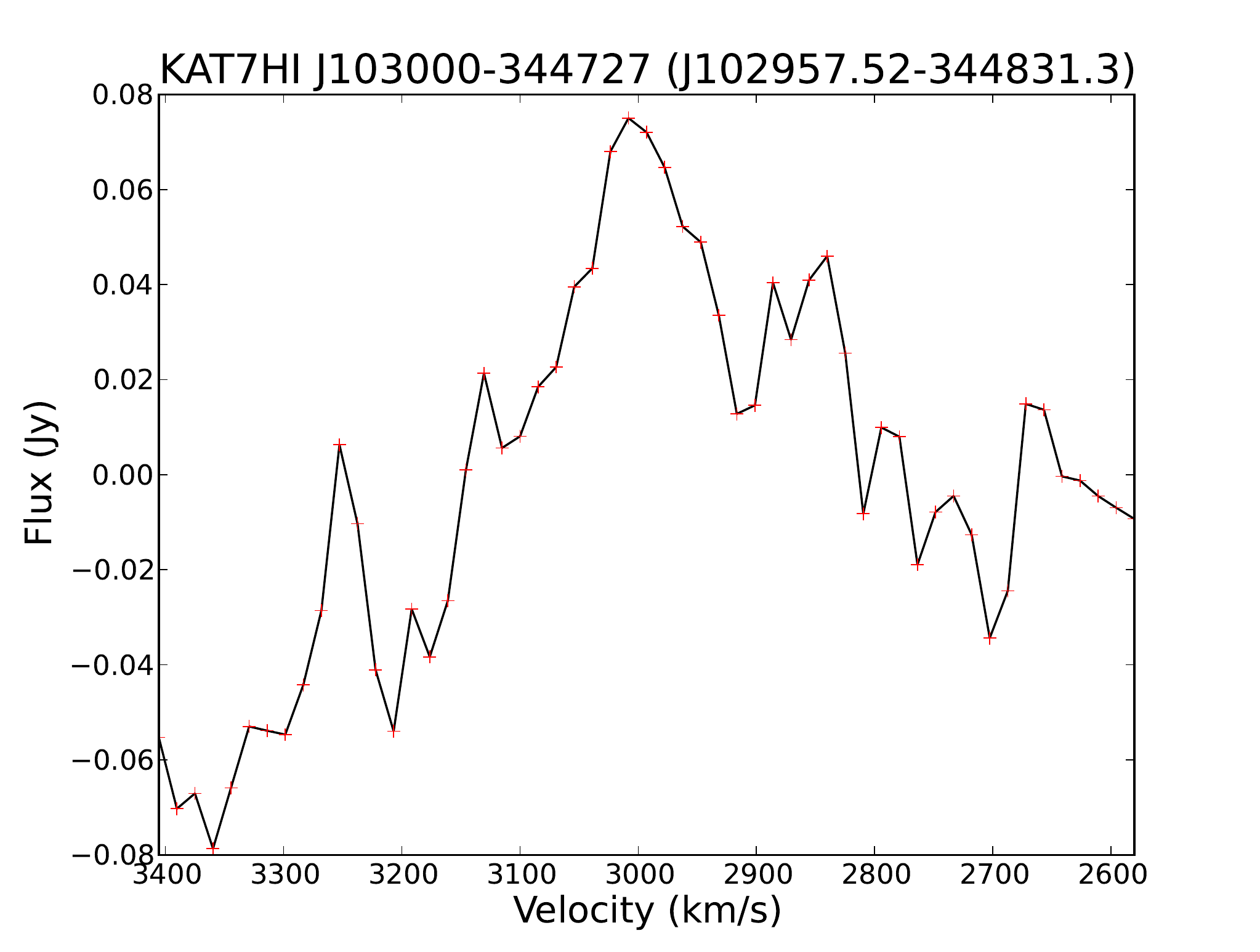} & \includegraphics[width=2.3in,trim=20 0 0 0]{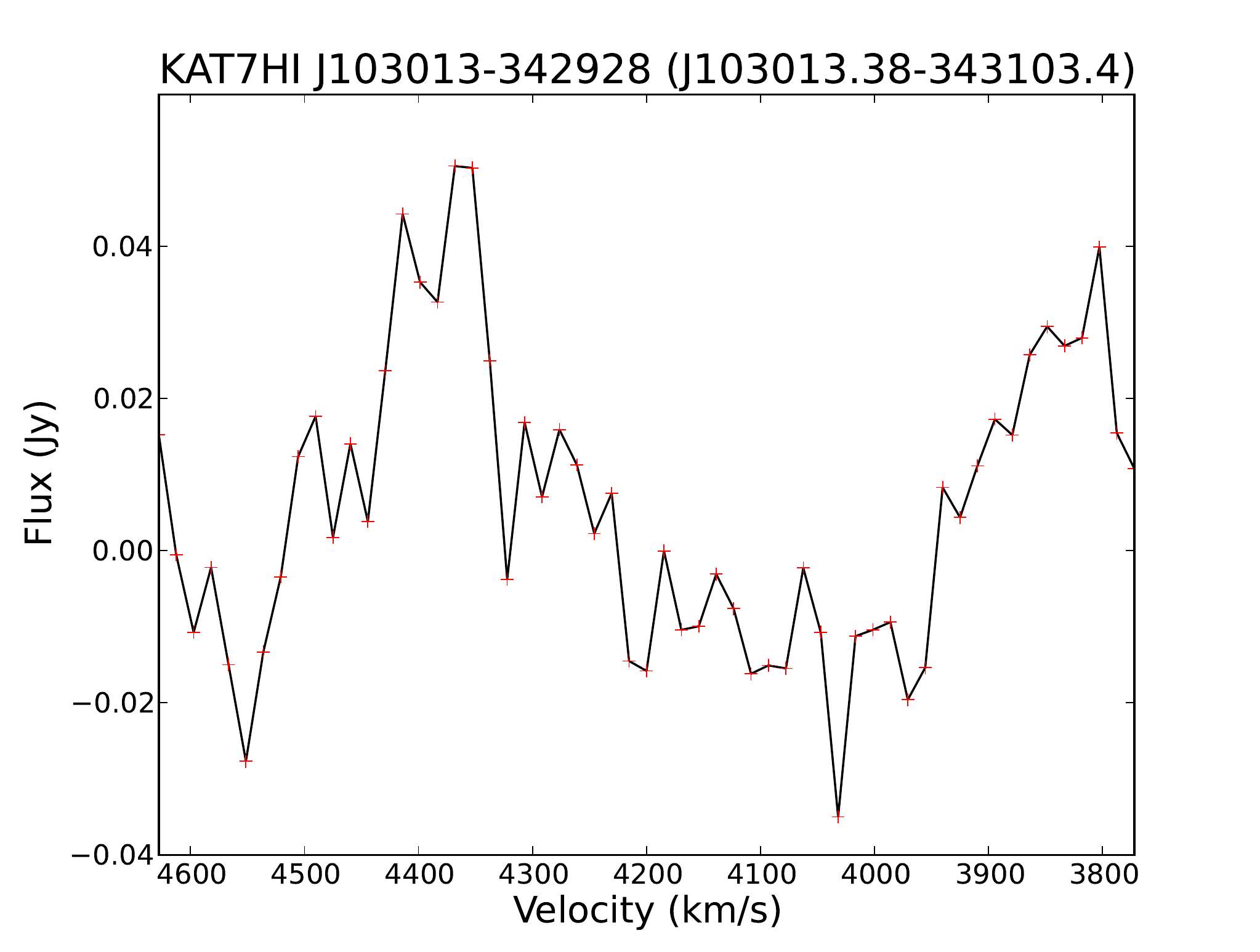} \\
\includegraphics[width=2.3in,trim=20 0 0 0]{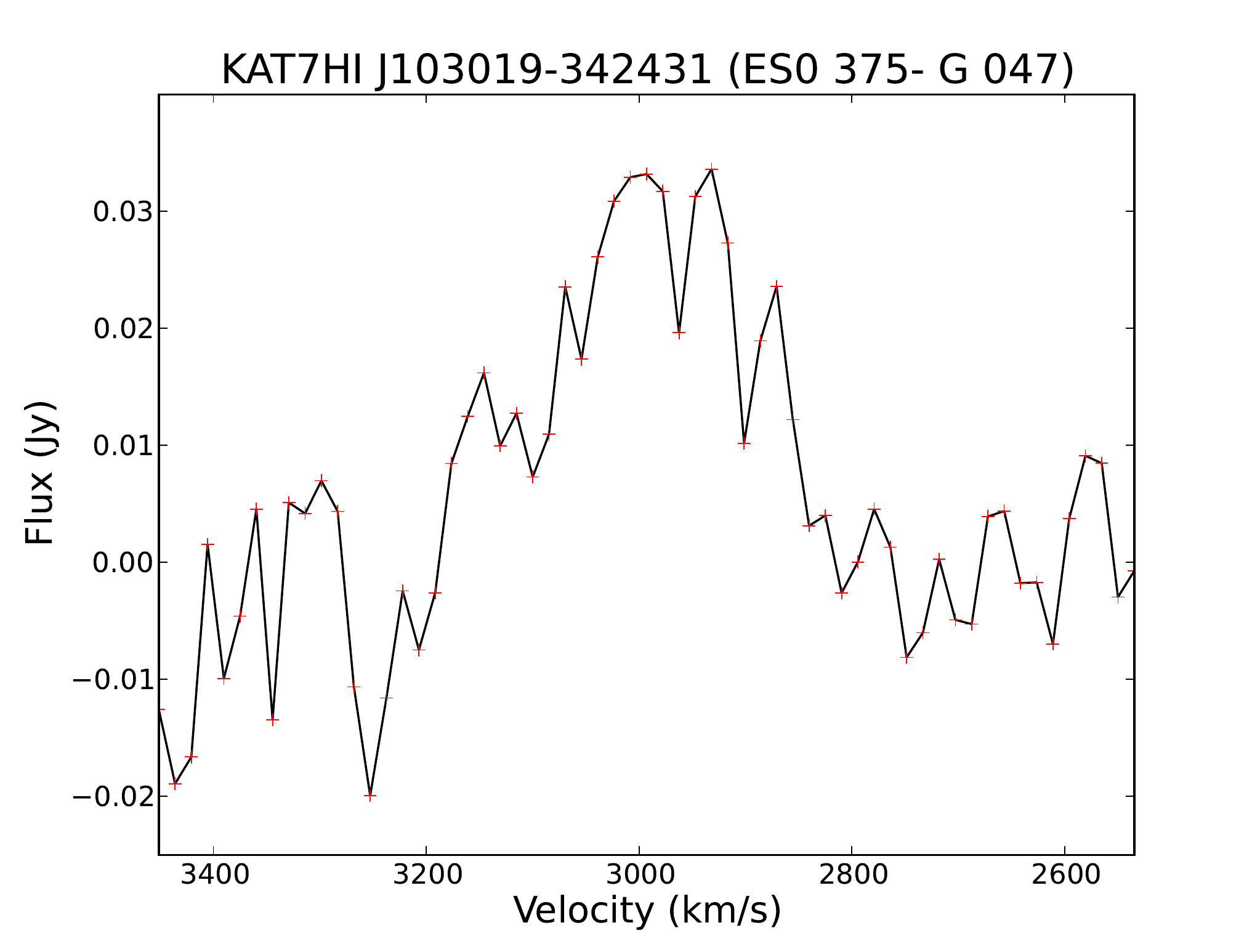} & \includegraphics[width=2.3in,trim=20 0 0 0]{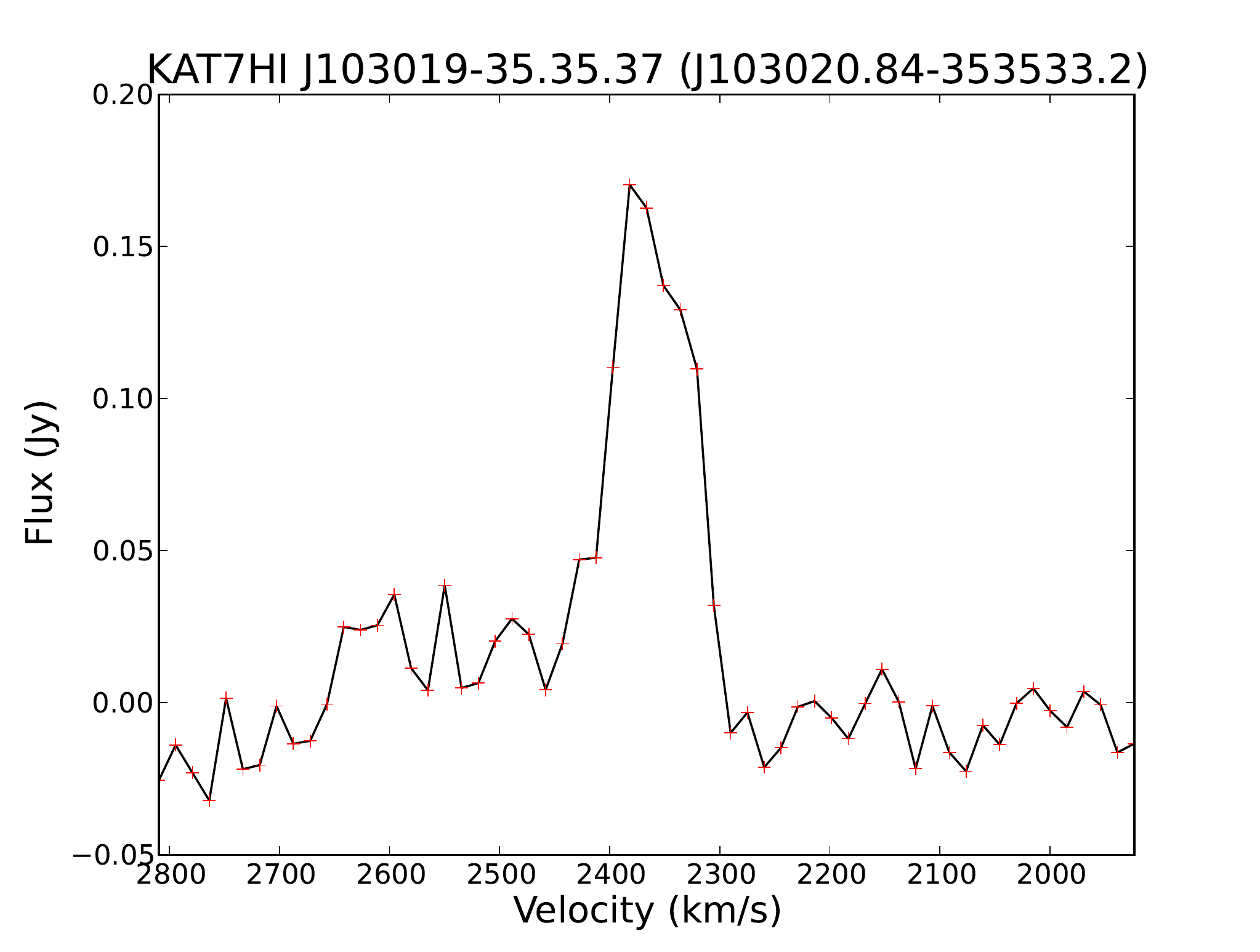} &
\includegraphics[width=2.3in,trim=20 0 0 0]{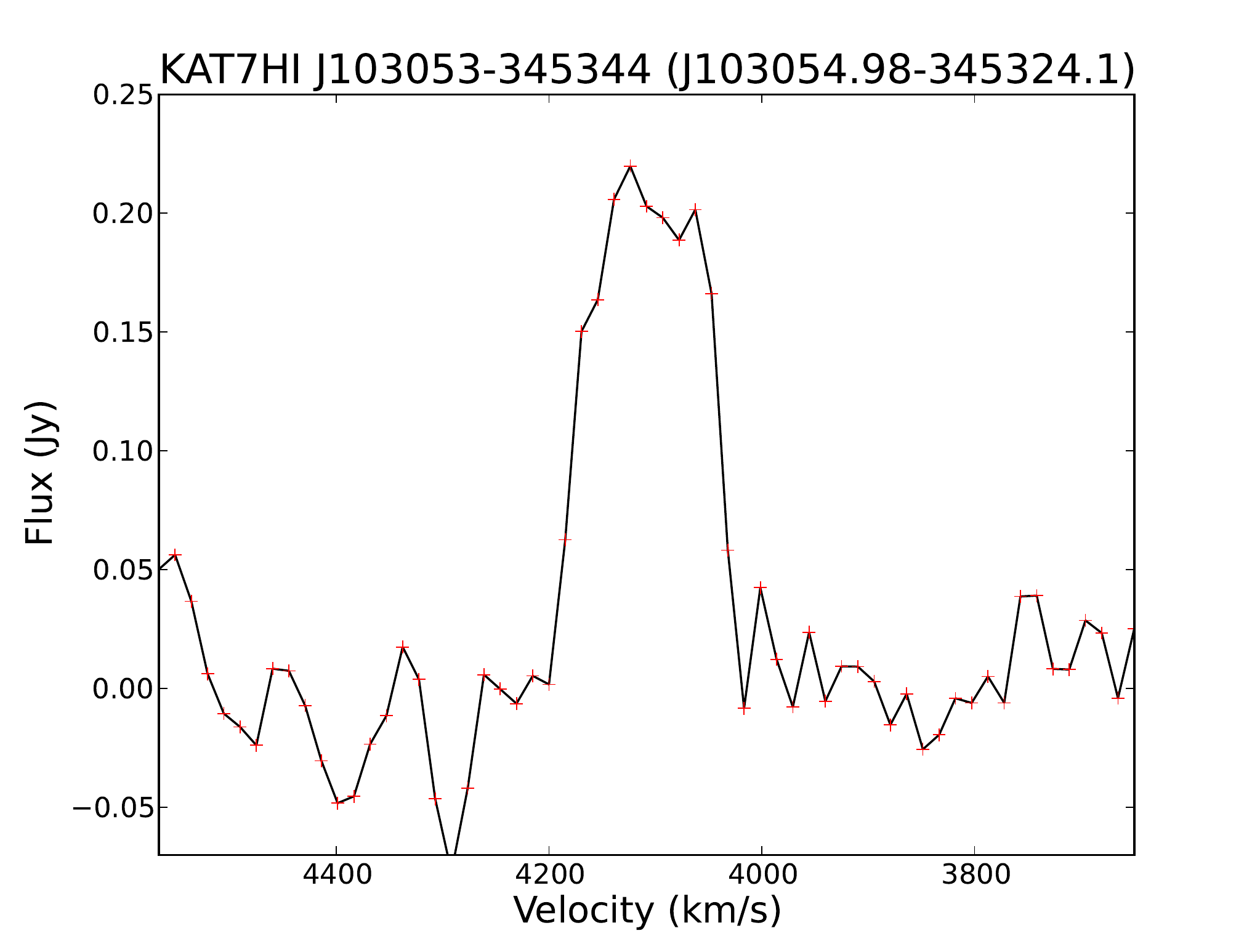} \\
\includegraphics[width=2.3in,trim=20 0 0 0]{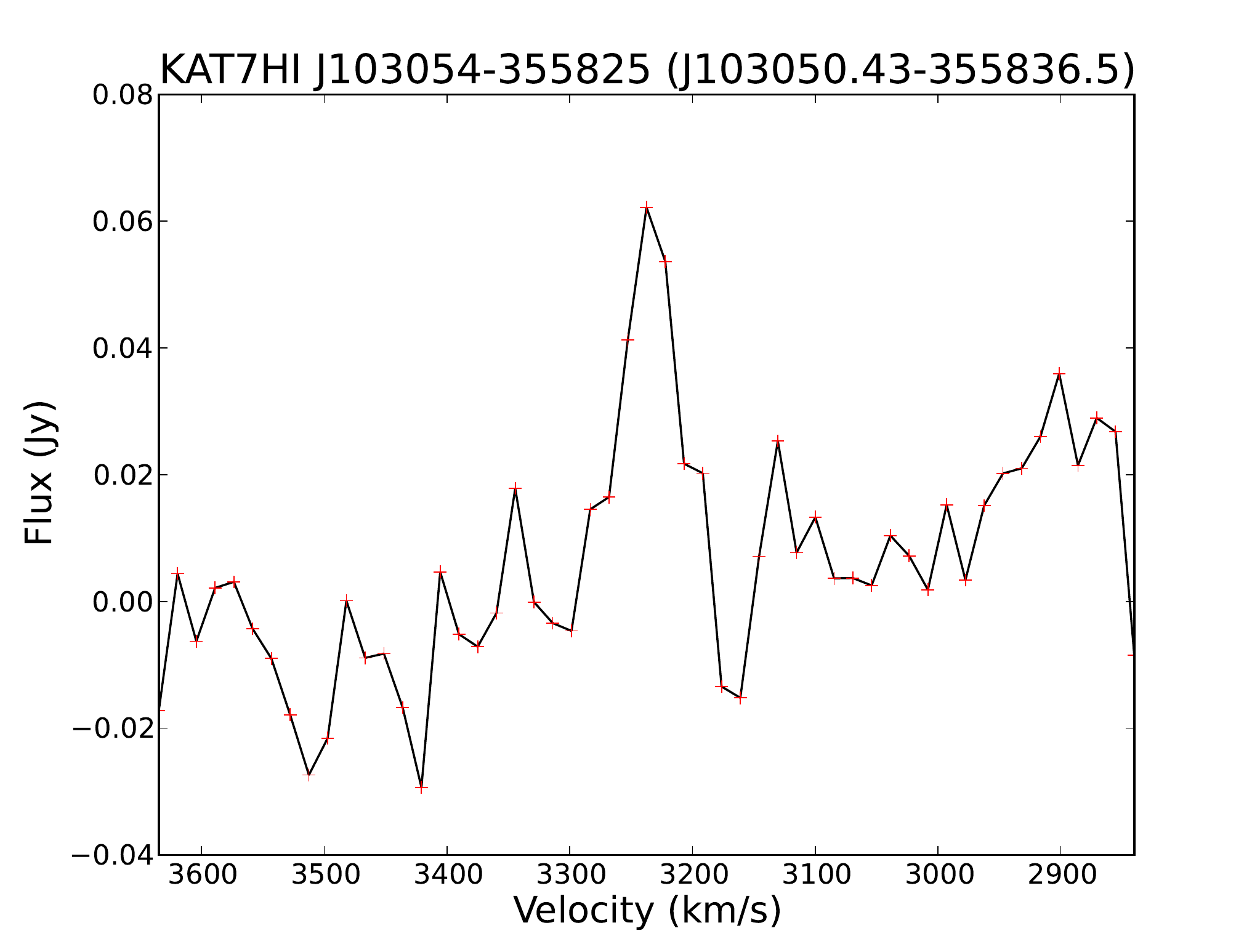} & \includegraphics[width=2.3in,trim=20 0 0 0]{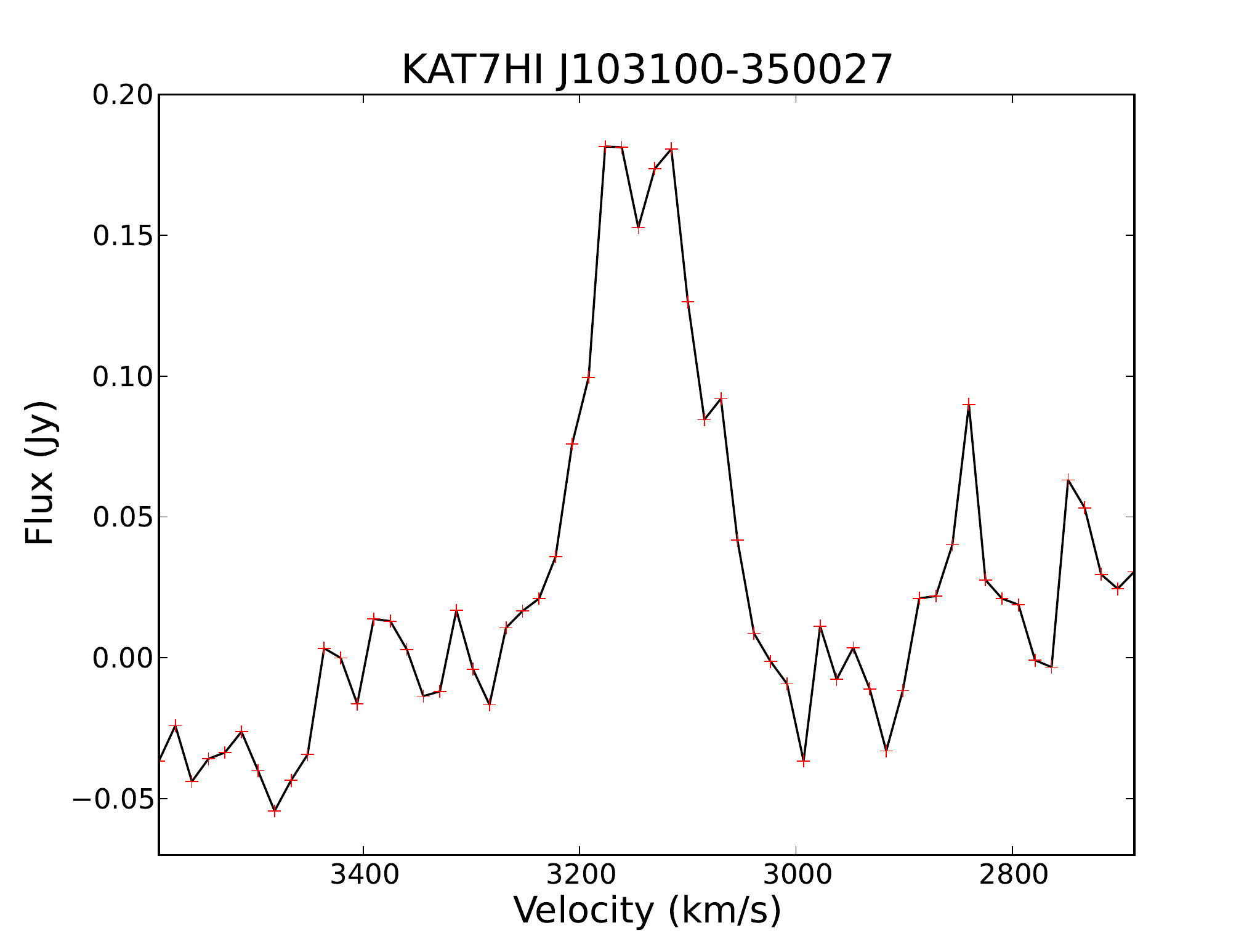} & \includegraphics[width=2.3in,trim=20 0 0 0]{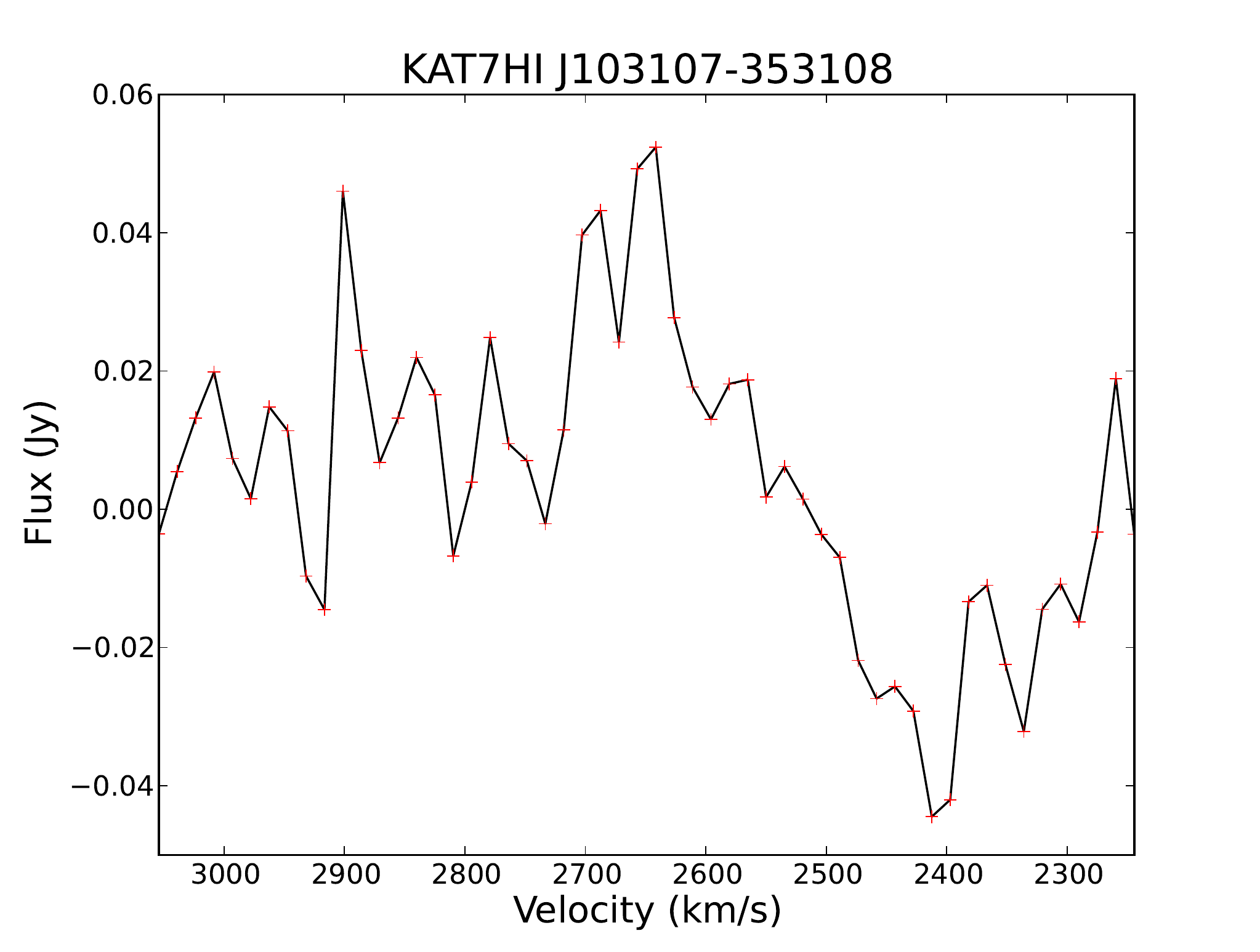} \\
\includegraphics[width=2.3in,trim=20 0 0 0]{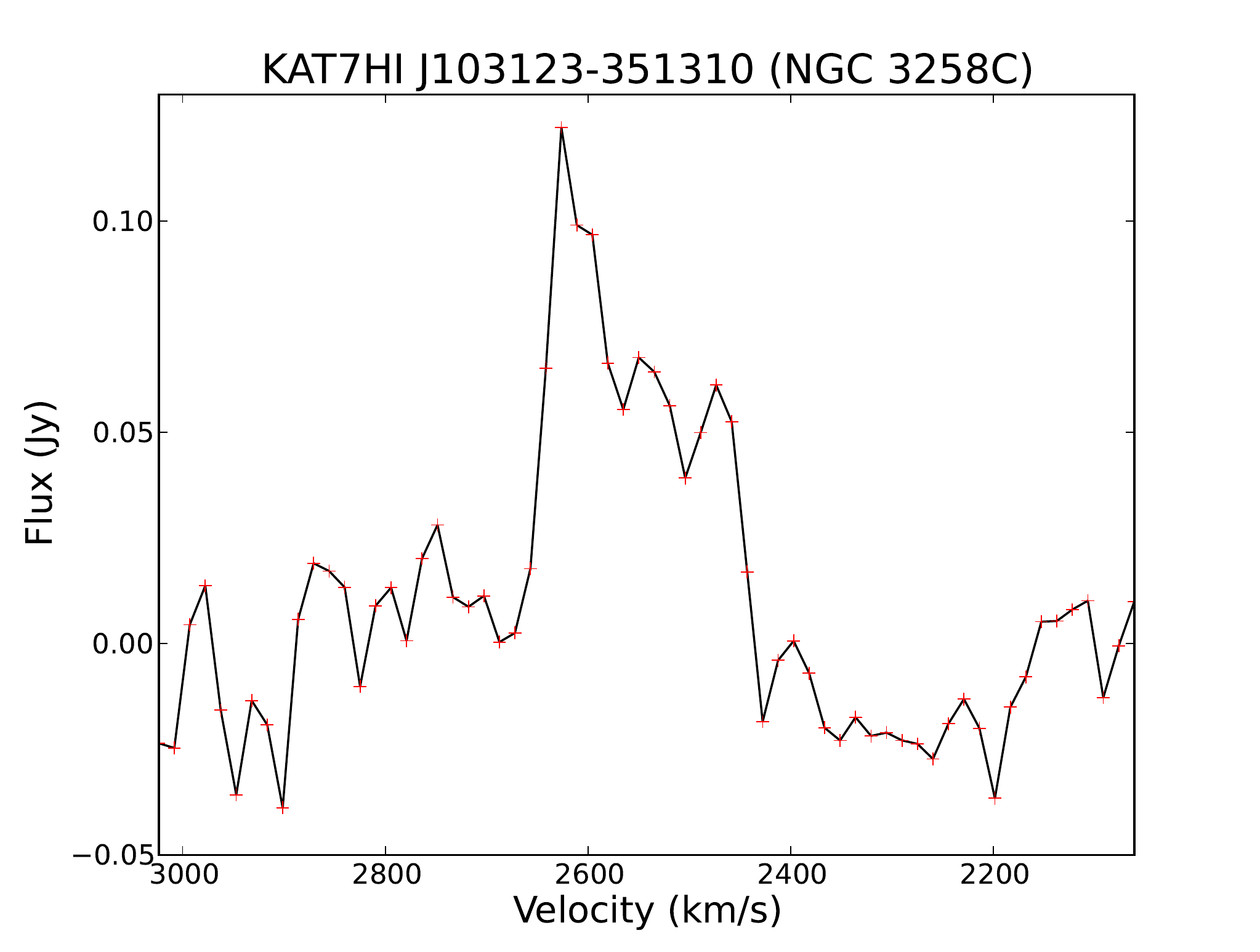} & \includegraphics[width=2.3in,trim=20 0 0 0]{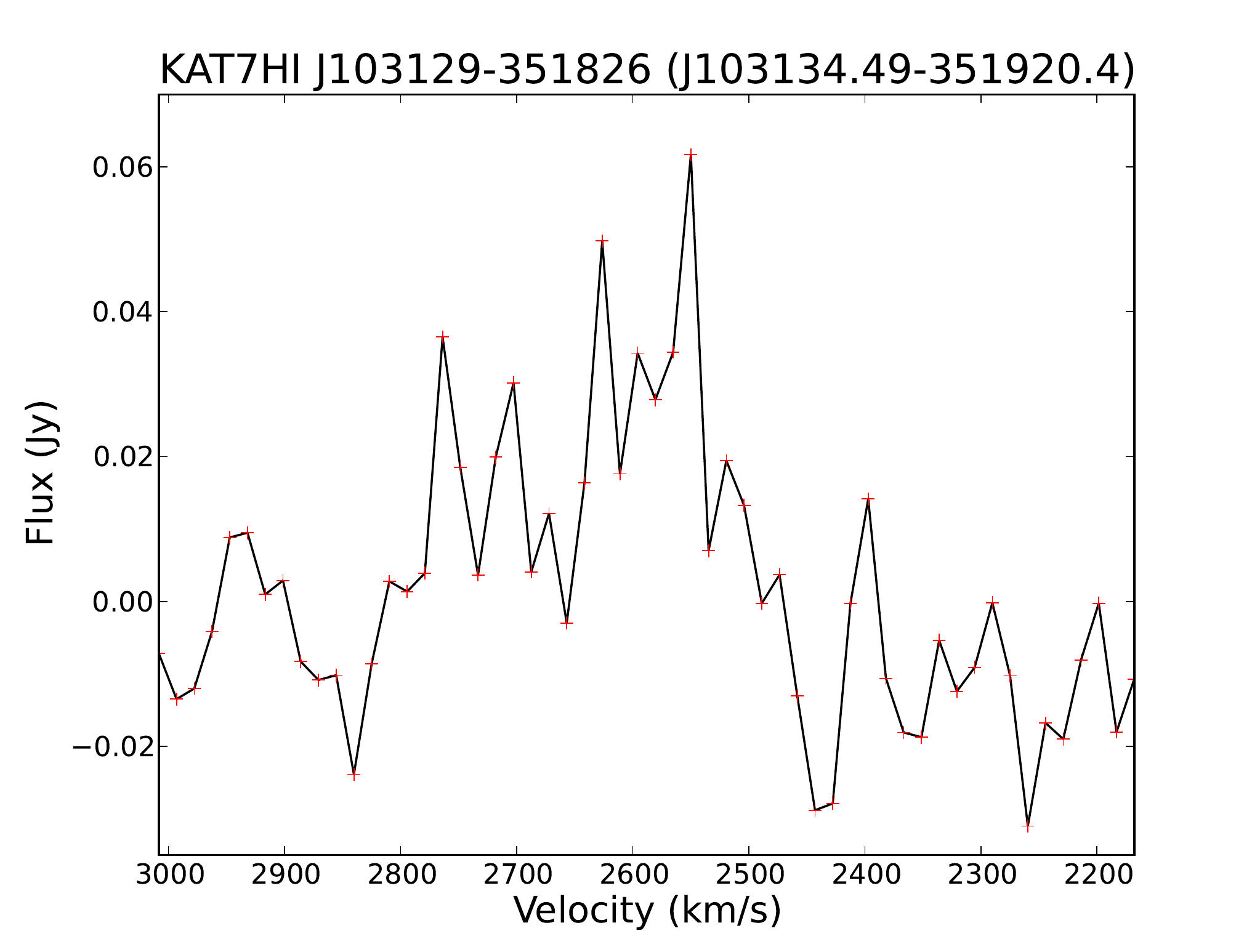} & \includegraphics[width=2.3in,trim=20 0 0 0]{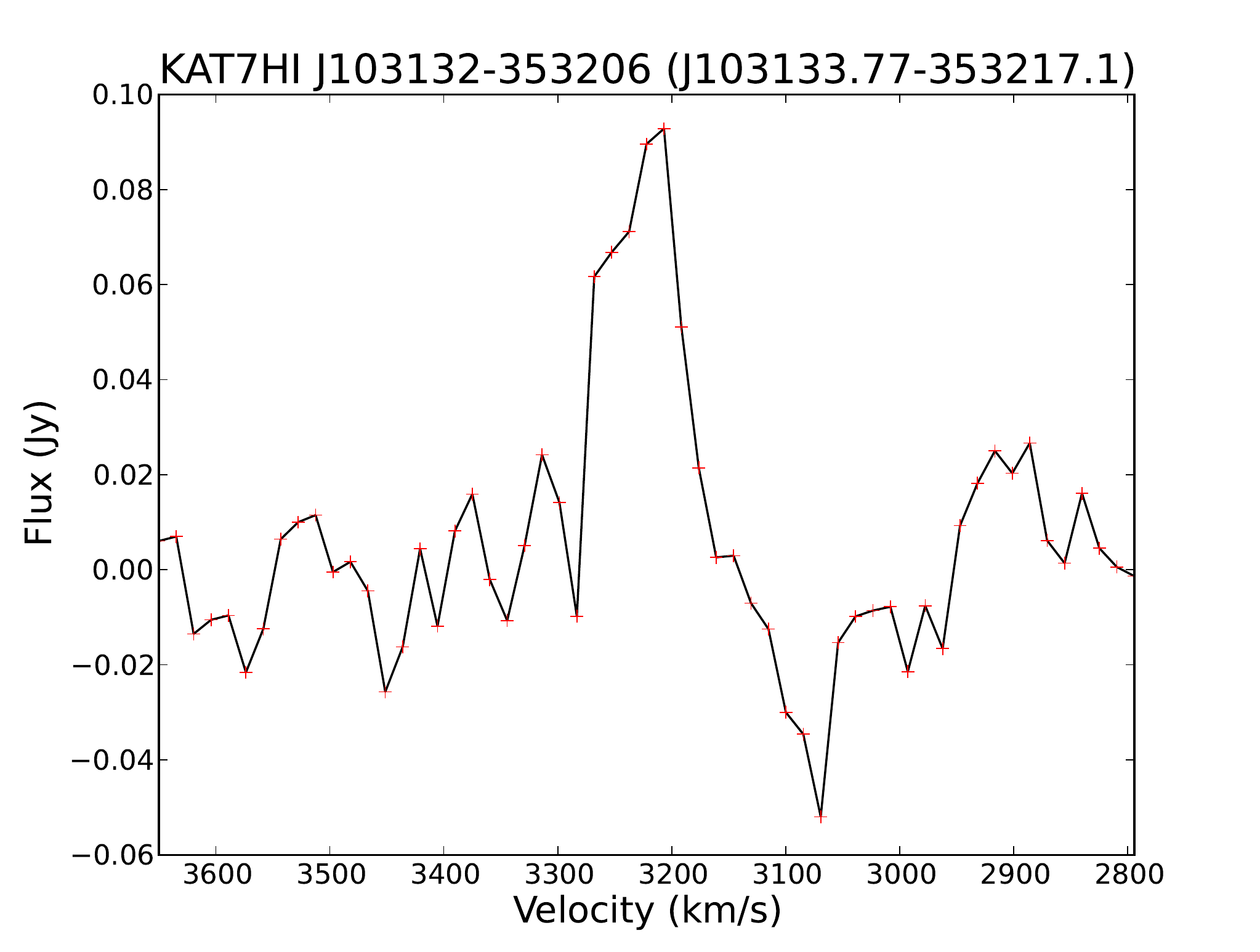} \\
\includegraphics[width=2.3in,trim=20 0 0 0]{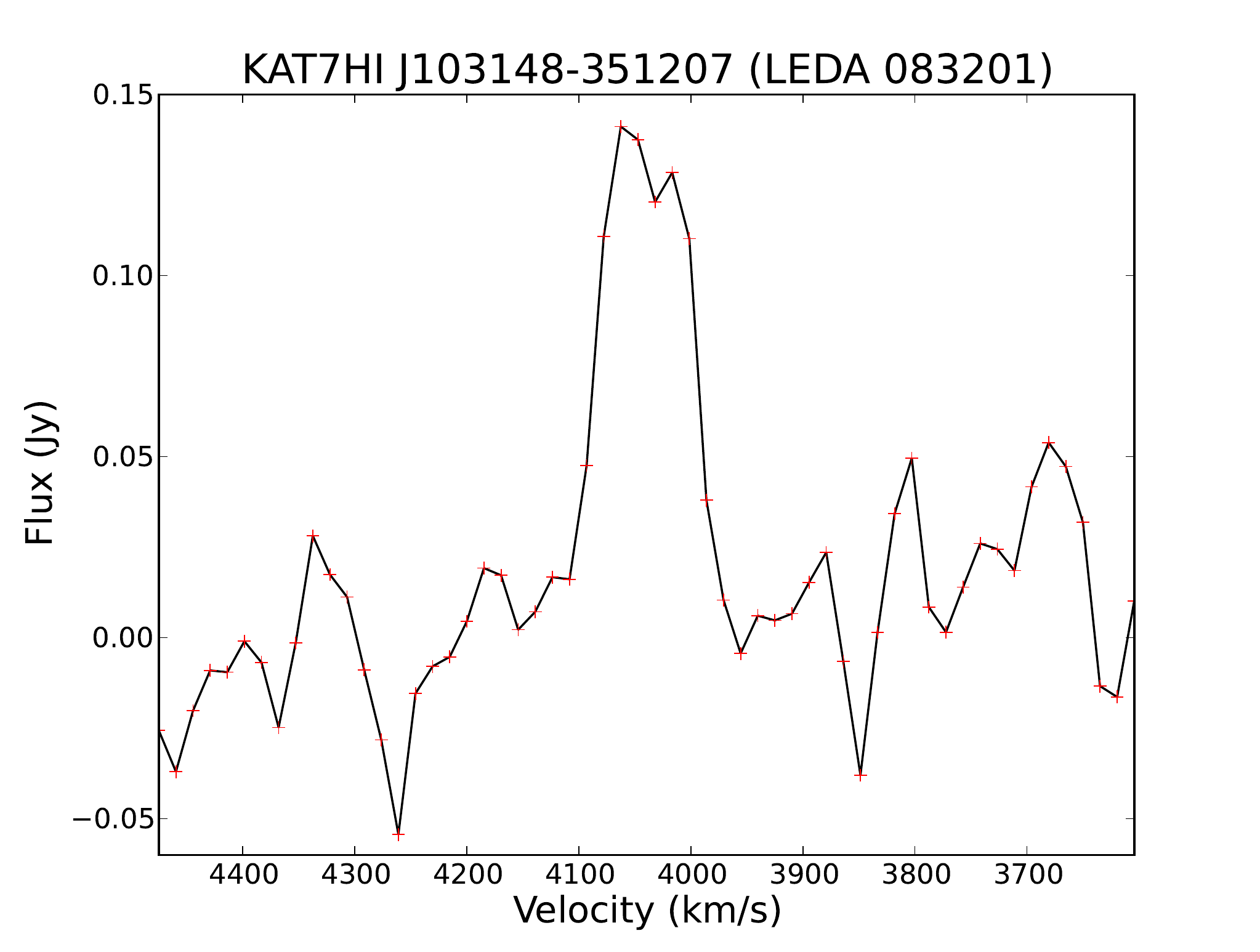} & \includegraphics[width=2.3in,trim=20 0 0 0]{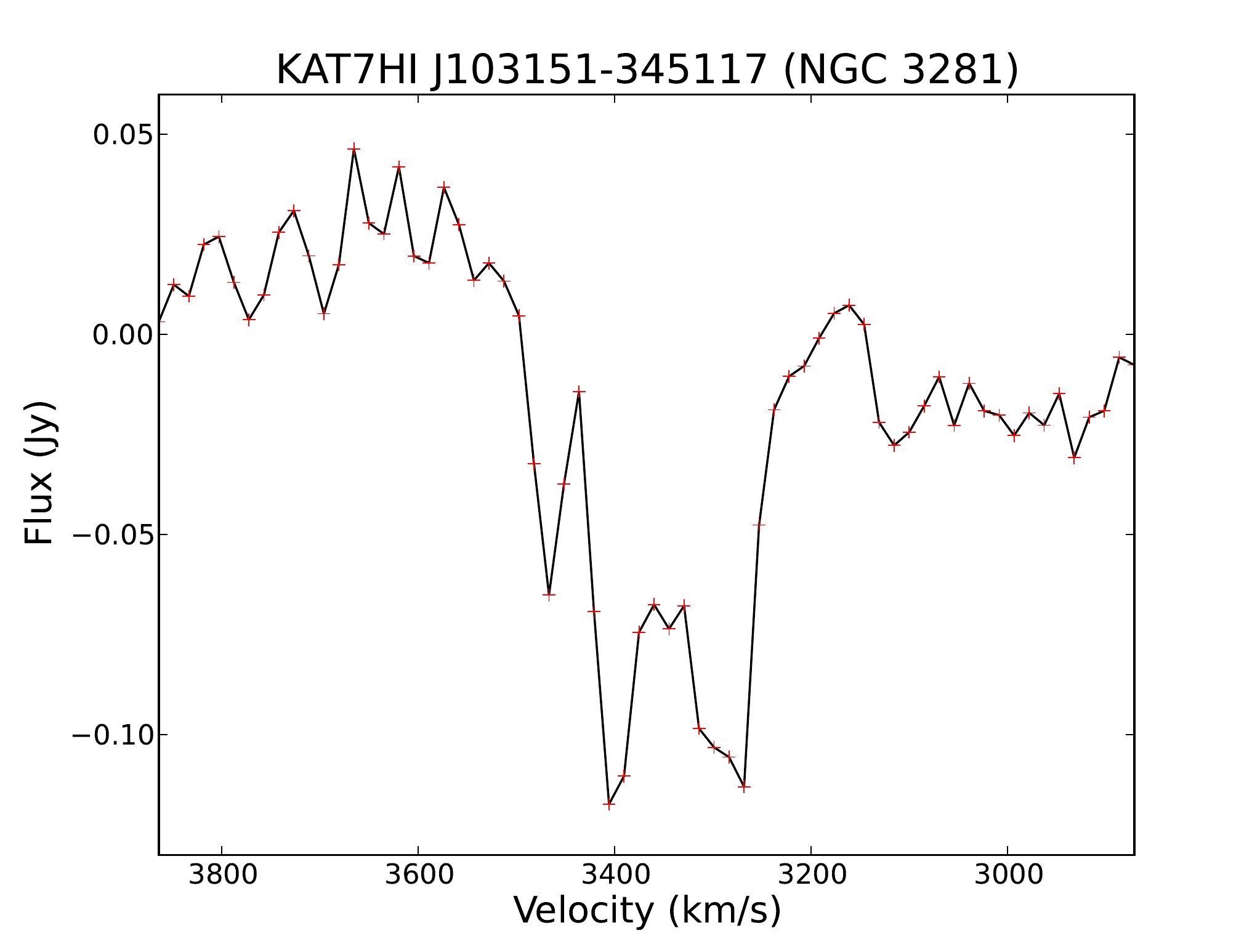} & \includegraphics[width=2.3in,trim=20 0 0 0]{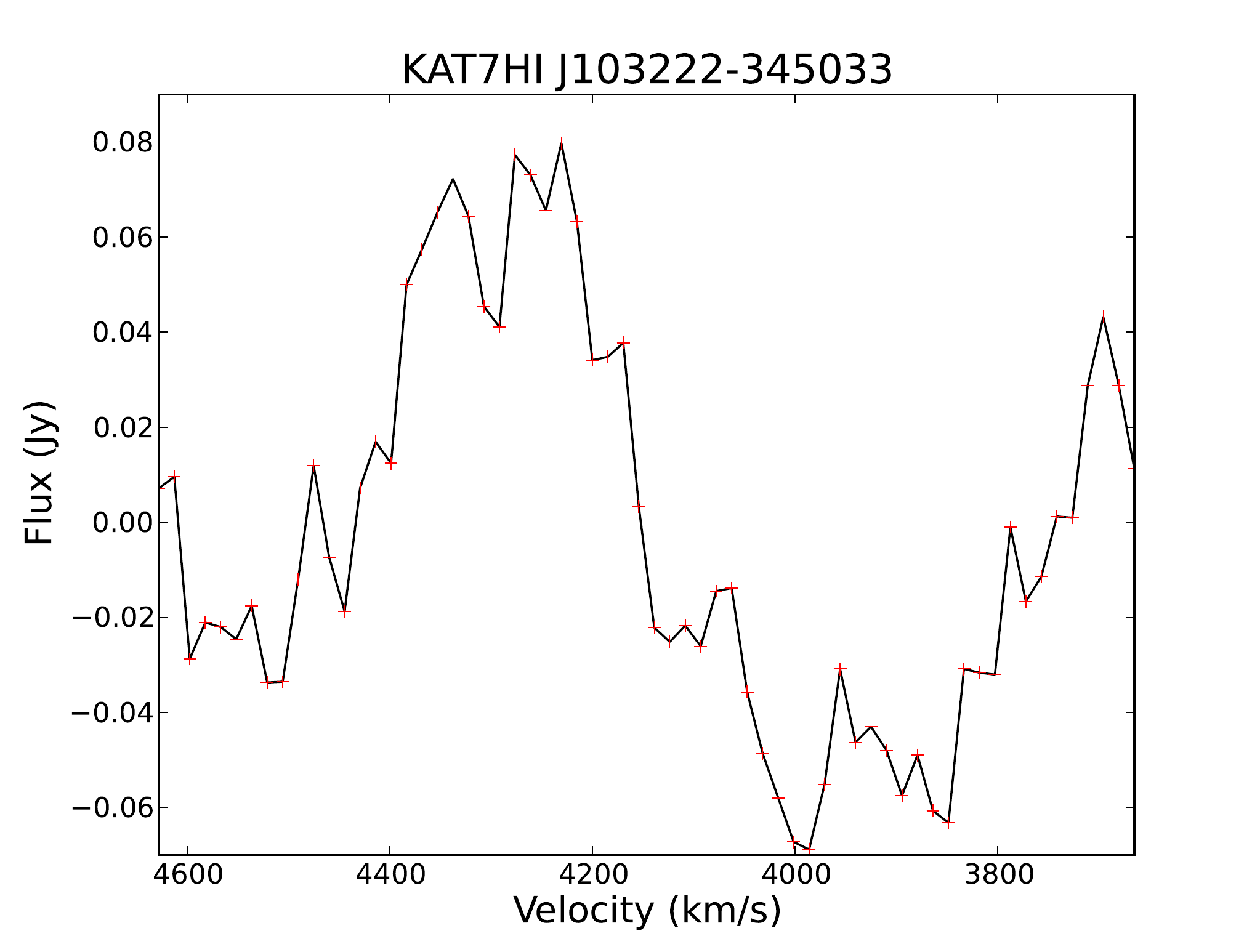}
\end{array}$
\end{center}
\end{figure*}

\begin{figure*}
\begin{center}$
\begin{array}{ccc}
\includegraphics[width=2.3in,trim=20 0 0 0]{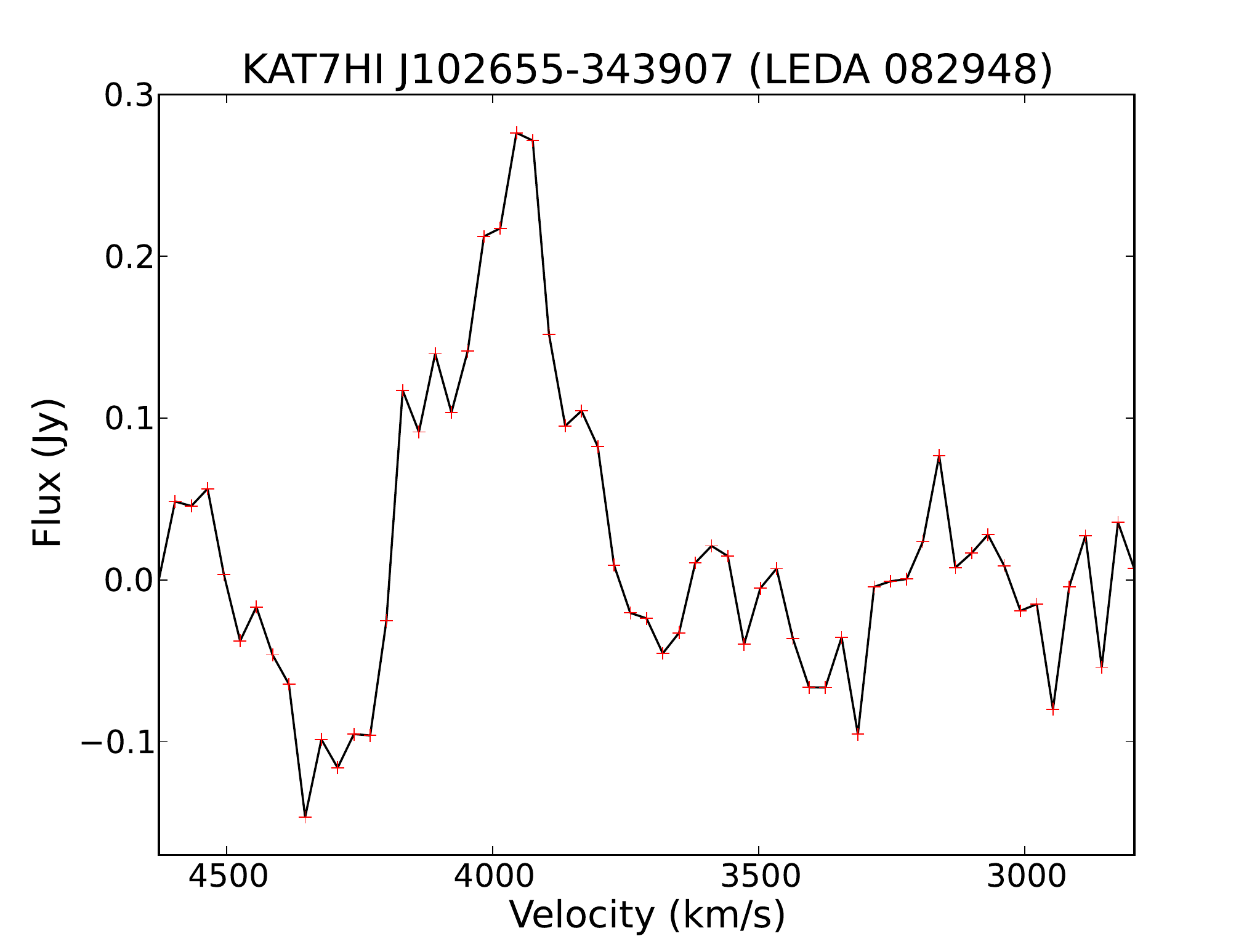} & \includegraphics[width=2.3in,trim=20 0 0 0]{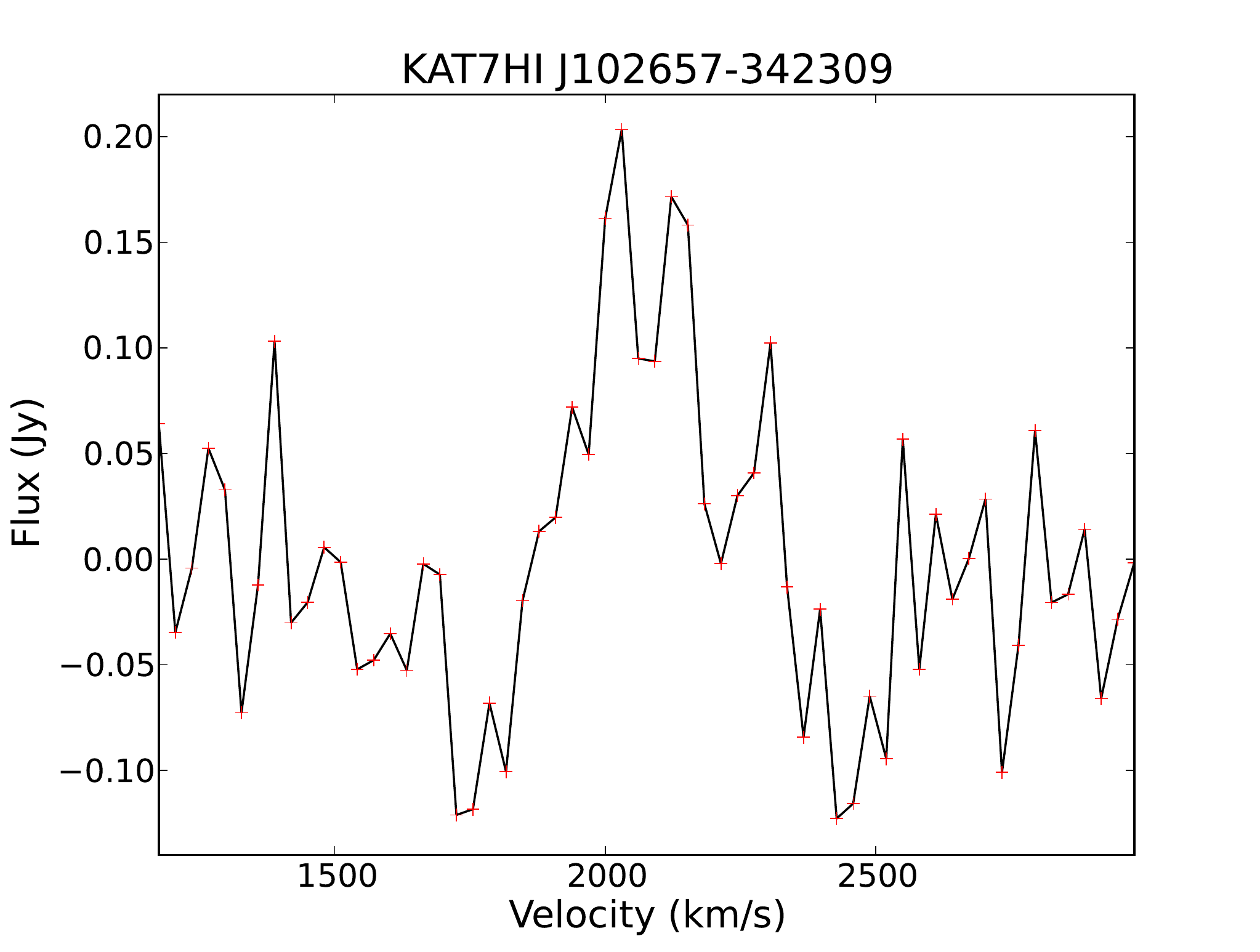} & \includegraphics[width=2.3in,trim=20 0 0 0]{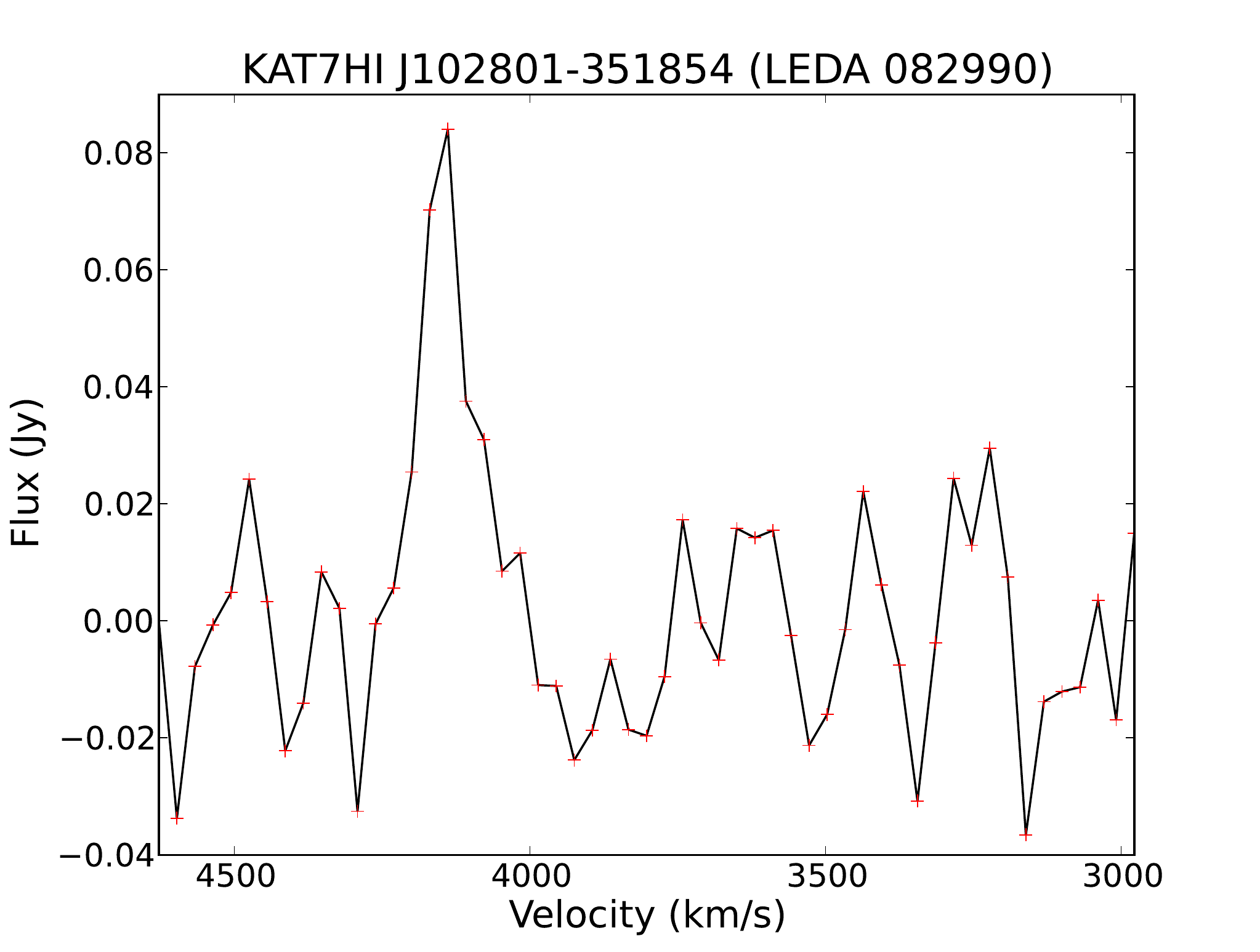} \\
\includegraphics[width=2.3in,trim=20 0 0 0]{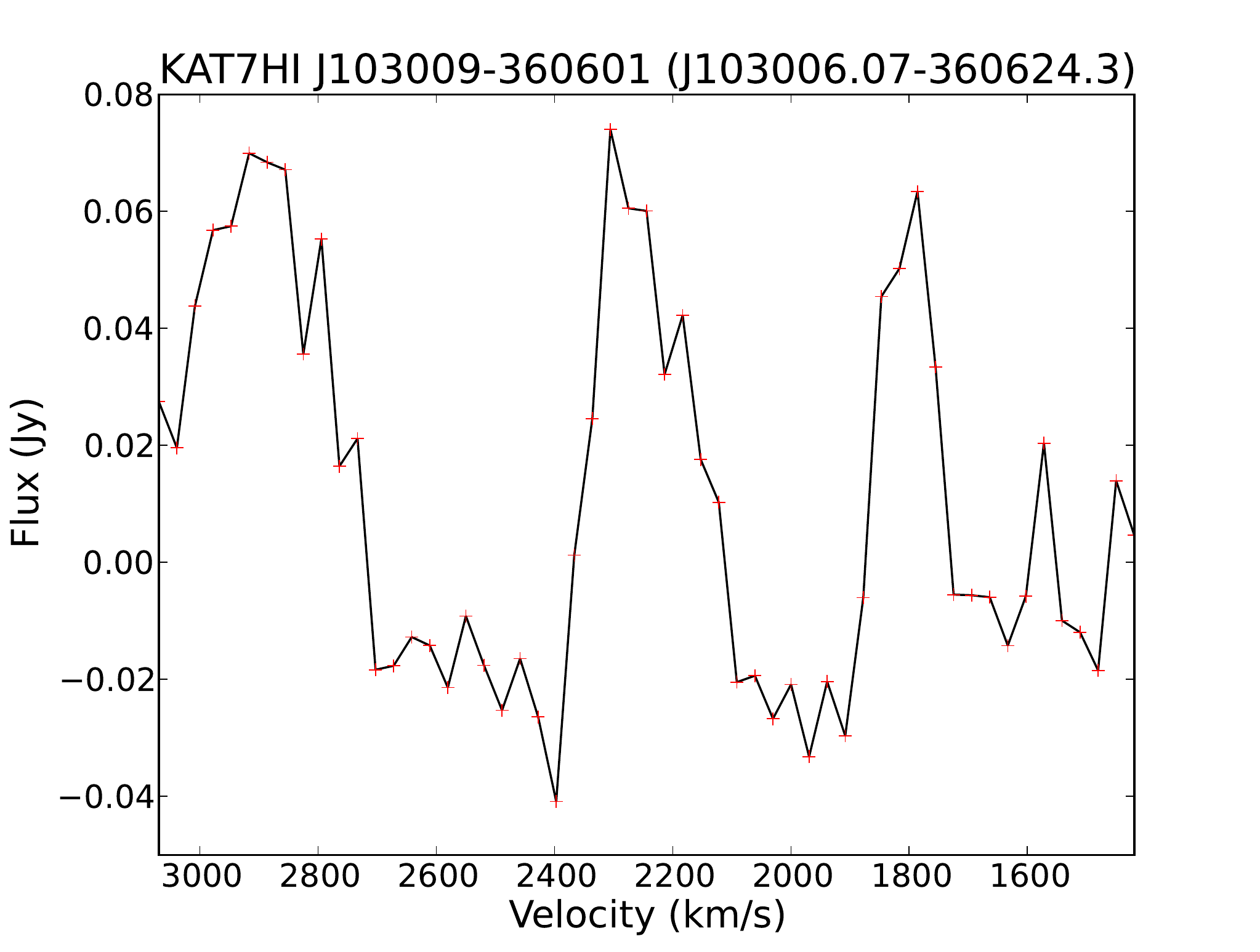} & \includegraphics[width=2.3in,trim=20 0 0 0]{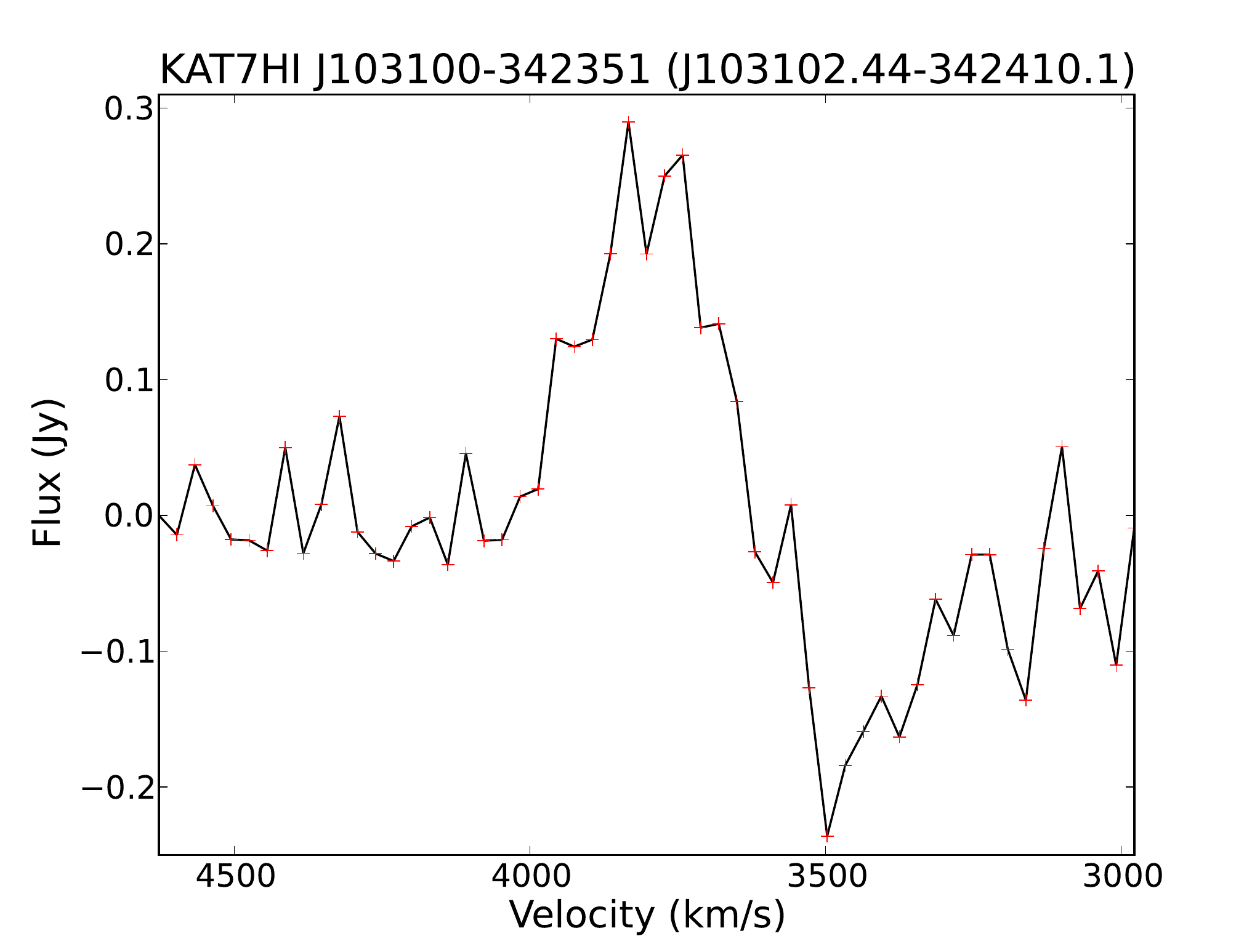} & \includegraphics[width=2.3in,trim=20 0 0 0]{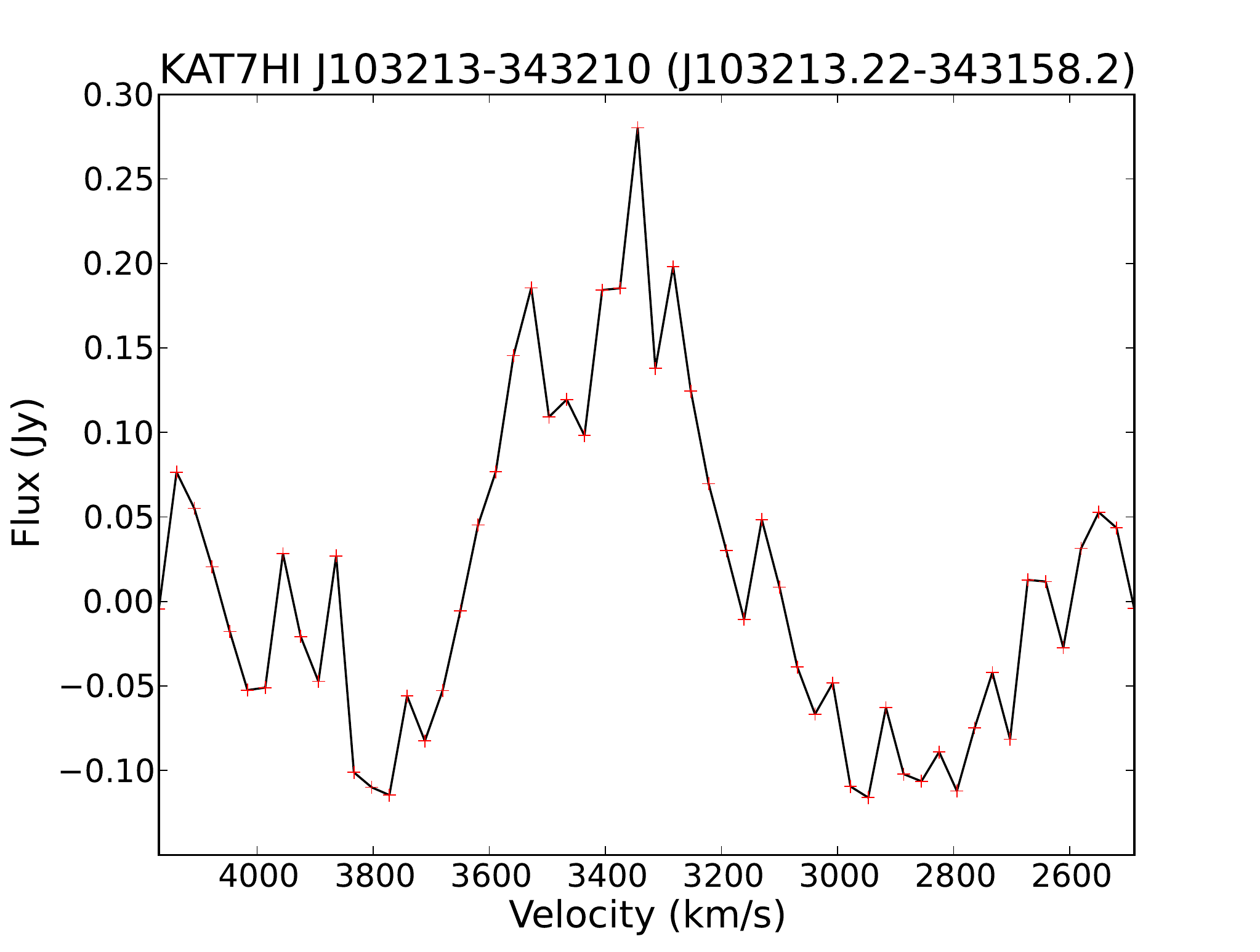} \\
\includegraphics[width=2.3in,trim=20 0 0 0]{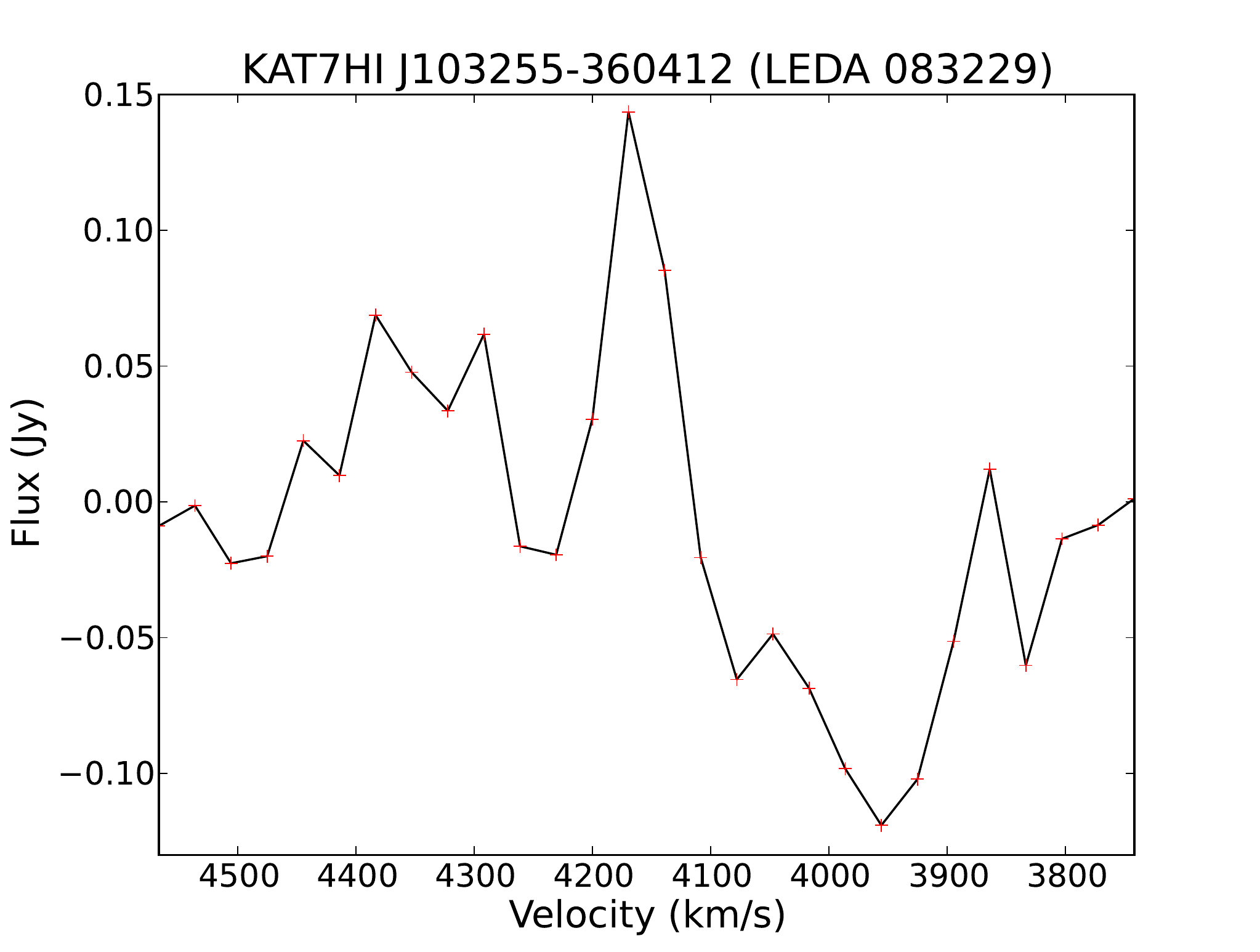} & & \\

\end{array}$
\end{center}
\caption{\hi\ profiles of the KAT-7 detections which were only detected in the 31\kms\ cube.  Note, in all but the last panel the x-axis range is twice that in the Figure \ref{hispecAppen}. Objects 31-37 appear in the same order as Table 2.}
\label{hispecAppen3}
\end{figure*}


\begin{thebibliography}{}
\bibitem[\protect\citeauthoryear{Adami et al.}{1998}]{Adami98} Adami, C., Biviano, A., \& Mazure, A.\ 1998, \aap, 331, 439 
\bibitem[\protect\citeauthoryear{Beers \& Geller}{1983}]{Beers83} Beers, T.~C., \& Geller, M.~J.\ 1983, \apj, 274, 491 
\bibitem[\protect\citeauthoryear{Beers et al.}{1991}]{Beers91} Beers, T.~C., Gebhardt, K., Forman, W., Huchra, J.~P., \& Jones, C.\ 1991, \aj, 102, 1581 
\bibitem[\protect\citeauthoryear{Berrier et al.}{2009}]{Berrier09} Berrier, J.~C., Stewart, K.~R., Bullock, J.~S., et al.\ 2009, \apj, 690, 1292
\bibitem[\protect\citeauthoryear{Binggeli et al.}{1987}]{Binggeli87} Binggeli, B., Tammann, G.~A., \& Sandage, A.\ 1987, \aj, 94, 251 
\bibitem[\protect\citeauthoryear{Bird}{1994}]{Bird94} Bird, C.~M.\ 1994, \aj, 107, 1637 
\bibitem[\protect\citeauthoryear{Blanton \& Berlind}{2007}]{Blanton07} Blanton, M.~R., \& Berlind, A.~A.\ 2007, \apj, 664, 791 
\bibitem[\protect\citeauthoryear{Bravo-Alfaro et al.}{2000}]{BravoAlfaro00} Bravo-Alfaro, H., Cayatte, V., van Gorkom, J.~H., \& Balkowski, C.\ 2000, \aj, 119, 580 
\bibitem[\protect\citeauthoryear{Brown et al.}{2014}]{Brown14} Brown, M.~J.~I., Jarrett, T.~H., \& Cluver, M.~E.\ 2014, arXiv:1411.5444 
\bibitem[\protect\citeauthoryear{Caldwell et al.}{1993}]{Caldwell93} Caldwell, N., Rose, J.~A., Sharples, R.~M., Ellis, R.~S., \& Bower, R.~G.\ 1993, \aj, 106, 473 
\bibitem[\protect\citeauthoryear{Carignan et al.}{2013}]{Carignan13} Carignan, C., Frank, B.~S., Hess, K.~M., et al.\ 2013, \aj, 146, 48 
\bibitem[\protect\citeauthoryear{Casoli et al.}{1991}]{Casoli91} Casoli, F., Boisse, P., Combes, F., \& Dupraz, C.\ 1991, \aap, 249, 359 
\bibitem[\protect\citeauthoryear{Chung et al.}{2009}]{Chung09} Chung, A., van Gorkom, J.~H., Kenney, J.~D.~P., Crowl, H., \& Vollmer, B.\ 2009, \aj, 138, 1741 
\bibitem[\protect\citeauthoryear{Cluver et al.}{2014}]{Cluver14} Cluver, M.~E., Jarrett, T.~H., Hopkins, A.~M., et al.\ 2014, \apj, 782, 90
\bibitem[\protect\citeauthoryear{Cohen et al.}{2014}]{Cohen14} Cohen, S.~A., Hickox, R.~C., Wegner, G.~A., Einasto, M., \& Vennik, J.\ 2014, \apj, 783, 136 
\bibitem[\protect\citeauthoryear{Colless \& Dunn}{1996}]{Colless96} Colless, M., \& Dunn, A.~M.\ 1996, \apj, 458, 435 
\bibitem[\protect\citeauthoryear{Condon et al.}{1998}]{Condon98} Condon, J.~J., Cotton, W.~D., Greisen, E.~W., et al.\ 1998, \aj, 115, 1693
\bibitem[\protect\citeauthoryear{Conselice et al.}{2001}]{Conselice01} Conselice, C.~J., Gallagher, J.~S., III, \& Wyse, R.~F.~G.\ 2001, \apj, 559, 791  %dEs in Virgo, great dyn analysis!
\bibitem[\protect\citeauthoryear{Courtois et al.}{2013}]{Courtois13} Courtois, H.~M., Pomar{\`e}de, D., Tully, R.~B., Hoffman, Y., \& Courtois, D.\ 2013, \aj, 146, 69 
\bibitem[\protect\citeauthoryear{Crowl et al.}{2005}]{Crowl05} Crowl, H.~H., Kenney, J.~D.~P., van Gorkom, J.~H., \& Vollmer, B.\ 2005, \aj, 130, 65 
\bibitem[\protect\citeauthoryear{Curran \& Whiting}{2010}]{Curran10} Curran, S.~J., \& Whiting, M.~T.\ 2010, \apj, 712, 303 
\bibitem[\protect\citeauthoryear{Cutri et al.}{2013}]{Cutri13} Cutri, R.~M., Wright, E.~L., Conrow, T., et al.\ 2013, Explanatory Supplement to the All\textit{WISE} Data Release Products, by R.~M.~Cutri et al.~, 1 
\bibitem[\protect\citeauthoryear{De Lucia et al.}{2012}]{deLucia12} De Lucia, G., Weinmann, S., Poggianti, B.~M., Arag{\'o}n-Salamanca, A., \& Zaritsky, D.\ 2012, \mnras, 423, 1277 
\bibitem[\protect\citeauthoryear{de Vaucouleurs et al.}{1991}]{deVaucouleurs91} de Vaucouleurs, G., de Vaucouleurs, A., Corwin, H.~G., Jr., et al.\ 1991, Third Reference Catalogue of Bright Galaxies.~Volume I: Explanations and references.~Volume II: Data for galaxies between 0$^{h}$ and 12$^{h}$.~ Volume III: Data for galaxies between 12$^{h}$ and 24$^{h}$., by de Vaucouleurs, G.; de Vaucouleurs, A.; Corwin, H.~G., Jr.; Buta, R.~J.; Paturel, G.; Fouqu{\'e}, P..~Springer, New York, NY (USA), 1991, 2091 p., ISBN 0-387-97552-7, Price US\$ 198.00.~ISBN 3-540-97552-7, Price DM 448.00.~ISBN 0-387-97549-7 (Vol.~I), ISBN 0-387-97550-0 (Vol.~II), ISBN 0-387-97551-9 (Vol.~III).,  
\bibitem[\protect\citeauthoryear{Dirsch et al.}{2003}]{Dirsch03} Dirsch, B., Richtler, T., \& Bassino, L.~P.\ 2003, \aap, 408, 929 
\bibitem[\protect\citeauthoryear{Donoso et al.}{2012}]{Donoso12} Donoso, E., Yan, L., Tsai, C., et al.\ 2012, \apj, 748, 80 
\bibitem[\protect\citeauthoryear{Dressler \& Shectman}{1988}]{Dressler88} Dressler, A., \& Shectman, S.~A.\ 1988, \aj, 95, 985 
\bibitem[\protect\citeauthoryear{Fassbender et al.}{2011}]{Fassbender11} Fassbender, R., B{\"o}hringer, H., Nastasi, A., et al.\ 2011, New Journal of Physics, 13, 125014 
\bibitem[\protect\citeauthoryear{Ferguson \& Sandage}{1990}]{Ferguson90} Ferguson, H.~C., \& Sandage, A.\ 1990, \aj, 100, 1 
\bibitem[\protect\citeauthoryear{Gandhi et al.}{2014}]{Gandhi14} Gandhi, P., Yamada, S., Ricci, C., et al.\ 2014, arXiv:1408.4453 
\bibitem[\protect\citeauthoryear{Gavazzi et al.}{1999}]{Gavazzi99} Gavazzi, G., Boselli, A., Scodeggio, M., Pierini, D., \& Belsole, E.\ 1999, \mnras, 304, 595 
\bibitem[\protect\citeauthoryear{Gavazzi et al.}{2008}]{Gavazzi08} Gavazzi, G., Giovanelli, R., Haynes, M.~P., et al.\ 2008, \aap, 482, 43 
\bibitem[\protect\citeauthoryear{Gereb et al.}{2014}]{Gereb14} Gereb, K., Morganti, R., \& Oosterloo, T.\ 2014, arXiv:1407.1799 
\bibitem[\protect\citeauthoryear{Gettings et al.}{2012}]{Gettings12} Gettings, D.~P., Gonzalez, A.~H., Stanford, S.~A., et al.\ 2012, \apjl, 759, LL23 
\bibitem[\protect\citeauthoryear{Huang et al.}{2012}]{Huang12} Huang, S., Haynes, M.~P., Giovanelli, R., \& Brinchmann, J.\ 2012, \apj, 756, 113 
\bibitem[\protect\citeauthoryear{Haynes \& Giovanelli}{1984}]{Haynes84} Haynes, M.~P., \& Giovanelli, R.\ 1984, \aj, 89, 758 
\bibitem[\protect\citeauthoryear{Heisler et al.}{1985}]{Heisler85} Heisler, J., Tremaine, S., \& Bahcall, J.~N.\ 1985, \apj, 298, 8 
\bibitem[\protect\citeauthoryear{Hern{\'a}ndez-Fern{\'a}ndez et al.}{2014}]{HernandezFernandez14} Hern{\'a}ndez-Fern{\'a}ndez, J.~D., Haines, C.~P., Diaferio, A., et al.\ 2014, \mnras, 438, 2186 
\bibitem[\protect\citeauthoryear{Hess}{2011}]{Hess11} Hess, K.~M.\ 2011, Ph.D.~Thesis, Univ. of Wisconsin-Madison
\bibitem[\protect\citeauthoryear{Hess \& Wilcots}{2013}]{Hess13} Hess, K.~M., \& Wilcots, E.~M.\ 2013, \aj, 146, 124 
\bibitem[\protect\citeauthoryear{Holt et al.}{2006}]{Holt06} Holt, J., Tadhunter, C., Morganti, R., et al.\ 2006, \mnras, 370, 1633 
\bibitem[\protect\citeauthoryear{Hou et al.}{2012}]{Hou12} Hou, A., Parker, L.~C., Wilman, D.~J., et al.\ 2012, \mnras, 421, 3594 
\bibitem[\protect\citeauthoryear{Hou et al.}{2014}]{Hou14} Hou, A., Parker, L.~C., \& Harris, W.~E.\ 2014, \mnras, 442, 406 
\bibitem[\protect\citeauthoryear{Hopp \& Materne}{1985}]{Hopp85} Hopp, U., \& Materne, J.\ 1985, \aaps, 61, 93 
\bibitem[\protect\citeauthoryear{Jaff{\'e} et al.}{2012}]{Jaffe12} Jaff{\'e}, Y.~L., Poggianti, B.~M., Verheijen, M.~A.~W., Deshev, B.~Z., \& van Gorkom, J.~H.\ 2012, \apjl, 756, LL28 
\bibitem[\protect\citeauthoryear{Jaff{\'e} et al.}{2015}]{Jaffe15} Jaff{\'e}, Y.~L., Smith, R., Candlish, G.~N., et al.\ 2015, \mnras, 448, 1715 
\bibitem[\protect\citeauthoryear{Jarrett et al.}{2011}]{Jarrett11} Jarrett, T.~H., Cohen, M., Masci, F., et al.\ 2011, \apj, 735, 112 
\bibitem[\protect\citeauthoryear{Jarrett et al.}{2012}]{Jarrett12} Jarrett, T.~H., Masci, F., Tsai, C.~W., et al.\ 2012, \aj, 144, 68 
\bibitem[\protect\citeauthoryear{Jarrett et al.}{2013}]{Jarrett13} Jarrett, T.~H., Masci, F., Tsai, C.~W., et al.\ 2013, \aj, 145, 6 
\bibitem[\protect\citeauthoryear{Jones et al.}{2004}]{Jones04} Jones, D.~H., Saunders, W., Colless, M., et al.\ 2004, \mnras, 355, 747 
\bibitem[\protect\citeauthoryear{Jones et al.}{2009}]{Jones09} Jones, D.~H., Read, M.~A., Saunders, W., et al.\ 2009, \mnras, 399, 683 
\bibitem[\protect\citeauthoryear{Kauffmann et al.}{2004}]{Kauffmann04} Kauffmann, G., White, S.~D.~M., Heckman, T.~M., et al.\ 2004, \mnras, 353, 713 
\bibitem[\protect\citeauthoryear{Kenney \& Young}{1989}]{Kenney89} Kenney, J.~D.~P., \& Young, J.~S.\ 1989, \apj, 344, 171 
\bibitem[\protect\citeauthoryear{Kenney et al.}{2004}]{Kenney04} Kenney, J.~D.~P., van Gorkom, J.~H., \& Vollmer, B.\ 2004, \aj, 127, 3361
\bibitem[\protect\citeauthoryear{Kraan-Korteweg et al.}{2002}]{KraanKorteweg02} Kraan-Korteweg, R.~C., Henning, P.~A., \& Schr{\"o}der, A.~C.\ 2002, \aap, 391, 887 
\bibitem[\protect\citeauthoryear{Lane et al.}{2007}]{Lane07} Lane, K.~P., Gray, M.~E., Arag{\'o}n-Salamanca, A., Wolf, C., \& Meisenheimer, K.\ 2007, \mnras, 378, 716
\bibitem[\protect\citeauthoryear{Lauberts}{1982}]{Lauberts82} Lauberts, A.\ 1982, Garching: European Southern Observatory (ESO), 1982
\bibitem[\protect\citeauthoryear{Maccagni et al.}{2014}]{Maccagni14} Maccagni, F.~M., Morganti, R., Oosterloo, T.~A., \& Mahony, E.~K.\ 2014, arXiv:1409.0566 
\bibitem[\protect\citeauthoryear{Maddox et al.}{2015}]{Maddox15} Maddox, N., Hess, K.~M., Obreschkow, D., Jarvis, M.~J., \& Blyth, S.-L.\ 2015, \mnras, 447, 1610 
\bibitem[\protect\citeauthoryear{Mahony et al.}{2013}]{Mahony13} Mahony, E.~K., Morganti, R., Emonts, B.~H.~C., Oosterloo, T.~A., \& Tadhunter, C.\ 2013, \mnras, 435, L58 
\bibitem[\protect\citeauthoryear{Martin}{1976}]{Martin76} Martin, W.~L.\ 1976, \mnras, 175, 633
\bibitem[\protect\citeauthoryear{McGee et al.}{2009}]{McGee09} McGee, S.~L., Balogh, M.~L., Bower, R.~G., Font, A.~S., \& McCarthy, I.~G.\ 2009, \mnras, 400, 937 
\bibitem[\protect\citeauthoryear{McGlynn et al.}{1996}]{McGlynn96} McGlynn, T., Scollick, K., White, N., SkyView: The Multi-Wavelength Sky on the Internet, McLean, B.~J.\ et al., New Horizons from Multi-Wavelength Sky Surveys, Kluwer Academic Publishers, 1996, IAU Symposium No.\ 179, 465
\bibitem[\protect\citeauthoryear{McMullin et al.}{2007}]{McMullin07} McMullin, J.~P., Waters, B., Schiebel, D., Young, W., \& Golap, K.\ 2007, Astronomical Data Analysis Software and Systems XVI, 376, 127 
\bibitem[\protect\citeauthoryear{Morganti et al.}{2005}]{Morganti05} Morganti, R., Tadhunter, C.~N., \& Oosterloo, T.~A.\ 2005, \aap, 444, L9 
\bibitem[\protect\citeauthoryear{Morganti et al.}{2009}]{Morganti09} Morganti, R., Peck, A.~B., Oosterloo, T.~A., et al.\ 2009, \aap, 505, 559 
\bibitem[\protect\citeauthoryear{Nakazawa et al.}{2000}]{Nakazawa00} Nakazawa, K., Makishima, K., Fukazawa, Y., \& Tamura, T.\ 2000, \pasj, 52, 623 
\bibitem[\protect\citeauthoryear{Oguri \& Lee}{2004}]{Oguri04} Oguri, M., \& Lee, J.\ 2004, \mnras, 355, 120 
\bibitem[\protect\citeauthoryear{Peck et al.}{1999}]{Peck99} Peck, A.~B., Taylor, G.~B., \& Conway, J.~E.\ 1999, \apj, 521, 103 
\bibitem[\protect\citeauthoryear{Pedersen et al.}{1997}]{Pedersen97} Pedersen, K., Yoshii, Y., \& Sommer-Larsen, J.\ 1997, \apjl, 485, L17 
\bibitem[\protect\citeauthoryear{Pranger et al.}{2013}]{Pranger13} Pranger, F., B{\"o}hm, A., Ferrari, C., et al.\ 2013, \aap, 557, A62 
\bibitem[\protect\citeauthoryear{Puche \& Carignan}{1988}]{Puche88} Puche, D., \& Carignan, C.\ 1988, \aj, 95, 1025 
\bibitem[\protect\citeauthoryear{Radburn-Smith et al.}{2006}]{RadburnSmith06} Radburn-Smith, D.~J., Lucey, J.~R., Woudt, P.~A., Kraan-Korteweg, R.~C., \& Watson, F.~G.\ 2006, \mnras, 369, 1131 
\bibitem[\protect\citeauthoryear{Ramella et al.}{2007}]{Ramella07} Ramella, M., Biviano, A., Pisani, A., et al.\ 2007, \aap, 470, 39 
\bibitem[\protect\citeauthoryear{Rines et al.}{2005}]{Rines05} Rines, K., Geller, M.~J., Kurtz, M.~J., \& Diaferio, A.\ 2005, \aj, 130, 1482 
\bibitem[\protect\citeauthoryear{Rubin et al.}{1985}]{Rubin85} Rubin, V.~C., Burstein, D., Ford, W.~K., Jr., \& Thonnard, N.\ 1985, \apj, 289, 81 
\bibitem[\protect\citeauthoryear{Sales et al.}{2011}]{Sales11} Sales, D.~A., Pastoriza, M.~G., Riffel, R., et al.\ 2011, \apj, 738, 109 
\bibitem[\protect\citeauthoryear{Sandage}{1978}]{Sandage78} Sandage, A.\ 1978, \aj, 83, 904
\bibitem[\protect\citeauthoryear{Sandage \& Tammann}{1987}]{Sandage87} Sandage, A., \& Tammann, G.~A.\ 1987, Carnegie Institution of Washington Publication, Washington: Carnegie Institution, 1987, 2nd ed.,  
\bibitem[\protect\citeauthoryear{Schawinski et al.}{2014}]{Schawinski14} Schawinski, K., Urry, C.~M., Simmons, B.~D., et al.\ 2014, \mnras, 440, 889 
\bibitem[\protect\citeauthoryear{Serna \& Gerbal}{1996}]{Serna96} Serna, A., \& Gerbal, D.\ 1996, \aap, 309, 65 
\bibitem[\protect\citeauthoryear{Simpson}{1998}]{Simpson98} Simpson, C.\ 1998, \apj, 509, 653 
\bibitem[\protect\citeauthoryear{Smith Castelli et al.}{2008}]{SmithCastelli08} Smith Castelli, A.~V., Bassino, L.~P., Richtler, T., et al.\ 2008, \mnras, 386, 2311 
\bibitem[\protect\citeauthoryear{Smith Castelli et al.}{2012}]{SmithCastelli12} Smith Castelli, A.~V., Cellone, S.~A., Faifer, F.~R., et al.\ 2012, \mnras, 419, 2472 
\bibitem[\protect\citeauthoryear{Solanes et al.}{2001}]{Solanes01} Solanes, J.~M., Manrique, A., Garc{\'{\i}}a-G{\'o}mez, C., et al.\ 2001, \apj, 548, 97
\bibitem[\protect\citeauthoryear{Scodeggio et al.}{1995}]{Scodeggio95} Scodeggio, M., Solanes, J.~M., Giovanelli, R., \& Haynes, M.~P.\ 1995, \apj, 444, 41 
\bibitem[\protect\citeauthoryear{Springel et al.}{2005}]{Springel05} Springel, V., White,  S.~D.~M., Jenkins, A., et al.\ 2005, \nat, 435, 629 
\bibitem[\protect\citeauthoryear{Stern et al.}{2012}]{Stern12} Stern, D., Assef, R.~J., Benford, D.~J., et al.\ 2012, \apj, 753, 30 
\bibitem[\protect\citeauthoryear{Storchi-Bergmann et al.}{1992}]{StorchiBergmann92} Storchi-Bergmann, T., Wilson, A.~S., \& Baldwin, J.~A.\ 1992, \apj, 396, 45 
\bibitem[\protect\citeauthoryear{Theureau et al.}{1998}]{Theureau98} Theureau, G., Bottinelli, L., Coudreau-Durand, N., et al.\ 1998, \aaps, 130, 333 
\bibitem[\protect\citeauthoryear{Ulvestad \& Wilson}{1989}]{Ulvestad89} Ulvestad, J.~S., \& Wilson, A.~S.\ 1989, \apj, 343, 659 
\bibitem[\protect\citeauthoryear{Vaduvescu et al.}{2014}]{Vaduvescu14} Vaduvescu, O., Kehrig, C., Bassino, L.~P., Smith Castelli, A.~V., \& Calder{\'o}n, J.~P.\ 2014, \aap, 563, AA118 
\bibitem[\protect\citeauthoryear{van Gorkom et al.}{1989}]{vanGorkom89} van Gorkom, J.~H., Knapp, G.~R., Ekers, R.~D., et al.\ 1989, \aj, 97, 708 
\bibitem[\protect\citeauthoryear{Vignali \& Comastri}{2002}]{Vignali02} Vignali, C., \& Comastri, A.\ 2002, \aap, 381, 834 
\bibitem[\protect\citeauthoryear{Wetzel et al.}{2012}]{Wetzel12} Wetzel, A.~R., Tinker, J.~L., \& Conroy, C.\ 2012, \mnras, 424, 232 
\bibitem[\protect\citeauthoryear{Winge et al.}{2000}]{Winge00} Winge, C., Storchi-Bergmann, T., Ward, M.~J., \& Wilson, A.~S.\ 2000, \mnras, 316, 1 
\bibitem[\protect\citeauthoryear{Wojtak \& {\L}okas}{2010}]{Wojtak10} Wojtak, R., \& {\L}okas, E.~L.\ 2010, \mnras, 408, 2442 
\bibitem[\protect\citeauthoryear{Wolfe \& Burbidge}{1975}]{Wolfe75} Wolfe, A.~M., \& Burbidge, G.~R.\ 1975, \apj, 200, 548 
\bibitem[\protect\citeauthoryear{Woo et al.}{2013}]{Woo13} Woo, J., Dekel, A., Faber, S.~M., et al.\ 2013, \mnras, 428, 3306 
\bibitem[\protect\citeauthoryear{Wright et al.}{2010}]{Wright10} Wright, E.~L., Eisenhardt, P.~R.~M., Mainzer, A.~K., et al.\ 2010, \aj, 140, 1868 
\bibitem[\protect\citeauthoryear{Zabludoff et al.}{1990}]{Zabludoff90} Zabludoff, A.~I., Huchra, J.~P., \& Geller, M.~J.\ 1990, \apjs, 74, 1 
\bibitem[\protect\citeauthoryear{Zabludoff \& Mulchaey}{1998}]{Zabludoff98} Zabludoff, A.~I., \& Mulchaey, J.~S.\ 1998, \apj, 496, 39 
\end{thebibliography}
\end{document}